\documentclass[preprint,3p]{elsarticle}
\usepackage{epsfig,amssymb,amsmath,natbib,listings,color}
\pdfoutput=1
\usepackage[toc,page]{appendix}
\usepackage{graphicx}
\usepackage{multicol}
\usepackage[outdir=./]{epstopdf}
\usepackage{psfrag}
\usepackage{amssymb}
\usepackage{amsthm}
\usepackage{lineno}
\usepackage{amsfonts}
\usepackage{amsmath}
\usepackage{caption}
\usepackage{subcaption}
\usepackage{bm}
\usepackage{framed}
\usepackage{color}
\usepackage{hyperref}
\usepackage[english]{algorithm2e} 
\usepackage[normalem]{ulem}
\usepackage{xcolor}
\hypersetup{colorlinks,breaklinks,
           linkcolor=blue,urlcolor=blue,
           anchorcolor=blue,citecolor=blue}
\usepackage{float}
\usepackage{comment}
\newtheorem{theorem}{Theorem}[section]

\newtheorem{remark}{Remark}
\graphicspath{{fig/}}

\newcommand{\lluis}[1]{\textcolor{black}{#1}}
\newcommand{\newlluis}[1]{\textcolor{black}{#1}}

\journal{arXiv}
\begin{document}
\begin{frontmatter}

\title{Bi-fidelity approximation for uncertainty quantification and sensitivity analysis of irradiated particle-laden turbulence}

\author[colorado1]{Hillary R. Fairbanks}
\ead{hillary.fairbanks@colorado.edu}
\author[stanford1]{Llu\'is Jofre}
\ead{jofre@stanford.edu}
\author[sandia1]{Gianluca Geraci}
\ead{ggeraci@sandia.gov}
\author[stanford1]{Gianluca Iaccarino}
\ead{jops@stanford.edu}
\author[colorado2]{Alireza Doostan\corref{cor1}}
\ead{alireza.doostan@colorado.edu}
\cortext[cor1]{Corresponding Author: Alireza Doostan}

\address[colorado1]{Applied Mathematics and Statistics, University of Colorado Boulder, Boulder, CO 80309, USA}
\address[stanford1]{Center for Turbulence Research, Stanford University, Stanford, CA 94305, USA}
\address[sandia1]{Center for Computing Research, Sandia National Laboratories, Albuquerque, NM 87185, USA}
\address[colorado2]{Smead Aerospace Engineering Sciences, University of Colorado Boulder, Boulder, CO 80309, USA}

\begin{abstract}

Efficiently performing predictive studies of irradiated particle-laden turbulent flows has the potential of providing significant contributions towards better understanding and optimizing, for example, concentrated solar power systems.
As there are many uncertainties inherent in such flows, uncertainty quantification is fundamental to improve the predictive capabilities of the numerical simulations. For large-scale, multi-physics problems exhibiting high-dimensional uncertainty, characterizing the stochastic solution presents a significant computational challenge as many methods require a large number of high-fidelity solves. This requirement results in the need for a possibly infeasible number of simulations when a typical converged high-fidelity simulation requires intensive computational resources. To reduce the cost of quantifying high-dimensional uncertainties, we investigate the application of a non-intrusive, bi-fidelity approximation to estimate statistics of quantities of interest associated with an irradiated particle-laden turbulent flow. This method relies on exploiting the low-rank structure of the solution to accelerate the stochastic sampling and approximation processes by means of cheaper-to-run, lower fidelity representations. The application of this bi-fidelity approximation results in accurate estimates of the QoI statistics while requiring a small number of high-fidelity model evaluations.
\end{abstract}

\begin{keyword}
Bi-fidelity approximation; Irradiated particle-laden turbulence; Low-rank approximation; Non-intrusive; Predictive computational science; Uncertainty quantification 
\end{keyword}

\end{frontmatter}


\section{Introduction}

The ability to quantitatively characterize and reduce uncertainties, in conjunction with model verification and validation (V\&V), plays a fundamental role in increasing the reliability of numerical simulations.
These types of studies are commonly encompassed within the field of uncertainty quantification (UQ), which has attracted increasing attention in the modeling and simulation community.
In this regard, the Predictive Science Academic Alliance Program (PSAAP) II at Stanford University~\cite{PsaapII-O}, focuses on advancing the state-of-the-art in large-scale, predictive simulations of irradiated particle-laden turbulence relevant to concentrated solar power (CSP) systems. To this end, physics-based models are developed and the numerical predictions are validated against data acquired from an in-house experimental apparatus designed to mimic a scaled-down particle-based solar energy receiver, and for which the quantification of uncertainties is of paramount importance. A significant challenge, and the scope of this work in particular, is investigating optimal UQ strategies for this complex, multi-physics flow when many sources of uncertainty are present. 

\subsection{Irradiated Particle-Laden Turbulent Flow}	\label{sec:irradiatedParticleLadenFlow}

Turbulent flow laden with inertial particles, or droplets, in the presence of thermal radiation is encountered in a wide range of natural phenomena and industrial applications.
For instance, it is well established that turbulence-driven particle inhomogeneity plays a fundamental role in determining the rate of droplet coalescence and evaporation in ocean sprays~\citep{Veron2015-A} and atmospheric clouds~\citep{Shaw2003-A}.
Another example is found when studying fires, in which turbulence, soot particles, and radiation are strongly interconnected resulting in very complex physical processes~\citep{Tieszen2001-A}.
From an industrial point of view, important applications include the atomization of liquid fuels in combustion chambers~\citep{Lasheras2000-A}, soot formation in rocket engines~\citep{Raman2016-A}, and more recently, volumetric particle-based solar receivers for energy harvesting~\citep{Ho2017-A}.

Even in the simplest configuration, e.g., homogeneous isotropic turbulence, particle-laden turbulent flow is known to exhibit complex interactions between the carrier and dispersed phases in the form of preferential concentration and turbulence modulation~\citep{Balachandar2010-A}.
Preferential concentration is the phenomenon by which heavy particles tend to avoid intense vorticity regions and accumulate in regions of high strain rate, while turbulence modulation refers to the alteration of fluid flow characteristics in the near-field region of particle clusters as a result of two-way coupling effects, e.g., enhanced dissipation, kinetic energy transfer, or formation of wakes and vortexes.
The physical complexity is further increased by the simple addition of solid walls as turbophoresis~\citep{Caporaloni1975-A}, i.e., tendency of particles to migrate towards regions of decreasing turbulence levels, becomes an important mechanism for augmenting the inhomogeneity in spatial distribution of the dispersed phase by driving particle accumulation at the walls.

As described above, characterization of particle-laden turbulent flow is a difficult problem; many experimental and numerical research studies have been devoted to this objective over the past decades, see, e.g., \cite{Squires1991-A,Wang1996-A,Sardina2012-A}.
However, the problem of interest in this work involves, in addition to particle-flow coupling, heat transfer from the particles to the fluid through radiation absorption.
The practical application motivating the study of this phenomenon is the improvement of energy harvesting in volumetric particle-based solar receivers.
At present, most CSP technologies use surface-based collectors to convert the incident solar radiation into thermal energy.
In this type of system, the energy is transferred to the working fluid downstream of the collection point via heat exchangers, typically resulting in large conversion losses at high temperatures.
By contrast, volumetric solar receivers continuously transfer the energy absorbed by particles directly to the operating fluid as they are convected through an environment exposed to thermal radiation.
This innovative technology is expected to increase the performance of CSP plants by avoiding the necessity of heat-exchanging stages, while requiring significantly high radiation-to-fluid energy transfer ratios.
This requirement imposes a very complex design constraint as the physical mechanisms governing irradiated particle-laden turbulent flow are still not fully comprehended, and therefore is a topic of intense research.
\newlluis{For example, Zamansky et al.~\cite{Zamansky2014-A} studied the interaction between radiation, particles and buoyancy in homogeneous isotropic turbulence (HIT) by means of point particle direct numerical simulation (PP-DNS) with the assumption of a constant radiation intensity in an optically thin environment.
The study revealed how non-uniformities of the gas-particle mixture resulted in inhomogeneities in heat absorption, and therefore in spatial temperature variations that induced large-scale fluid motion by local gas expansion and buoyancy.
The resulting baroclinic vorticity production triggered a feedback loop that generated larger non-uniformities by centrifuging the inertial particles leading to new non-uniformities in heat absorption.}
\newlluis{In the context of buoyancy effects in HIT, Frankel et al.~\cite{Frankel2016-A} investigated the impact of heating on the settling of particles, finding that, contrary to non-heated particles which enhance small- and large-scale turbulent motions when their settling velocity is sufficiently high compared to the Kolmogorov velocity, the heating of particles resulted in a significant reduction of the mean settling velocity caused by rising buoyant plumes in the vicinity of particle clusters and which affected all scales of turbulence.}
\newlluis{However, in regimes relevant to particle-based solar receivers, the background turbulence dominates over the buoyancy effects as the Froude number of the systems are typically very large.}
\newlluis{This problem was studied by Pouransari and Mani~\cite{Pouransari2017-A} by considering a convective flow of gas and particles subject to radiation and in the absence of buoyancy effects.}
\newlluis{In their study, it is shown how different particle concentration patterns, resulting from different Stokes numbers, have a strong influence on the heat transfer rates between particles and gas.}
\newlluis{A similar flow configuration, but considering polydispersity of particles, has been recently analyzed by Rahmani et al.~\cite{Rahmani2018-A}, finding also that the effective heat transfer rate between the two phases is significantly impacted by particle clustering in addition to polydispersity.}

\subsection{Uncertainty Quantification for Complex, Large-Scale, High-Dimensional Systems}

The complexity of constructing predictive models of CSP systems is furthered by the fact that there are many sources of uncertainty inherent in the underlying physical processes, for instance, turbulence models, particle properties or input radiation. This often high-dimensional uncertainty, in conjunction with large computational demands of high-fidelity (HF) simulation of irradiated particle-laden turbulence, necessitates cost-efficient, non-intrusive (i.e., sampling-based) UQ methods that accurately estimate the statistics of the quantities of interest (QoIs). Many widely-used non-intrusive methods, such as stochastic collocation \cite{Mathelin03,Xiu05} and polynomial chaos expansions (PCEs)~\cite{Ghanem03,Xiu02,Doostan11a}, suffer from a rapid (up to exponential) growth of computational cost as a function of the number of input variables characterizing the uncertainty. On the other hand, the cost of standard Monte Carlo (MC) sampling methods, while formally independent of the number of input variables, may be prohibitive when the QoI exhibits large variance and is expensive to evaluate. Much recent research has targeted developing cost reduction techniques to improve MC sampling methods.


As a form of cost reduction, there has been growing interest in multilevel and multi-fidelity methods, that is, methods relying on multiple models with varying levels of accuracy and cost, with the aim of accurately estimating the QoI statistics in a computationally efficient manner. Relative to (accurate) HF models, these models of reduced cost and accuracy are referred to, generally, as low-fidelity (LF) models. Inspired by multigrid methods~\cite{Brandt77,Briggs00}, multilevel techniques have evolved to not only include UQ methods exploiting coarser grid resolutions, but also multi-fidelity methods, where levels may correspond to a broader class of modeling schemes, e.g., simplified physics or reduced time-stepping (the interested reader is directed to~\cite{Peherstorfer16,Fernandez16} for a review of multi-fidelity methods). Relating to this work, interest is greatest with respect to multi-fidelity MC methods as we focus on high-dimensional uncertain inputs.

The control variates method \cite{Asmussen07} is a cost reduction technique that introduces a second, easily simulated, and correlated variable (to the original QoI) as a means to reduce the variance of the MC estimator of the QoI's expected value. This reduced variance results in requiring fewer simulations of the HF model to meet a desired mean square error. 
A specific extension of the control variates approach is the multilevel Monte Carlo (MLMC) method developed first in \cite{Heinrich01} and extended in \cite{Giles08}. MLMC estimates the expectation of the target QoI from the coarsest grid (temporal and/or spatial) solutions as well as differences between each two consecutive grid solutions in a telescoping sum fashion. When the differences corresponding to finer grids become smaller, fewer MC realizations of finer grid solutions are needed, thus leading to an overall reduced cost. Applications of MLMC to numerical partial differential equations, as in \cite{Cliffe11,Barth11, Teckentrup13}, 
show the success of the method for simple mathematical systems, making it ideal for high-dimensional, large-scale problems in which there exists convergence analysis with regards to the discretization scheme. In the last decade, several other types of control variates, in the form of multilevel and multi-fidelity, have been studied which rely on LF models, many of which do not adhere strictly to the coarsening of the spatial or temporal discretization schemes, e.g., \cite{Speight09, Nobile15, Vidal15, Geraci15, Fairbanks17}. Importance sampling~\cite{Asmussen07} is another approach to variance reduction that has been utilized in the context of multi-fidelity UQ, see, e.g., \cite{Peherstorfer16b,Peherstorfer16}. All of these cost reduction methods for MC have been shown to significantly improve the computational cost in comparison to standard MC simulation of the HF model.

Earlier work on multi-fidelity modeling of parametric/uncertain problems is based on Gaussian process regression, a.k.a kriging or co-kriging in the multivariate case,~\cite{Kennedy00,Qian06,Laurenceau08}. In particular, the seminal work in~\cite{Kennedy00} builds a Gaussian process approximation of the QoI based on an autoregressive model trained from nested observations of multiple, less expensive models. Each model of the sequence is related to the lower-fidelity model via a multiplicative constant and an additive Gaussian process correction term that are estimated from the lower-fidelity model evaluations as well as fewer realizations of the model itself. 

The recent work in \cite{Narayan14,Zhu14,Doostan16,Skinner17,Hampton2018practical} builds a reduced basis (or low-rank) approximation of the HF QoI using LF model evaluations and a {\it small} set of selected HF samples. The HF reduced basis -- consisting of realizations of the HF (vector-valued) solution at selected input samples -- as well as an interpolation rule in this basis are determined from LF realizations. A practical error estimate of this bi-fidelity (BF) approach is presented in~\cite{Hampton2018practical}, which can be used to determine if a given pair of low- and high-fidelity models will lead to an accurate BF approximation. In the present study, we adopt the BF approach and error estimate of~\cite{Hampton2018practical} to illustrate their efficacy in UQ of a CSP systems as an instance of a complex, multi-physics engineering application.  

\subsection{Objectives and Organization of the Work}	
\label{sec:objectives}

The system studied in this work is based on a small-scale apparatus\lluis{~\cite{Villafane2017-A}} designed to \newlluis{enable the reproduction of the physical mechanisms encountered in volumetric particle-based solar receivers}.
\lluis{Physics-based modeling of irradiated particle-laden turbulent flow -- as detailed in Section~\ref{sec:problem_description} -- and its numerical investigation and validation against data obtained from the experimental apparatus -- description given in Section~\ref{sec:solar_receiver} -- are difficult tasks that intrinsically require several model assumptions, selection of coefficient and parameter values, and characterization of initial and boundary conditions.
These steps, even if performed carefully, result in potential sources of uncertainty that can impact the quantities of interest (QoIs).}
\lluis{Examples of such uncertainties include the incomplete description of particle diameters~\cite{Rahmani2018-A} and thermal radiative properties~\cite{Frankel2017-A}, variability of the incident radiation and its complex interaction with boundaries, and structural uncertainty inherent in the approximations utilized, e.g., turbulence modeling~\cite{Jofre2018-A}.}
In addition to the large number of uncertainty sources, accurate predictions of the complex interaction of particle-laden turbulent flow with radiative heat transfer demand the utilization of expensive HF numerical simulations.
As an example, the cost of a medium-scale HF calculation of this problem requires approximately $500$k core-hours per sample on the Mira supercomputer (ALCF)~\cite{Mira-O}.
Therefore, if brute-force UQ techniques, e.g., MC simulation with ${\cal O}(10^{3})$ samples, are to be performed, the total cost is of the order of $500$M core-hours, thus motivating the need for accelerated UQ strategies.
In this regard, the objective of this work is to investigate the BF approximation UQ strategy on large-scale, multi-physics applications based on the PSAAP II solar receiver. In particular, it can be shown that this BF approximation provides accurate estimates of the QoI statistics, while maintaining a reduced cost similar to that of LF models for simulations with ${\cal O}(10^{3})$ samples.

The paper is organized as follows.
In Section~\ref{sec:problem_description}, the physical models and numerical method utilized to simulate irradiated particle-laden turbulent flow are described.
The particle-based solar receiver studied is detailed next, in Section~\ref{sec:solar_receiver}, in terms of computational setup, input uncertainties, and QoIs considered.
The BF approximation strategy is presented in Section~\ref{sec:bi-fidelity}, as well as a brief discussion of the associated error bound. The performance of this BF approximation, with regards to both accuracy and cost is investigated in Section~\ref{sec:results}. From these results, comparisons are made between this approximation and alternative LF models.
Finally, the work is concluded and future directions are proposed in Section~\ref{sec:conclusions}.

\section{Physics Modeling and Numerical Method}	\label{sec:problem_description} 
%
%

\begin{figure}[t]
  \centering
  \includegraphics[width=0.7\textwidth]{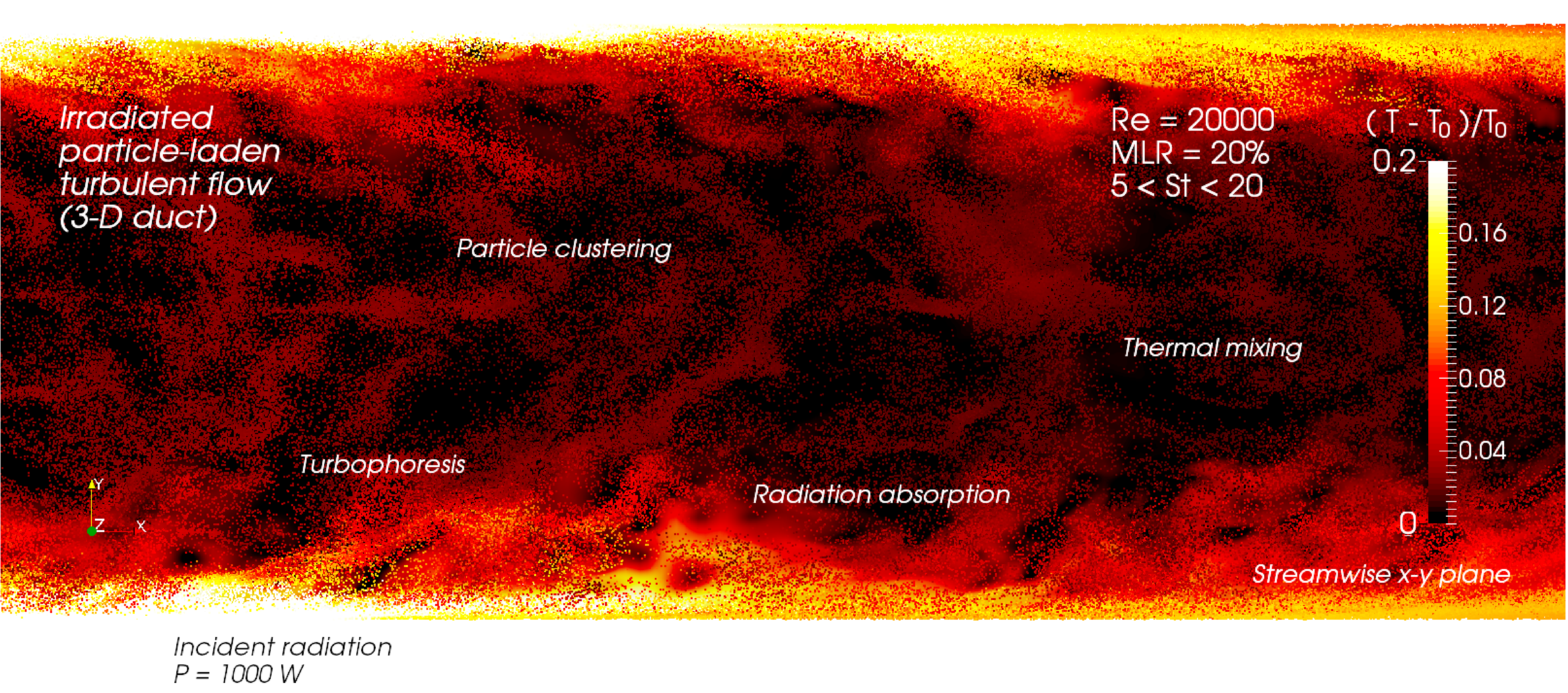}  
  \caption{Irradiated particle-laden turbulent flow in a square duct (instantaneous, streamwise $x$-$y$ plane snapshot). The quantity represented is the normalized temperature increment ($T_{0}=300$ K) of fluid and particles. It is interesting to note (i) the presence of elongated particle clusters resulting from preferential concentration, (ii) the accumulation of particles at the walls by turbophoretic effects, (iii) the absorption of radiation by particles (illuminated from below), and (iv) the transfer of energy from the particles to the fluid and the subsequent thermal mixing enhanced by turbulence.}		\label{fig:irradiated_particle_laden_turbulent_flow}
\end{figure}

\newlluis{The problem under study comprises the physical mechanisms depicted in Figure~\ref{fig:irradiated_particle_laden_turbulent_flow}.
The instantaneous snapshot, extracted from a single $\textrm{HF}$ simulation, corresponds to a slice of the normalized temperature increment, $(T - T_{0})/T_{0}$, of fluid and particles in the $x$-$y$ plane along the streamwise direction.
The fluid-particle mixture flows from a development square duct, with statistically steady-state distribution of particles (clustering + turbophoresis) at temperature $T_{0} = 300$ K, into a windowed section with the same dimensions but transversely illuminated by lasers from below.
The result is that the micron-sized, nickel particles absorb thermal radiation as they are advected through the duct; predominantly near the walls where they accumulate due to turbophoretic effects.
The thermal energy is then transferred to the fluid in the vicinity of particles by thermal exchange, and subsequently mixed into the bulk of the fluid by diffusion processes interacting with the cascade of turbulent scales.}
The equations describing this type of flow are continuity, Navier-Stokes in the low-Mach-number limit, conservation of energy assuming ideal-gas behavior, Lagrangian particle transport, and radiative heat transfer. The modeling of these three physical processes -- turbulent flow, particle transport, and radiative heat transfer -- and their couplings, are sequentially described in the subsections below.

\subsection{Variable-Density Turbulent Flow}	\label{sec:flow}

The volumetric particle-based solar receiver operates at atmospheric pressure conditions in which air, the carrier fluid, is assumed to follow the ideal-gas equation of state (EoS), $P_{th}=\rho_{g} R_{air}T_{g}$, where $P_{th}$ is the thermodynamic pressure, $\rho_{g}$ is the density, $R_{air}$ is the specific gas constant for air, and $T_{g}$ is the temperature.
As indicated by the EoS, density varies with temperature.
However, the Mach number of the flow $M\!a=u/c$, with $u$ the local flow velocity and $c$ the speed of sound of the medium, is small ($M\!a < 0.03$) for the range of velocities and temperatures considered.
Therefore, the low-Mach-number approximation is utilized to separate the hydrodynamic part, $p \ll P_{th}$, from the total pressure, $P_{tot}=P_{th} + p$. This decomposition results in the following equations of fluid motion
\begin{align}
  & \frac{\partial \rho_{g}}{\partial t} + \nabla \cdot \left(\rho_{g} \textbf{u}_{g} \right) = 0,	\label{eq:continuity} \\
  & \frac{\partial \left( \rho_{g} \textbf{u}_{g} \right)}{\partial t} + \nabla \cdot \left( \rho_{g} \textbf{u}_{g} \otimes \textbf{u}_{g} \right) = -\nabla p + \nabla \cdot \left[ \mu_{g} \left( \nabla \textbf{u}_{g} + \nabla \textbf{u}^{\intercal}_{g} \right) - \frac{2}{3}\mu_{g}(\nabla \cdot \textbf{u}_{g})\boldsymbol{I} \right] + \left( \rho_{g} - \rho_{0} \right)\textbf{g} + \textbf{f}_{TWC},	\label{eq:momentum} \\
  & \frac{\partial \left( \rho_{g} C_{v,g} T_{g} \right)}{\partial t} + \nabla \cdot \left( \rho_{g} C_{p,g} T_{g} \textbf{u}_{g} \right) = \nabla \cdot\left( \lambda_{g} \nabla T_{g} \right) + S_{TWC},	\label{eq:energy}  
\end{align}
where $\textbf{u}_{g}$ is the gas velocity, $\rho_{0}$ is an ambient reference density, $\boldsymbol{I}$ is the identity matrix, $\textbf{g}$ is the gravitational acceleration, and $\mu_{g}$ and $\lambda_{g}$ are the dynamic viscosity~\cite{Sutherland1893-A} and thermal conductivity~\cite{Weast1989-B}, respectively. Additionally, $C_{v,g}$ and $C_{p,g}$ are, respectively, the isochoric and isobaric specific heat capacities, and $\textbf{f}_{TWC}$ and $S_{TWC}$ are two-way coupling terms representing the effect of particles on the fluid and, respectively, approximated as point sources in the forms
\begin{align}
  & \textbf{f}_{TWC} = \sum_{p} m_{p}\frac{\textbf{u}_{p} - \textbf{v}_{p}}{\tau_{p}}\delta \left(\textbf{x} - \textbf{x}_{p} \right),	\label{eq:momentumTWC} \\
  & S_{TWC} = \sum_{p} \pi d_{p}^{2} h\left( T_{p} - T_{g} \right)\delta \left(\textbf{x} - \textbf{x}_{p} \right).	\label{eq:energyTWC}
\end{align}
Here, $m_{p} = \rho_{p}\pi d_{p}^{3}/6$ and $\textbf{v}_{p}$ are the particle mass and velocity, respectively, $\textbf{u}_{p}$ is the gas velocity at the particle location, $\tau_{p}=\rho_{p}d_{p}^{2}/(18 \mu_{g})$ is the particle relaxation time, $d_{p}$ is the particle diameter, and $\delta \left(\textbf{x} - \textbf{x}_{p} \right)$ is the Dirac delta function concentrated at the particle location $\textbf{x}_{p}$.
\lluis{The expression for the fluid-particle convection coefficient is $h = Nu \lambda_{f}/d_{p}$ with $Nu$ the particle Nusselt number. In the problem studied, particles are assumed to be isothermal as the Biot number is small, $Bi = h d_{p}/\lambda_{p} \ll 1$.}

\subsection{Lagrangian Particle Transport}	\label{sec:particles}

The carrier fluid is transparent to the incident radiation.
Hence, micron-sized nickel particles, i.e., the dispersed phase, are seeded into the gas to generate a non-transparent gas-particle mixture that absorbs and transfers the incident radiation from the particles to the gas phase.
The diameters of the particles are several orders of magnitude smaller than the smallest significant (Kolmogorov) turbulent scale $\eta$, and the density ratio between particles and gas is $\rho_{p}/\rho_{g} \gg 1$.
As a result, particles are modeled following a Lagrangian point-particle approach with Stokes' drag as the most important force~\cite{Maxey1983-A}.
Their description in terms of position, velocity and temperature is given by
\begin{align}
  & \frac{d \textbf{x}_{p}}{d t} = \textbf{v}_{p},	\label{eq:position} \\
  & \frac{d \textbf{v}_{p}}{d t} = \frac{\textbf{u}_{p} - \textbf{v}_{p}}{\tau_{p}} + \textbf{g}, 	\label{eq:velocity} \\
  & \frac{d \left( m_{p} C_{v,p} T_{p} \right)}{d t} = \frac{\pi d_{p}^{2}\left(1 - \omega \right)}{4}\int_{4\pi}\left( I - \frac{\sigma T_{p}^{4}}{\pi} \right)d\Omega - \pi d_{p}^{2} h \left(T_{p} - T_{g} \right),	\label{eq:temperature}
\end{align}
where $C_{v,p}$ is the particle specific isochoric heat capacity, $\omega = Q_{s}/\left( Q_{a} + Q_{s} \right)$ is the scattering albedo with $Q_{a}$ and $Q_{s}$ the absorption and scattering efficiencies, respectively, $I$ is the radiation intensity, $\sigma$ is the Stefan-Boltzmann constant, and $d\Omega=\mbox{sin}\thinspace\theta d\theta d\phi$ is the differential solid angle.
In the conservation equation for particle temperature,~(\ref{eq:temperature}), the first term on the right-hand-side accounts for the amount of radiation absorbed by a particle, while the second term represents the heat transferred to its surrounding fluid. 

In the point-particle approximation, particle-wall and particle-particle interactions are typically described by one-dimensional collision models based on the balance of total momentum and energy.
In the case of collisions involving two objects, $A$ and $B$, the velocities after impact, $v_{A}$ and $v_{B}$, are given by
\begin{align}
  & v_{A} = \left[ m_{A}u_{A} + m_{B}u_{B} + m_{B} C_{R}\left(u_{B} - u_{A} \right) \right]/\left(m_{A} + m_{B}\right),	\label{eq:velocity_A} \\
  & v_{B} = \left[ m_{A}u_{A} + m_{B}u_{B} + m_{A} C_{R}\left(u_{A} - u_{B} \right) \right]/\left( m_{A} + m_{B} \right),	\label{eq:velocity_B}
\end{align}
where $u_{A}$ and $u_{B}$ are the velocities of the objects before impact, $m_{A}$ and $m_{B}$ are the mass of the objects, and $0 \leq C_{R} \leq 1$ is the restitution coefficient.
The limits of $C_{R}$ correspond to the cases in which the objects coalesce at impact ($0$, perfectly inelastic collision) and rebound with the same relative speed as before impact ($1$, perfectly inelastic collision); intermediate values represent inelastic collisions in which kinetic energy is dissipated.
The above equations simplify to $v_{A}=-C_{R}u_{A}$ and $v_{B} = 0$ when object $B$ is a static wall.

\subsection{Radiative Heat Transfer}	\label{sec:radiation}

In the problem under consideration, the flow and particle timescales are orders of magnitude larger than the radiation timescale, which is related to the speed of light.
As a consequence, it can be assumed that the radiation field changes instantaneously with respect to temperature and particle distributions; i.e., radiation field is quasi-steady.
Under this assumption, and considering that air is transparent at all wavelengths and that absorption and scattering are determined solely by the presence of particles and solid boundaries, the radiative heat transfer equation becomes
\begin{equation}
  \hat{\textbf{s}}\cdot \nabla I = -\sigma_{e}I + \sigma_{a}\frac{\sigma T_{p}^{4}}{\pi} + \frac{\sigma_{s}}{4 \pi}\int_{4\pi}I \Phi d\Omega,	\label{eq:radiation}
\end{equation}
where $\hat{\textbf{s}}$ is the direction vector, $\sigma_{e} = \sigma_{a} + \sigma_{s}$ is the total extinction coefficient with $\sigma_{a}$ and $\sigma_{s}$ the absorption and scattering coefficients, respectively, and $\Phi$ is the scattering phase function that describes the directional distribution of scattered radiation.
The total extinction coefficient can also be defined in terms of absorption and scattering efficiencies as $\sigma_{e} = \left( Q_{a} + Q_{s} \right)\pi d_{p}^{2} n_{p} / 4$ with $n_{p}$ the local number density of particles.
Moreover, assuming gray radiation $Q_{a} + Q_{s} \approx 1$, which leads to $\omega \approx Q_{s}$ (see (\ref{eq:temperature})), and as a result $\sigma_{a} \approx Q_{a}\pi d_{p}^{2} n_{p} / 4$ and $\sigma_{s} \approx Q_{s} \pi d_{p}^{2} n_{p} / 4$.

\subsection{Numerical Method}	\label{sec:numericalMethod}

The equations of fluid motion,~(\ref{eq:continuity})-(\ref{eq:energy}), are solved following an Eulerian finite-volume discretization implemented in an in-house solver that is second-order accurate in space and suitable to non-uniform, \newlluis{Cartesian} meshes.
A fourth-order Runge-Kutta scheme is used for integrating the equations in time, together with a fractional-step method for imposing conservation of mass~\cite{Esmaily2018-A}.
Integration in time of the Lagrangian position, velocity, and temperature of particles,~(\ref{eq:position})-(\ref{eq:temperature}), is fully coupled with the advancement of the flow equations to ensure fourth-order accuracy.
The transfer of radiative heat,~(\ref{eq:radiation}), is calculated by means of an in-house discrete ordinates method (DOM) interfaced to the flow solver via an Eulerian representation of the particles distribution.

\section{Description of the Particle-Based Solar Receiver}	\label{sec:solar_receiver} 

\subsection{Computational Setup and Physical Parameters}	\label{sec:setup_physics}

\begin{figure}[t]
  \centerline{\includegraphics[width=\textwidth]{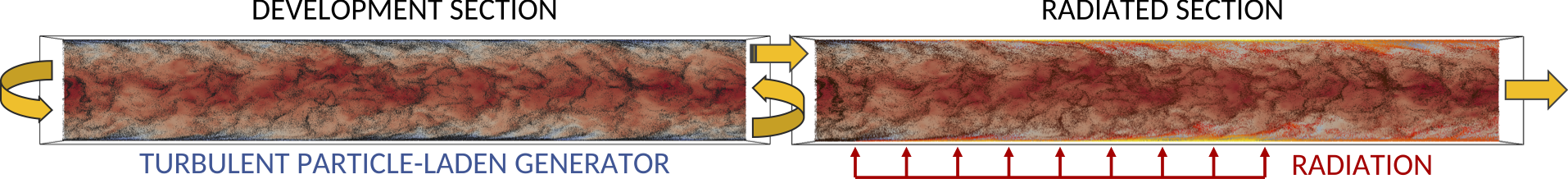}}
  \caption{Computational setup of the PSAAP II volumetric particle-based solar energy receiver. An isothermal periodic section (left domain) is utilized to generate fully developed particle-laden turbulent flow, which is used as inflow conditions for the second section (right domain) where the gas-particle mixture is irradiated perpendicularly to the flow direction from one the sides.}	\label{fig:problemSetup}
\end{figure}

Numerical simulations of the volumetric particle-based solar receiver are performed on the computational setup depicted in Fig.~\ref{fig:problemSetup}.
Two square duct domains, with dimensions $1.7L\times W\times W$ ($L = 0.16$ m, $W = 0.04$ m) in the streamwise ($x$-axis) and wall-normal directions ($y$- and $z$-axis), are utilized to mimic the development and radiated sections of the experimental apparatus.
The development section (left domain) is an isothermal, $T_{0}$, periodic particle-laden turbulent flow generator that provides inlet conditions for the inflow-outflow radiated section (right domain).
The solid boundaries of the development section ($y$- and $z$-sides) are considered smooth, no-slip, adiabatic walls.
Regarding the radiated section, the same boundary conditions are imposed except for the radiated region in which the $y$- and $z$-boundaries are modeled as non-adiabatic walls accounting for heat fluxes due to the radiation energy absorbed by the glass windows.

The bulk Reynolds number of the gas phase at the development section is $Re_{b} = \rho_{g} u_{b} L/\mu_{g} = 20000$, with $u_{b}$ the gas bulk velocity.
\newlluis{As characterized from the experiments, the particle size distribution is approximated by 5 different classes (uniformly sampled) with Kolmogorov Stokes numbers $St_{\eta} = 6, 8, 10, 12, 17$ (diameters listed in Table~\ref{tab:air_particles}) and with a total mass loading ratio (MLR) of $\textrm{MLR} = n_{p}m_{p}/\rho_{g} \approx 20$\%.}
Detailed values of the development section flow conditions and material properties are listed in Table~\ref{tab:air_particles}.
The gas-particle mixture, as depicted in Fig.~\ref{fig:probe_sketch}, is volumetrically irradiated through an $L \times W$ glass window starting at $\Delta x=0.1 L$ from the beginning of the radiated section.
The radiation source consists of an array of diodes mounted on a vertical support placed $\Delta y \approx 3 W$ from the radiated window and aligned with the streamwise direction of the flow.
The diodes generate a total power of $P\approx 1$ kW approximately uniform within a $18^{\circ}$ cone angle.

\begin{table}[t]
\centering
\tabcolsep7pt\begin{tabular}{llll}
\hline
Parameter & Value & Parameter & Value \\
\hline
$u_{b}$    & $8$ m/s                    & $T_{0}$   & $300$ K              \\
$P_{th}$   & $101325$ Pa                & $R_{air}$ & $287$ J/(kg$\cdot$K) \\
$C_{p,g}$  & $1012$ J/(kg$\cdot$K)      & $C_{v,g}$ & $723$ J/(kg$\cdot$K) \\
$\rho_{p}$ & $8900$ $\mathrm{kg/m^{3}}$ & $C_{v,p}$ & $450$ J/(kg$\cdot$K) \\
$d_{p}$ & $8.4$, $9.8$, $11.2$, $12.2$, $14.6$ $\mathrm{\mu m}$ & $\textbf{g}$ & $\left(9.81,0,0\right)$ $\mathrm{m/s^{2}}$ \\
\hline
\end{tabular}
\caption{Flow conditions at development section and physical properties.}	\label{tab:air_particles}
\end{table}

\subsection{Uncertainties and Quantities of Interest}	\label{sec:uqs_qois}

In our UQ study, we consider $14$ input variables to describe various experiment and \lluis{parameter} uncertainties, as shown in Table \ref{tab:uncertainties}. 
\newlluis{These correspond to incertitude in particle restitution coefficient for the different particle classes ($\xi_1-\xi_5$), Eqs.~(\ref{eq:velocity_A}) and~(\ref{eq:velocity_B}), correction to Stokes' drag law ($\xi_6$), Eq.~(\ref{eq:velocity}), particle Nusselt number ($\xi_7$), Eqs.~(\ref{eq:energyTWC}) and~(\ref{eq:temperature}), mass loading ratio ($\xi_8$), Eqs.~(\ref{eq:momentumTWC}) to~(\ref{eq:energyTWC}) and~(\ref{eq:radiation}), particle absorption and scattering efficiencies ($\xi_9-\xi_{10}$), Eq.~(\ref{eq:radiation}), incident radiation flux ($\xi_{11}$), Eq.~(\ref{eq:radiation}), and heat fluxes from the walls to the fluid ($\xi_{12}-\xi_{14}$), Eq.~(\ref{eq:energy}).}

\begin{figure}[t]
  \centering
  \includegraphics[width=0.25\textwidth]{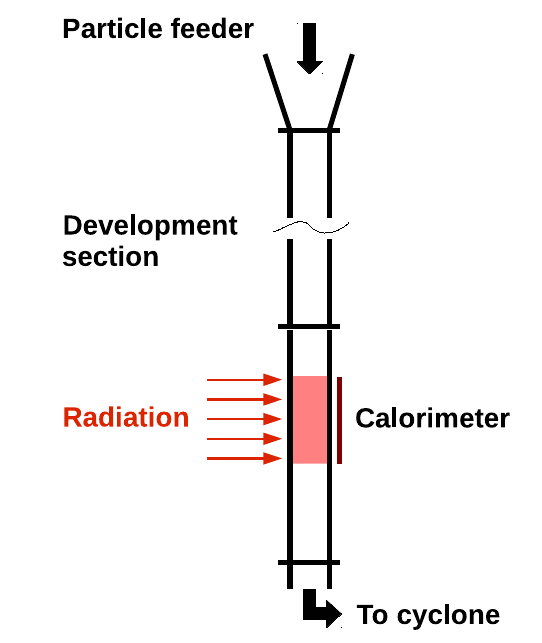}
  \hspace{2mm}
  \includegraphics[width=0.60\textwidth]{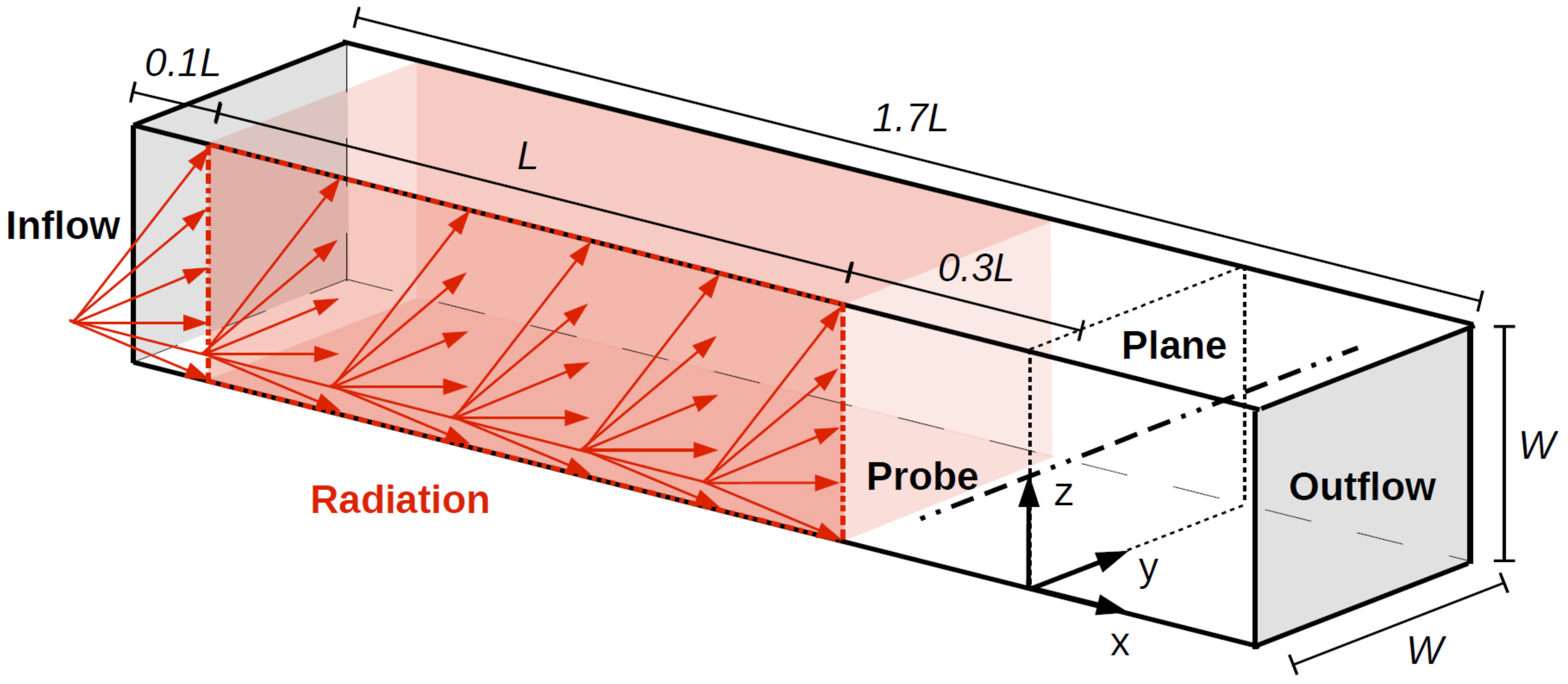}
  \caption{\lluis{Schematic representation of the experimental and computational setups. Open loop duct and radiation source (left). Detailed configuration of the illuminated computational section (right). The test segment is $1.7L$ long and $W$ width, with a radiated volume of $L \times W \times W$ starting at $0.1 L$ from the beginning of the section, and with a probe perpendicular to the flow and located $0.3L$ downstream.}} \label{fig:probe_sketch}
\end{figure}

The intervals of the input variables listed in Table~\ref{tab:uncertainties} have been carefully characterized on the basis of information provided by the team responsible for conducting the experiments, and by taking into consideration results and conclusions extracted from published studies. In particular, the intervals of the particle restitution coefficients follow the trend observed in experimental investigations by Yang \& Hunt~\cite{Yang2006-A} in which $C_{R}$ increases with Stokes number.
The expression for Stokes' drag force correction and its coefficient interval is based on the theoretical work by Brenner~\cite{Brenner1962-A}.
\lluis{The particle Nusselt number range is extracted from numerical studies of heated particles performed by Ganguli \& Lele~\cite{Ganguli2018a-A,Ganguli2018b-A}.}
The intervals for particle absorption and scattering efficiencies are obtained from Mie scattering theory and take into account sensitivity to shape deformation as investigated by Farbar et al.~\cite{Farbar2017-A}.
\lluis{The long development section of the experiment is modeled in the simulations by means of a periodic domain. This approach reduces the computational cost noticeably, but at the expense of not having direct control of the mass loading ratio. Consequently, the interval for this quantity is characterized based on preliminary comparisons with experimental data for a periodic particle-laden duct flow case without thermal radiation at steady-state conditions.}
\lluis{Similarly, the heat fluxes from the walls to the fluid are not fully resolved in the simulations and are instead directly modeled as boundary conditions; their calculation would require the solution of a complex conjugate convective heat transfer problem. In order to properly represent the uncertainty associated with this approximation, comparisons against the experiment were performed for an irradiated turbulent flow case without particles. This was done in a manner that the thermal response of the system was fully driven by the fluxes from the walls to the fluid and the intervals for these uncertainties could be obtained.}
\newlluis{Finally, the variability of the incident radiation is based on the uncertainty in emission intensity of the laser diodes reported by the manufacturer and the uncertainty in absorptivity of the glass walls.}

\lluis{The performance of the BF approximation is focused on thermal QoIs at the probe location.}
As detailed in Fig.~\ref{fig:probe_sketch}, the probe is located $\Delta x = 0.3 L$ downstream from the radiated perimeter, and is perpendicular to the flow direction along the $y$-axis at $z=W/2$.
\lluis{Of particular interest in this study is the time-averaged, normalized increment of gas temperature along the $y$-axis profile, i.e., $Q = (\langle T \rangle - T_{0})/T_{0} = \Delta T/T_{0}$, and the average heat flux over the plane at the probe location, i.e., $Q = \int C_{p,g} \langle \rho_{g} \textbf{u}_{g} \rangle \Delta T d\textbf{S}$.}

\begin{table}[t]
\centering
\tabcolsep7pt\begin{tabular}{llll}
\hline
Variable & Interval & Variable & Interval \\
\hline
$\xi_1:$ Prt. rest. coeff. 1 & [0.0 : 0.6] & $\xi_8:$ Mass load. ratio & [18 : 22]\%   \\
$\xi_2:$ Prt. rest. coeff. 2 & [0.1 : 0.7] & $\xi_9:$ Prt. abs. eff.    & [0.37 : 0.41] \\
$\xi_3:$ Prt. rest. coeff. 3 & [0.2 : 0.8] & $\xi_{10}:$ Prt. scatt. eff. & [0.69 : 0.76] \\
$\xi_4:$ Prt. rest. coeff. 4 & [0.3 : 0.9] &  $\xi_{11}:$  Radiation          & [1.8 : 2.0] $\mathrm{MW/m^{2}}$ \\
$\xi_5:$ Prt. rest. coeff. 5 & [0.4 : 1.0] & $\xi_{12}:$  Radiated wall      & [1.6 : 6.4] $\mathrm{kW/m^{2}}$ \\
$\xi_6:$ Stokes' drag corr.   & [1.0 : 1.5] & $\xi_{13}:$  Opposite wall      & [1.2 : 4.7] $\mathrm{kW/m^{2}}$ \\
$\xi_7:$ Prt. Nusselt num.   & [1.5 : 2.5] & $\xi_{14}:$  Side $x$-$y$ walls & [0.1 : 0.2] $\mathrm{kW/m^{2}}$ \\
\hline
\end{tabular}
\caption{List of random inputs $\xi_i$, $i=1,\dots,14$, with the corresponding ranges. All inputs are assumed to be uniformly distributed.}
\label{tab:uncertainties}
\end{table}

\subsection{Simulation Strategy}	\label{sec:strategy_ensemble}

In the experimental apparatus, fully developed conditions are achieved using a long duct \lluis{($\approx 7$m)} with an aspect ratio of order hundred.
To reduce the computational cost of simulating a long duct with inflow-outflow boundary conditions, the computational setup is divided in two domains as described in Section~\ref{sec:setup_physics}.
First, randomly distributed particles are seeded into the development section with an initial fully developed turbulent velocity field.
This system is then evolved in time for $20$ flow through times (FTTs), defined as $\mathrm{FTT}=L/u_{b}$, to achieve fully developed turbulent particle-laden flow conditions as in the experiments. 
After $20$ FTTs, the instantaneous Eulerian and Lagrangian solutions are copied into the second domain, radiative illumination is activated in the radiated region, and the total system (two domains) is integrated in time for $5$ additional FTTs intended to flush the thermal transient ($1$ FTT) and collect statistics ($4$ FTTs).
This procedure is repeated independently for each realization.


%
%
\subsection{Description of the High- and Low-Fidelity Models}	\label{sec:fidelities}

Three model fidelities have been designed to perform the UQ study: one HF model and two LF representations, denoted LF1 and LF2. All three models use the same description of uncertainty as described in Section \ref{sec:uqs_qois}.
The HF model corresponds to a point particle direct numerical simulation (PP-DNS) with sufficient resolution ($\approx55$M cells/section) to capture all the significant (integral to Kolmogorov) turbulent scales, while approximating the particles as Lagrangian points ($\approx15$M particles/section) with nonzero mass.
The flow grid is uniform in the streamwise direction with spacings in wall units equal to $\Delta x^{+} \approx 12$, while stretched in the wall-normal directions with the first grid point at $y^{+}, z^{+} \approx 0.5$ and with resolutions in the range $0.5 < \Delta y^{+}, \Delta z^{+} < 6$.
The radiative heat transfer equation is solved on a uniform DOM mesh of $270 \times 160 \times 160$ grid points ($\approx7$M cells) with $350$ quadrature points (discrete angles) \lluis{per element}. 

\begin{figure}[t]
  \centerline{\includegraphics[width=0.8\textwidth]{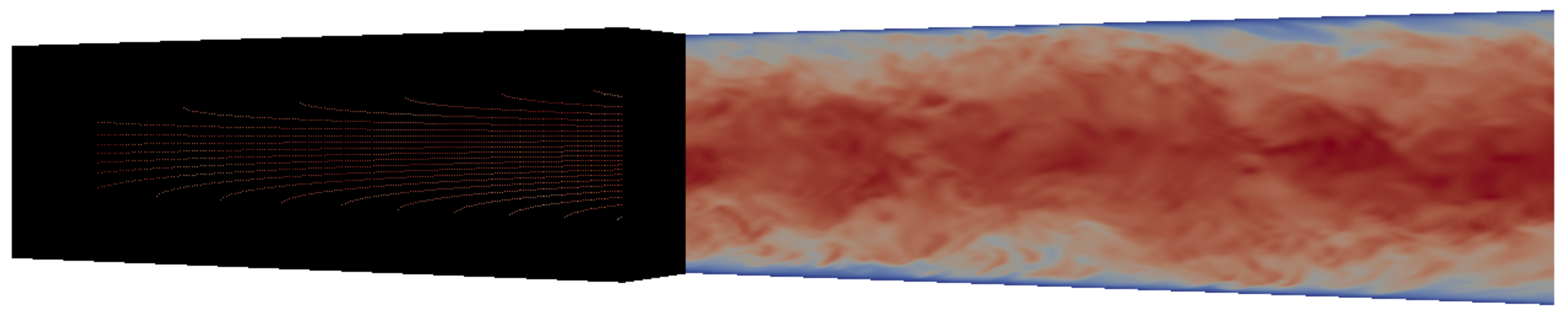}}
  
  \centerline{\includegraphics[width=0.4\textwidth]{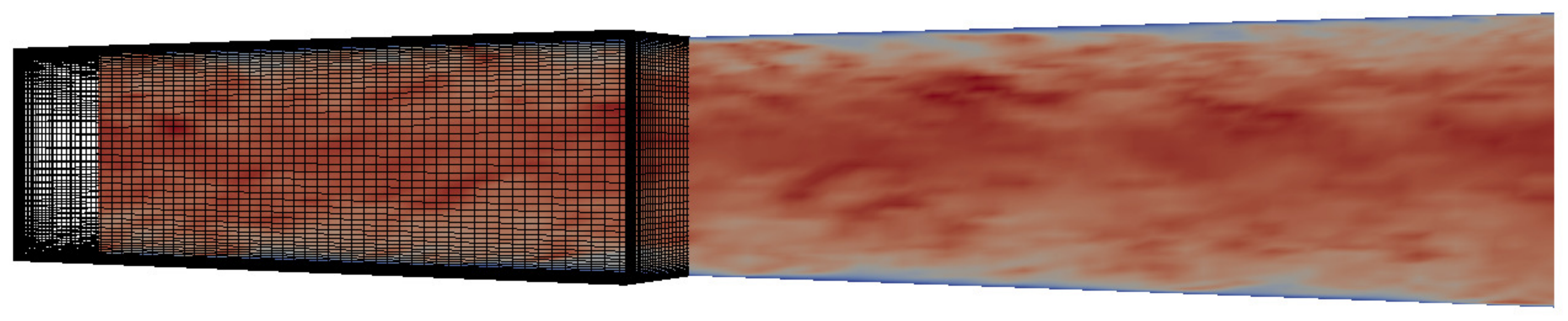}
              \includegraphics[width=0.4\textwidth]{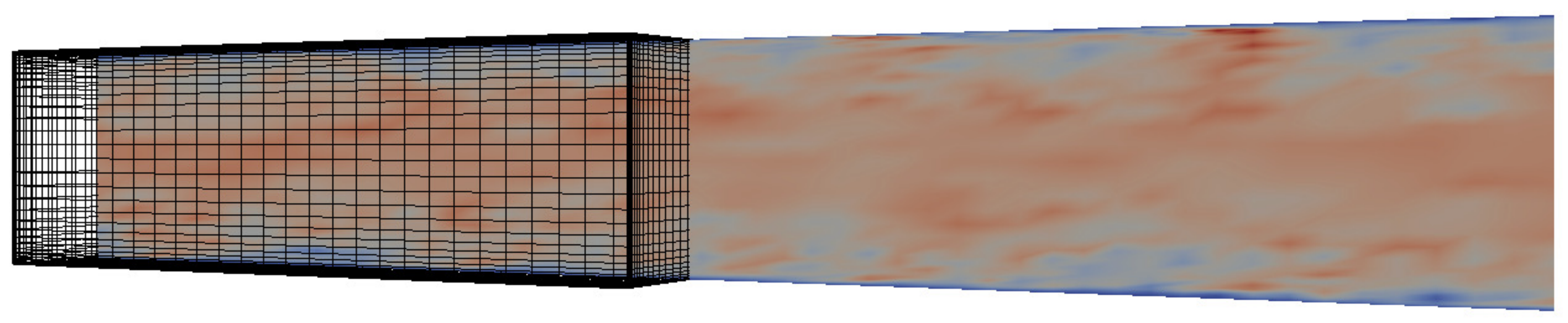}}
  \caption{Turbulent gas phase Eulerian resolution of the fidelity levels designed; gas velocity as background. HF $540 \times 320 \times 320$ gridpoints (top), LF1 $108 \times 64 \times 64$ gridpoints (bottom left), LF2 $54 \times 32 \times 32$ gridpoints (bottom right).}	
  \label{fig:fidelity_levels}
\end{figure}

Based on this HF model, two LF models have been constructed by coarsening the Eulerian and Lagrangian resolutions, resulting in the LF1 and LF2 representations that are, respectively, $\approx170\times$ and $\approx1300\times$ cheaper to simulate (per sample) than the HF model.
The flow and radiation meshes, and quadrature points are uniformly coarsened in each direction by a factor of $5$ ($\textrm{LF1}$) and $10$ ($\textrm{LF2}$), which additionally allows for larger time steps;
\lluis{a schematic of the different fidelity models for the turbulent flow phase is shown in Fig.~\ref{fig:fidelity_levels}.}
The discrete phase is coarsened by reducing the number of point particles in the calculation by $5\times$ ($\textrm{LF1}$) and $10\times$ ($\textrm{LF2}$).
To preserve the dimensionless parameters of the problem, this is efficiently accomplished by grouping physical particles into parcels representing their total effect~\cite{Subramaniam2013-A}, viz. surrogate particles~\cite{Amsden1989-TR}.
In the case of uniform parcels, the evolution of the $N_{s}$ surrogate particles can be described with the same set of Lagrangian equations used for the $N_{p}$ physical particles,~(\ref{eq:position})-(\ref{eq:temperature}), with the only modification of multiplying the physics coupling terms by the ratio $W_{s} = N_{p}/N_{s}$ as
\begin{align}
  & \textbf{f}_{TWC} = \sum_{p} W_{s} m_{p}\frac{\textbf{u}_{p} - \textbf{v}_{p}}{\tau_{p}}\delta \left(\textbf{x} - \textbf{x}_{p} \right), \quad S_{TWC} = \sum_{p} W_{s} \pi d_{p}^{2} h\left( T_{p} - T_{f} \right)\delta \left(\textbf{x} - \textbf{x}_{p} \right),\label{eq:surrogate_momentum_energy_twc}
  \\
  & \sigma_{a} \approx W_{s} Q_{a}\pi d_{p}^{2} n_{p} / 4 \quad \mbox{and} \quad \sigma_{s} \approx W_{s} Q_{s} \pi d_{p}^{2} n_{p} / 4.	\label{eq:surrogate_radiation_coefficients}
\end{align}
%

\section{Bi-fidelity Approximation Strategy}\label{sec:bi-fidelity}
\subsection{Construction of Bi-fidelity Approximation}\label{sec:bf_formation}

The BF approximation of this work follows the approach of~\cite{Doostan16,Hampton2018practical,Skinner17}, which seeks to identify a low-rank representation of the HF QoI assisted by the realizations of its LF counterpart. While the QoIs in this work are scalar-valued, this BF approximation relies on vector-valued quantities $\bm u$ that describe the scalar-valued QoI. For example, when the QoI is the time- and spatial-averaged gas temperature along the probe line (see Fig.~\ref{fig:probe_sketch}), the elements of $\bm u$ may be the time-averaged estimates of the gas temperature along this profile.


\begin{figure}[H]
    \centering
        \includegraphics[width=.7\textwidth]{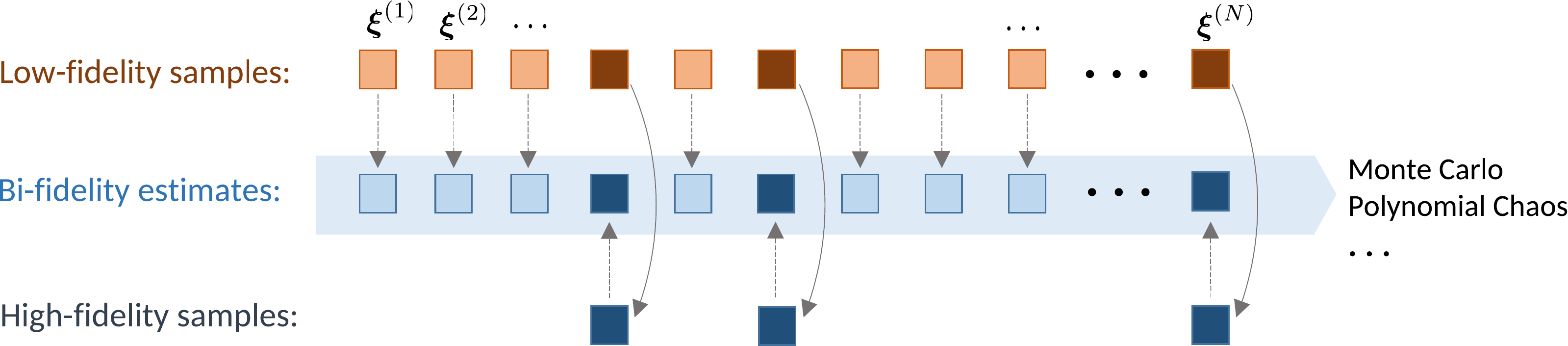}
   \caption{Schematic for the formation of the BF approximation.\label{fig:bf_schematic}}
\end{figure}

The construction of the BF approximation is completed in four main steps, as illustrated in Fig. \ref{fig:bf_schematic}. In the first step, $N$ independent samples of the inputs $\bm\xi$, each denoted by $\bm \xi ^{(i)}$, are generated according to their joint probability density function (PDF) and used to form $N$ LF realizations of $\bm u$, each referred to as $\bm u _L ^{(i)}\in\mathbb{R}^{m}$ (see Fig. \ref{fig:bf_schematic}, orange boxes, top row). Here, $m$ is the number of (spatial or temporal) degrees of freedom describing $\bm u_{L}$. These realizations are then organized in an $m \times N$ LF data matrix $\bm U_L$ such that
\begin{equation}
\bm U_L :=
\begin{bmatrix}
\bm u _L ^{(1)} & \bm u _L ^{(2)} & \cdots & \bm u _L ^{(N)}
\end{bmatrix}.
\end{equation}

Typically $N$ is large, as obtaining many LF samples is computationally feasible, and the value of $m$ corresponds to the spatial degrees of freedom of the numerical solver. In the second step, we seek to identify a subset of realizations of $\bm u _L$ of size $r$ (and the associated input samples of $\bm\xi$) that form a basis for the range space of $\bm U_L$. To this end, we perform a rank $r\ll N$ factorization of $\bm U_L$ via matrix interpolative decomposition (MID) \cite{Gu96,Cheng05,Martinsson11}. Specifically, the rank $r$ approximation $\hat{\bm{U}}_{L}$ is obtained via column-pivoted (truncated) QR factorization
\begin{align}
\label{eqn:QR1}
\bm U_L \bm P
&\approx
\bm Q \begin{bmatrix}  
\bm R _{11} & \bm R _{12}
\end{bmatrix}\\
\label{eqn:QR2}
&=\bm Q\bm R _{11}
\begin{bmatrix}  
\bm I & \bm R _{11}^{\dagger} \bm R _{12}
\end{bmatrix},
\end{align} 
where $\bm Q \in \mathbb{R}^{m \times r}$ has $r$ orthogonal columns, $\bm R _{11} \in \mathbb{R}^{r \times r}$ is an upper triangular matrix, $\bm R _{12} \in \mathbb{R}^{r \times N-r}$, $\bm P\in \mathbb{R}^{N \times N}$ is a permutation matrix, and $^\dagger$ indicates the pseudoinverse. It is straight-forward to show that the product $\bm Q\bm R _{11}$ in (\ref{eqn:QR2}) is equal to the left $r$ columns of $\bm U_L$, denoted by
\begin{equation}\label{eq:LF_colskel}
{\bm U}^c_L =
\begin{bmatrix}
\bm u _L ^{(i_1)} & \bm u _L ^{(i_2)} & \cdots & \bm u _L ^{(i_r)}
\end{bmatrix}
\end{equation}
and referred to as the {\it column skeleton} of $\bm U_L$. In (\ref{eq:LF_colskel}), the column indices $i_k$, $k=1,\dots,r$, are identified from the permutation matrix $\bm P$ in (\ref{eqn:QR1}) and indicate input samples $\bm \xi ^{(i_k)}$ corresponding to the LF realizations $\bm u _L ^{(i_k)}$ (see Fig. \ref{fig:bf_schematic}, dark orange boxes, top row). From (\ref{eqn:QR2}), setting $\bm C_L = [\bm I \quad \bm R _{11}^{\dagger} \bm R _{12}] \bm P$ to be the LF {\it coefficient matrix}, the rank $r$ MID approximation of $\bm U_L$ is given as
\begin{equation}\label{eq:LF_mid}
\hat{\bm U}_L :=
{\bm U}_L^c
\bm C_L,
\end{equation} 
indicating that $\bm U_L^c$, or equivalently the LF realizations $\bm u _L ^{(i_k)}$, act as a reduced basis for the range of the LF matrix $\bm U_L$. 

In the third step, the HF model is simulated at the inputs $\bm \xi ^{(i_k)}$ corresponding to the LF reduced basis (see Fig. \ref{fig:bf_schematic}, dark blue boxes, bottom row). This results in the associated HF column skeleton
\begin{equation}\label{eq:HF_colskel}
{\bm U}^c_H =
\begin{bmatrix}
\bm u _H ^{(i_1)} & \bm u _H ^{(i_2)} & \cdots & \bm u _H ^{(i_r)}
\end{bmatrix}.
\end{equation}
Note that ${\bm U}^c_H$ is $M\times r$, where $M\geq m$ as it corresponds to the number of HF spatial degrees of freedom. For HF models with a finer mesh resolution than the LF model it follows that $M> m$.
\lluis{However, as is the case in this work, data values may be extracted at equivalent coordinates by using extraction points independent of the grid tessellation, resulting in $M=m$.} 
In the final step the BF approximation is formed by taking the product of the HF column skeleton in~(\ref{eq:HF_colskel}) and the LF coefficient matrix
\begin{equation}\label{eq:bf}
\hat{\bm U}_H :=
{\bm U}_H^c
\bm C_L,
\end{equation} 
where the $i$-th column of $\hat{\bm U}_H$, denoted $\hat{\bm u}^{(i)}_H$, is the BF approximation to ${\bm u}_H$ at $\bm\xi^{(i)}$ (see Fig. \ref{fig:bf_schematic}, blue boxes, middle row). Once formed, approximate QoI realizations are calculated directly from the BF realizations $\hat{\bm u}^{(i)}_H$.

With regards to computational cost, the BF approximation requires $N$ LF simulations, performing rank-revealing QR on $\bm U_L$ with $\mathcal{O}(rmN)$ floating point operations for basis identification and the coefficient matrix computation, and only $r$ HF simulations to form ${\bm U}_H^c$. Typically, obtaining HF solutions is a bottleneck, and thus limiting the number of HF simulations to small $r$ is of great value and a fundamental component of this approximation.

\begin{remark}\label{rem:identical_at_basis}
By construction of MID, the columns of $\hat{\bm U}_L$ with indices $i_k$, $k=1,\dots,r$, identified by the permutation matrix $\bm P$ in (\ref{eqn:QR1}), are exactly the same as the corresponding columns of $\bm U_L$. Stated differently, an $r\times r$ sub-matrix of $\bm C_L$ is a permutation of the identity matrix. This, therefore, implies that the columns of the BF matrix $\hat{\bm U}_H$ with indices $i_k$  (identified from MID of $\bm U_L$) are the same as the $r$ HF realizations forming the columns skeleton $\bm U^c_H$. 
\end{remark}
\begin{remark}
While the optimal rank for this BF approximation is not known {\it a priori}, the rank of the LF data matrix $\bm U_L$ may indicate a good selection range. Theoretical results of \cite{Gu96,Cheng05,Martinsson11} show the error of the rank $r$ LF approximation in~(\ref{eq:LF_mid}) is bounded by a scaling of the $(r+1)$-th largest singular value of $\bm U_L$. This suggests that the decay of singular values of $\bm U_L$, specifically, where a significant drop occurs, to be a worthwhile consideration for the value of $r$.
\end{remark}
\subsection{Theoretical Error Estimation of Bi-Fidelity Approximation}\label{sec:theory}

This section briefly presents the theoretical results of \cite{Hampton2018practical} that provide a practical error bound on the spectral norm of $\Vert {\bm U}_H-\hat{\bm U}_H\Vert$. This error bound may be used with small cost to assess the suitability of a given pair of low- and high-fidelity models to produce an accurate BF approximation. Instances of successful BF approximation via $\hat{\bm U}_H$ in (\ref{eq:bf}) have been reported in a number of recent studies \cite{Narayan14,Zhu14,Doostan16,Skinner17, Hampton2018practical}.

Following \cite{Hampton2018practical}, the BF estimate $\hat{\bm U}_H$ is accurate as long as there exists an $M\times m$ matrix $\bm T$, with bounded norm, such that $\bm U_{H}\approx\bm T\bm U_{L}$. For a given pair of low- and high-fidelity problems, the existence of such a mapping is not always guaranteed. However, \cite{Hampton2018practical} shows that such a mapping can be constructed when  
\begin{equation}\label{eq:eig1}
\epsilon(\tau) = \lambda _{\max} (  {\bm U}_{H}  ^T {\bm U}_{H} - \tau   {\bm U}_{L} ^T  {\bm U}_{L}),\qquad \tau\ge 0,
\end{equation}
is small enough and that $\bm U_L$ (similarly $\bm U_H$) has fast decaying singular values. In (\ref{eq:eig1}), $\lambda _{\max}(\cdot)$ denotes the largest eigenvalue of a matrix, and ${\bm U}_{H}  ^T {\bm U}_{H}$ and ${\bm U}_{L} ^T  {\bm U}_{L}$ are the Gramians of the HF and LF matrices, respectively.
%
%
The following theorem from \cite{Hampton2018practical} provides a bound on the BF approximation error. In particular we note the dependence of the error bound on $\tau$ and $\epsilon(\tau)$. 

\begin{theorem}[Theorem 1 of \cite{Hampton2018practical}]
Let $\hat{\bm U}_H$ be the rank $r$ BF approximation, as in~(\ref{eq:bf}), to the HF data matrix $\bm U_H$. Let $\hat{\bm U}_L$ be the rank $r$ MID approximation, given in (\ref{eq:LF_mid}), to the LF data matrix ${\bm U}_L$, where $\bm C_L$ is the corresponding coefficient matrix. For $\epsilon (\tau)$, as defined in~(\ref{eq:eig1}), and $\Vert \cdot \Vert$ the spectral norm, it follows that the BF error may be bounded as
\begin{equation}\label{eq:bound}
\Vert {\bm U}_H-  \hat{\bm U}_H\Vert \leq \min_{\substack{k< \text{rank}({\bm U}_{\ell}) \\ \tau \geq 0}}
\left((1 + \Vert \bm C_L \Vert) \sqrt{\tau \sigma_{k+1}^2 + \epsilon (\tau)} +
\Vert {\bm U}_L-  \hat{\bm U}_L \Vert \sqrt{\tau + \epsilon (\tau)\sigma_k^{-2}}\right),
\end{equation}
where $\sigma _k$ and $\sigma _{k+1}$ are the $k$-th and $(k+1)$-th largest singular values of $\bm U_L$, respectively.
\end{theorem}


Evaluating the bound in (\ref{eq:bound}) requires the computation of $\epsilon(\tau)$ for multiple values of $\tau$. Following (\ref{eq:eig1}), this in turn requires access to the entire HF data matrix $\bm U_H$, which is not possible as $\bm U_H$ is never generated in practice. As an alternative, estimates of $\epsilon(\tau)$ may be calculated using a subset of $R\ll N$ HF and LF samples via 
\begin{equation}\label{eq:eig}
\hat{\epsilon}(\tau) = \frac{N}{R}\lambda _{\max} (  ({\bm U}_{H} ^{R})^T {\bm U}_{H} ^{R} - \tau   ( {\bm U}_{L}^{R}) ^T  {\bm U}_{L}^{R}),
\end{equation}
where the superscript $R$ indicates the number of columns of $\bm U_H$ and the corresponding columns of $\bm U_L$ used to set ${\bm U}_{H} ^{R}$ and ${\bm U}_{L} ^{R}$, respectively. Stated differently, estimates of $\epsilon(\tau)$ may be obtained using $R$ HF samples, instead of $N$. To evaluate the remainder of the bound in~(\ref{eq:bound}), MID is applied to the LF data matrix to obtain values for $\Vert {\bm U}_{L} - \hat{\bm U}_{L} \Vert $ and $\Vert {\bm C}_{L}\Vert $. Combining these estimates and minimizing over identified values of $(\tau, \hat{\epsilon}(\tau) )$ and the singular values of ${\bm U}_{L}$ results in an approximate bound. The numerical results of \cite{Hampton2018practical} show empirically that $R$ slightly larger than the approximation rank $r$ is sufficient to estimate the optimal pair $(\tau,\epsilon(\tau))$. This therefore suggests the efficacy of (\ref{eq:bound}) in estimating the BF approximation error or in suitability of a given pair of low- and high-fidelity models for BF modeling. 

\subsection{Using Bi-Fidelity Approximation to Estimate QoI Statistics}\label{sec:bfpce}
%


Our discussion so far has focused on the construction of $N$ BF estimates of a vector-valued QoI using the corresponding $N$ LF realizations, along with $r$ selected HF realizations. We next turn our attention to how these $N$ BF estimates may be utilized for the purpose of UQ or sensitivity analysis.  

When the BF approximation achieves the desired accuracy, standard methods such as MC sampling, stochastic collocation, or sparse PCE may be employed on $\hat{\bm U}_H$ (more specifically, the columns of $\hat{\bm U}_H$) instead of $\bm U_H$ to estimate the moments and PDF of the QoI, and perform global sensitivity analysis. When the dimension $d$ of the random inputs is not high, methods such as sparse PCE or stochastic collocation may be employed. Otherwise, MC sampling methods are preferred. Achieving small sampling errors, requires having access to large enough number of BF -- and therefore LF -- samples $N$. On the other hand, when the BF approximation does not meet the accuracy requirements but leads to estimates well correlated with the HF data, the BF approximation may serve as a control variate to MC in a single-level \cite{Asmussen07} or in a multilevel setting as in \cite{Fairbanks17}. 

In the numerical results of Section \ref{sec:results}, we use the $\ell_1$-minimization approach of \cite{Doostan11a,Hampton15a} to build sparse PCEs of the QoIs as approximate maps between the random inputs $\bm \xi$ and the QoIs. The resulting PCEs are, in turn, used to estimate the statistics, here, histogram, of the QoIs and perform global sensitivity analysis. The QoI statistics can be either computed directly via the PC coefficients, e.g., for the mean and variance, or by sampling the PCE itself in a MC fashion, e.g., for histogram. We follow the latter approach in the experiments of Section \ref{sec:results} to generate histograms of the QoIs. For global sensitivity analysis, we perform variance decomposition to compute the so-called Sobol' indices \cite{Sobol01}, which provide a means to quantitatively describe the importance of input parameters, by calculating each parameter's contribution to the total variance of the output QoI. Following the work of \cite{Sudret08}, we compute the Sobol' indices directly from the PCE coefficients. 


%
%
\section{Numerical Results of Bi-Fidelity Approximation}
\label{sec:results}

To investigate the performance of the BF approximation, $256$ LF2, $128$ LF1, and $26$ HF simulations were performed, such that the 26 HF simulations correspond to the first 26 simulations of LF1 and LF2, and the 128 LF1 simulations correspond to the first 128 of the LF2 simulations. From the two LF models, two BF approximations are formed: bi-fidelity 1 (BF1) approximation and bi-fidelity 2 (BF2) approximation. The BF1 approximation is formed from $N=128$ LF1 samples and $r$ HF samples, and the BF2 approximation is formed from $N=256$ LF2 samples and $r$ HF samples. The number of HF simulations is left as $r$, as the selection of this value will be discussed in the following results.

The motivation of this approximation is to form a BF model that accurately predicts the HF data, and the goal of these results is to investigate whether or not there is an improvement over the performance of the the LF models. For these tests, two primary time-averaged, \lluis{i.e., $\langle \cdot \rangle$}, thermal QoIs are considered: (i) heat flux through the plane at the probe location, $Q = \int C_{p,g} \langle \rho_{g} \textbf{u}_{g} \rangle \Delta T d\textbf{S}$, ($\Delta x = 0.3L$ downstream from the radiated perimeter) \lluis{normalized by the mean QoI estimated from the HF data}, and (ii) \lluis{normalized temperature increment}, $\Delta T /T_{0}= (\langle T \rangle-T_{0})/T_{0}$, values along the profile at the probe location ($\Delta x = 0.3L$ downstream from the radiated perimeter at $z=0.5W$), where focus is placed on spatially-averaged $\Delta T/T_{0}$, and point estimates at $y/W=0.5$, $y/W=0.1$, and $y/W=0.05$.  

For each QoI, five primary tasks are considered: 
(i) BF rank identification as to best optimize accuracy and computational cost of the approximation, (ii) BF error bound estimation, to verify the accuracy of the approximation for a fixed rank, (iii) QoI approximation via available data, (iv) estimation of statistics via sparse PCE of LF and BF models, and (v) cost analysis to compare approximate core-hours needed to obtain converged simulations from each model.


For a given matrix $\bm U$ of LF or BF data, we report the relative spectral error, with respect to its HF counterpart $\bm U_H$, defined as 
\begin{equation}\label{eq:rel_spect_error}
\text{relative spectral error} = \frac{\Vert \bm U_H - \bm U \Vert }{\Vert \bm U_H \Vert},
\end{equation}
For scalar QoIs, we report the relative $\ell_2$ (root mean-square error),
\begin{equation}\label{eq:rel_l2_error}
\text{relative $\ell_2$ error} = \frac{\sqrt{\sum\limits_{i=1}^N (Q_H^{(i)} - Q^{(i)})^2}}{\sqrt{\sum\limits_{i=1}^N (Q_H^{(i)})^2}},
\end{equation}
where $Q^{(i)}$ is a LF or BF simulated QoI and $Q_H^{(i)}$ is the corresponding HF realization. 

\subsection{QoI \#1: Heat Flux Through the $\Delta x=0.3L$ Plane}\label{sec:qoi_hf}

To estimate the statistics of the heat flux QoI, the BF approximation is formed from realizations $\bm u$ of the heat flux values over the entire $\Delta x=0.3L$ plane. For ease of comparison, these values are extracted as a $500 \times 500$ uniform grid of points for both HF and LF simulations (thus here $m=M$). The scalar-valued QoI estimates $Q$ are taken to be the \lluis{heat flux} determined from all of the elements of $\bm u$, normalized by the mean QoI estimated from the HF data. 

\begin{figure}[h]
    \centering
    \begin{subfigure}[b]{0.48\textwidth}
    \centering
        \includegraphics[width=.9\textwidth]{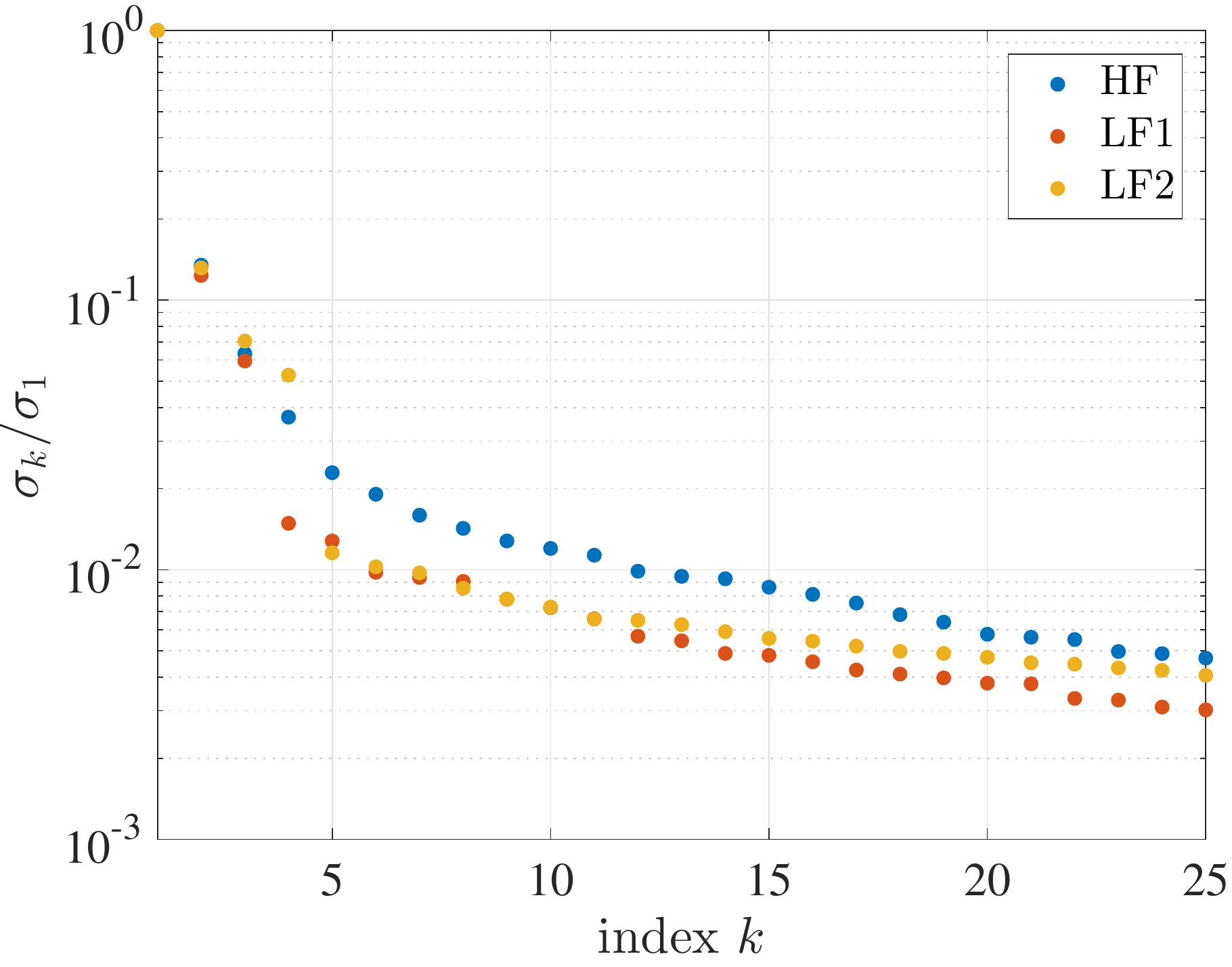}
        \caption{}
    \end{subfigure}
    \begin{subfigure}[b]{0.48\textwidth}
    \centering
        \includegraphics[width=.9\textwidth]{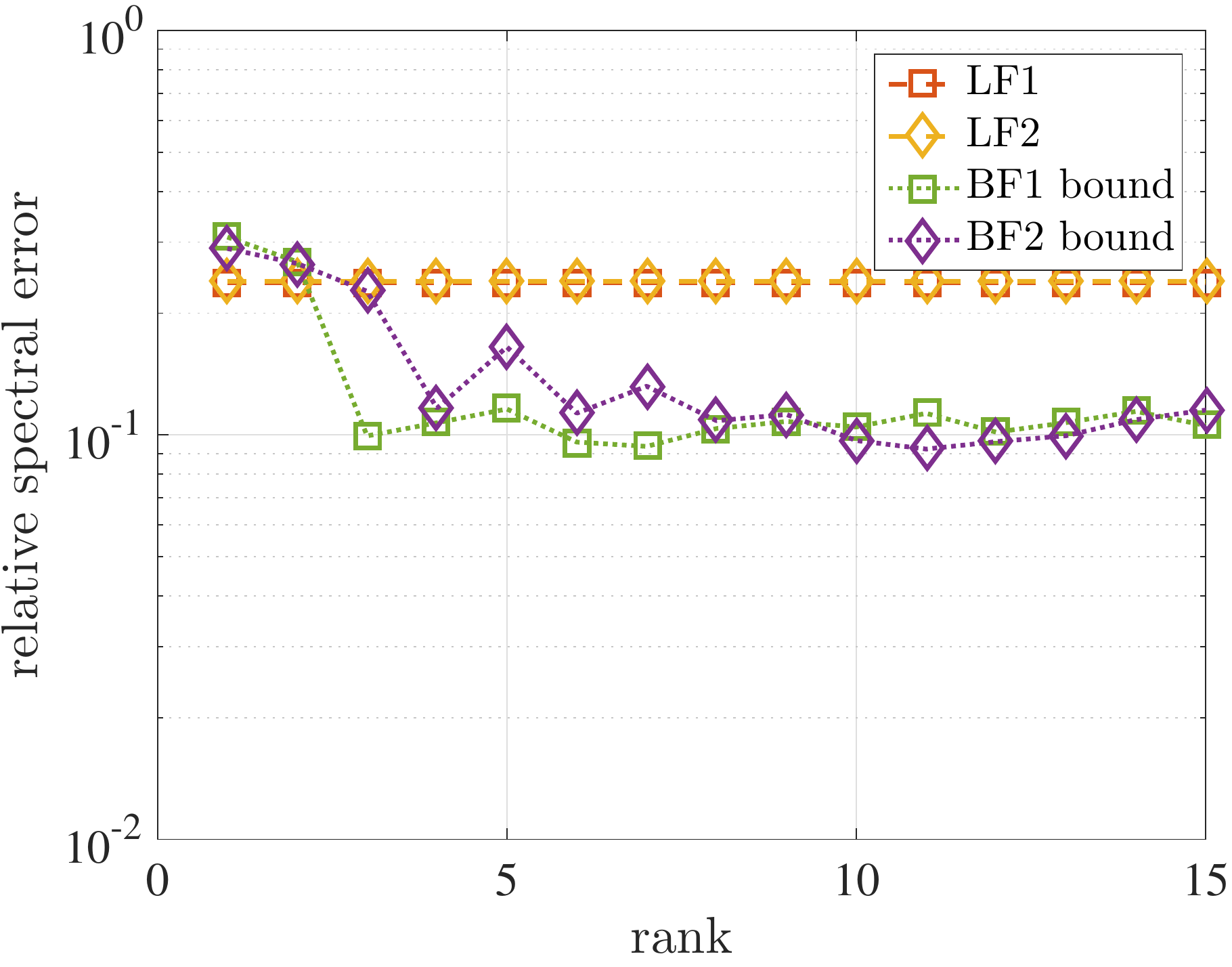}
        \caption{}
    \end{subfigure}
   \caption{(a) Decay of normalized singular values of LF and HF matrices using available data. (b) Error bound estimates for both BF approximations as a function of rank $r$. For comparison, relative spectral error of the LF data are provided. \label{fig:exhf_sing}}
\end{figure}

When forming the BF approximation, the first task is to identify the approximation rank. This is critical as it will dictate both the computational cost -- corresponding to $r$ HF realizations --  and accuracy of the approximation. As a reduced cost relies on minimal number of HF samples, the selected rank must be as small as possible. However, excessively small approximation ranks may lead to inaccurate BF solutions. To aid in rank selection, consider the results of Fig. \ref{fig:exhf_sing}. Fig. \ref{fig:exhf_sing} (a) provides the decay of singular values of LF and HF matrices. The magnitudes of these singular values decay most rapidly within the first six indices, indicating that six realizations of heat flux capture most of the information of the LF and HF data.
Fig. \ref{fig:exhf_sing} (b) displays the error bound estimate of the BF approximation matrix as a function of rank $r$. 
Note, the number of simulations to calculate $\hat{\epsilon} (\tau)$ is set to $R=r+2$; a more thorough assessment of the error bound will be discussed shortly. 
The error bound estimate levels out for a conservative value of $r= 6$, which is the BF approximation rank we use for the rest of the results. 
For comparison, the LF1 and LF2 spectral errors are provided (see~(\ref{eq:rel_spect_error})), indicating that very few HF samples are needed to observe an improvement over the LF models. The presentation of the corresponding cost analysis is postponed until Section~\ref{sec:cost}.

\begin{figure}[H]
    \centering
    \begin{subfigure}[b]{0.48\textwidth}
    \centering
        \includegraphics[width=.9\textwidth]{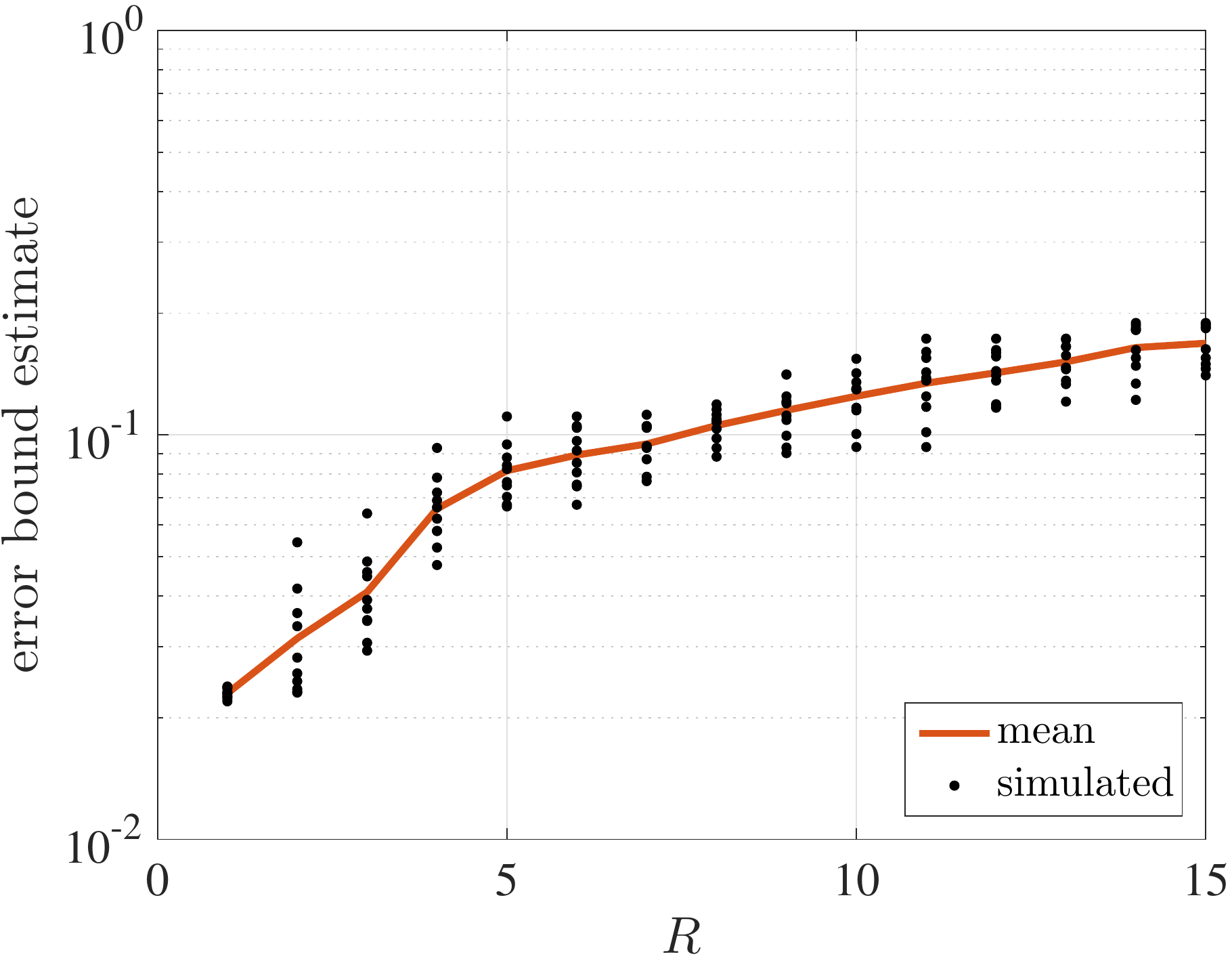}
        \caption{BF1}
    \end{subfigure}
    \begin{subfigure}[b]{0.48\textwidth}
    \centering
        \includegraphics[width=.9\textwidth]{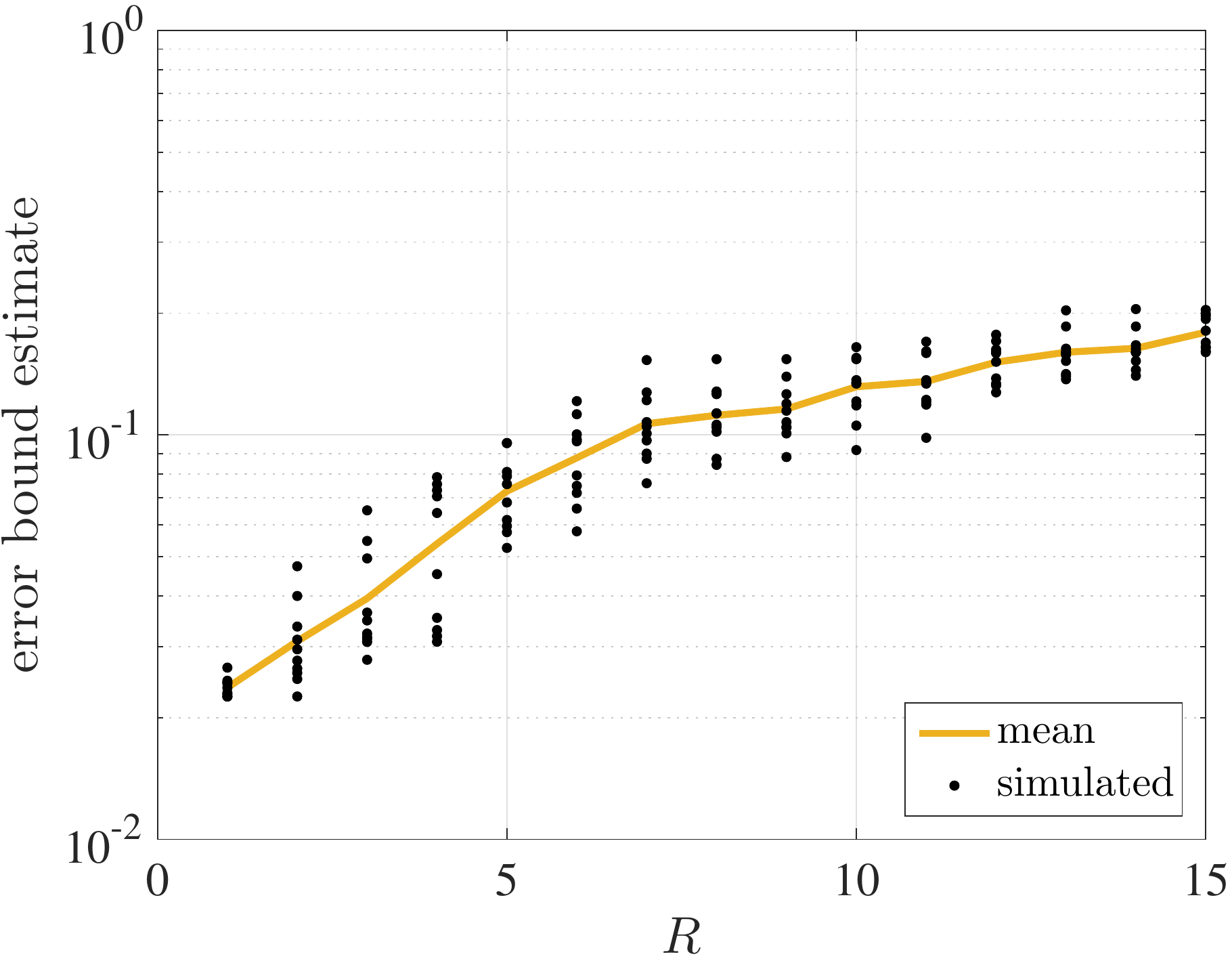}
        \caption{BF2}
    \end{subfigure}
   \caption{Error bound estimate from~(\ref{eq:bound}) using $\hat{\epsilon}(\tau)$ and rank $r=6$ for varying values of $R$, for (a) BF1 and (b) BF2 models. Values are based on $10$ different sets of $R$ columns, where column selection not fully independent due to the small number of available HF samples.\label{fig:exhf_bound}}
\end{figure}

While the previous results indicate that the rank $r=6$ BF approximation more accurately describes the HF data, the calculated error must be verified by more thoroughly investigating the theoretical error bound estimates from Section \ref{sec:theory}. Specifically, the bound must be estimated for multiple values of $R$, the number of samples used to estimate $\hat{\epsilon}(\tau)$.
Fig. \ref{fig:exhf_bound} provides the error bound estimates as a function of $R$ for the BF1 and BF2 models. Each point represents an error bound estimate calculated from $R$ random columns of $\bm U_H$ (out of $26$ total columns) and the corresponding columns of $\bm U_L$. The solid line represents the average value of the points at each value of $R$. Numerical results of \cite{Hampton2018practical} suggest a value of $R\approx 2r$ will provide a true error bound. With rank $r=6$, these results estimate the error bound to be $0.12$ for both BF models.
Recall from Fig. \ref{fig:exhf_sing} (b) that this error bound estimate is smaller than that of either LF model, indicating that improvement in accuracy may be estimated without knowledge of the true BF error.  

\begin{figure}[H]
  \centering
  \includegraphics[width=.42\textwidth]{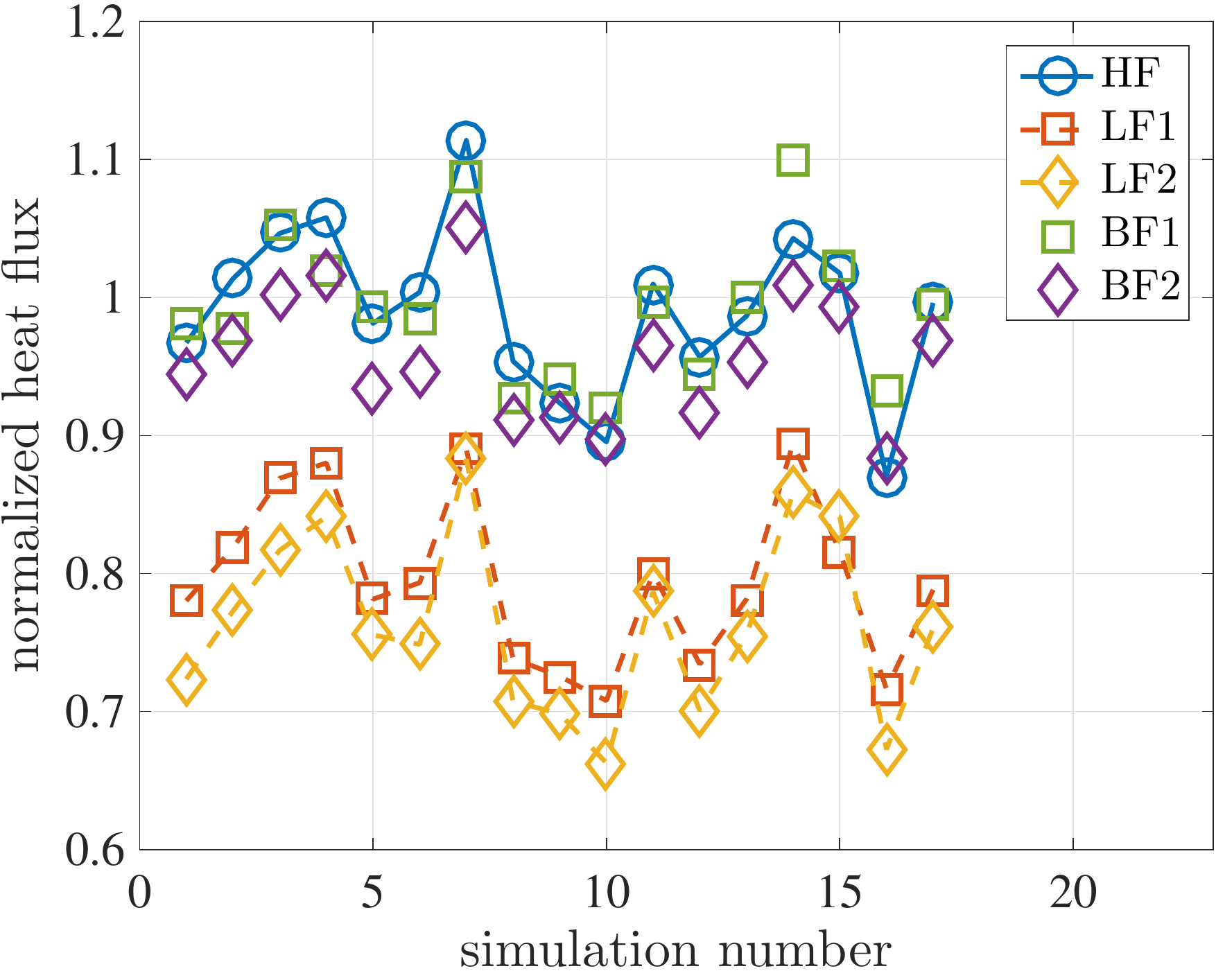}
  \caption{Normalized total heat flux values for $17$ independent simulations, from the five different models, where the BF approximation is of rank $r=6$. The BF approximations are more accurate than either LF QoI data. This data excludes simulations corresponding to the HF realizations used in the BF approximations. \label{fig:exhf_simulations}}
\end{figure}

To compare the performances of the LF and BF approximations with regards to the heat flux QoI, consider the results of Fig. \ref{fig:exhf_simulations}, where $17$ simulated values of the heat flux using the HF, LF and BF models are provided. Given the discussion of Remark \ref{rem:identical_at_basis}, we exclude the heat flux values corresponding to the $r$ HF realizations used to form the BF approximation. For each simulated value, the BF approximations are significantly more accurate than their LF counterparts. With regards to the relative $\ell_2$ error, the BF QoIs are about $7\times$ more accurate than the LF QoIs, with both BF1 and BF2 having errors of $0.03$, LF1 an error of $0.20$, and LF2 an error of $0.22$. 

%

To estimate the moments and PDF, as well as perform sensitivity analysis, a sparse PCE surrogate is formed from available data, as discussed in Section \ref{sec:bfpce}. Table \ref{tab:pce_hf} provides the QoI mean and coefficient of variation (CoV) values determined from the approximate coefficients of each LF and BF surrogate model, as well as the relative error between each QoI mean and the HF QoI mean. Note the HF QoI mean is calculated directly from available data. These results show that the BF1 mean is about $100\times$ more accurate than the LF mean values and the BF2 mean is about $10\times$ more accurate than the LF mean values, where the BF1 model predicts within $0.15\%$ of the HF QoI mean. The LF models, on the other hand, predict to within only $20-22\%$ of the HF QoI mean. The CoV is provided for all models as well. The QoI variation for each fidelity model is comparable to that of the HF model, with that of BF1 most closely describing the HF data. Since the BF2 CoV is of the same order of magnitude as the respective error, it cannot be completely relied on as an estimate for the HF CoV. The BF1 error, on the other hand, is significantly smaller than the CoV, and thus the CoV estimate can be trusted, indicating that the BF1 approximation most closely represents the HF data. 

\begin{table}[H]
\centering  
\begin{tabular}{| l | ll | l |} 
\hline            
\textbf{Model} & \textbf{Mean} & \textbf{Rel.} & \textbf{CoV}   \\ [0.5ex]     
\textbf{Fidelity} & \textbf{QoI} & \textbf{Error} &   \\ [0.5ex] 
\hline                 
  \hline 
 LF1 &$0.80$ & $20$\%  &$0.10$  \\ [0.5ex] 
BF1&$0.99$  &   $0.15$\% &$0.07$\\ [0.5ex] 
LF2 &$0.78$ & $22$ \% &$0.11$\\ [0.5ex] 
BF2 &$0.97$ & $2.7$\%&$0.06$\\ [0.5ex] 
\hline
\hline
HF &$1.0$   &  -           & $0.08$\\ [0.5ex] 
\hline
\end{tabular}
\caption{\label{tab:pce_hf} Comparison of the mean and CoV the of heat flux estimated by the LF1, LF2, BF1, and BF2 models. These statistics are computed via a sparse PCE surrogate as discussed in Section \ref{sec:bfpce}.}
\end{table}

\begin{figure}[h]
    \centering
    \begin{subfigure}[b]{0.48\textwidth}
    \centering
        \includegraphics[width=.9\textwidth]{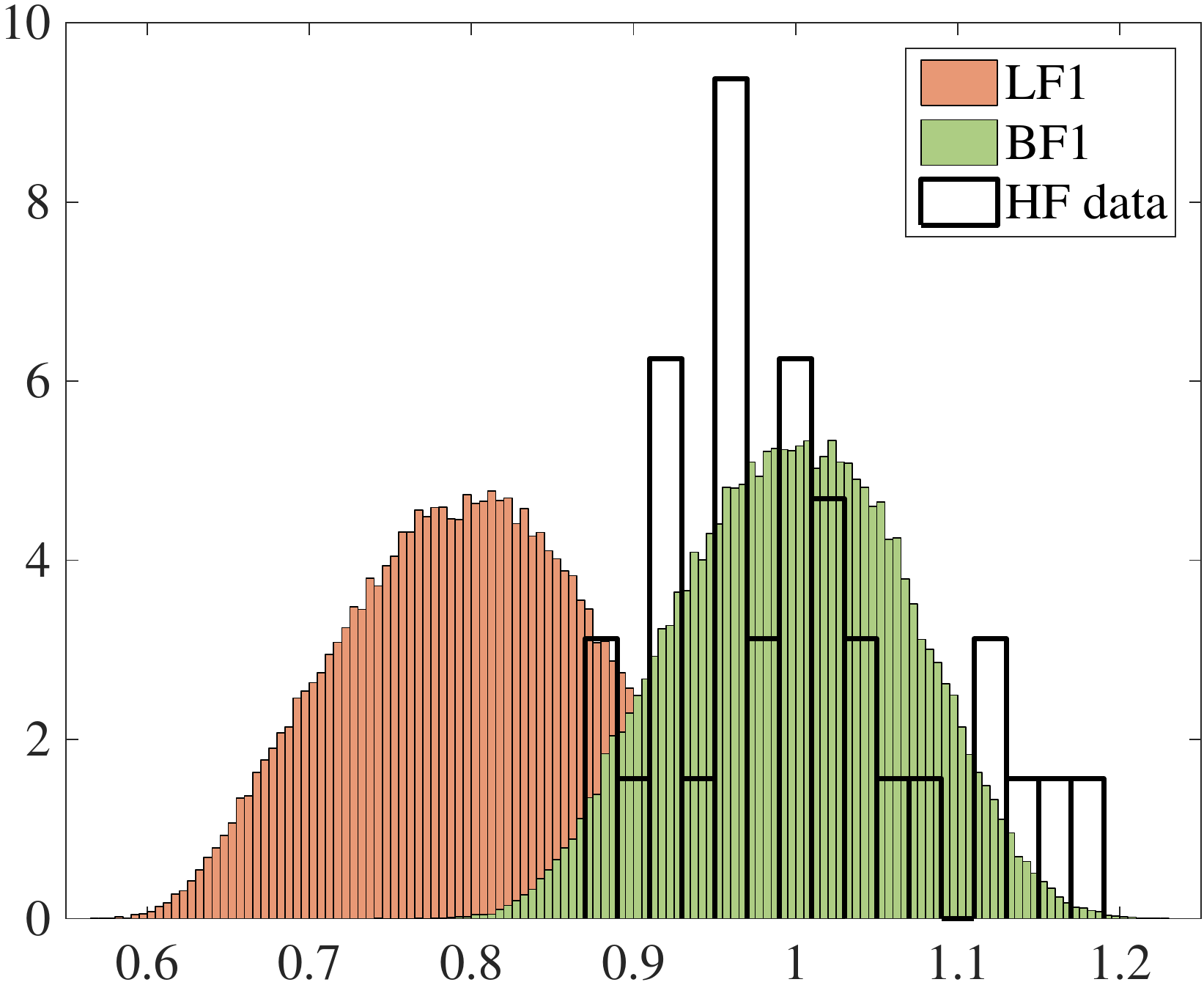}
        \caption{LF1 and BF1}
    \end{subfigure}
    \begin{subfigure}[b]{0.48\textwidth}
    \centering
        \includegraphics[width=.9\textwidth]{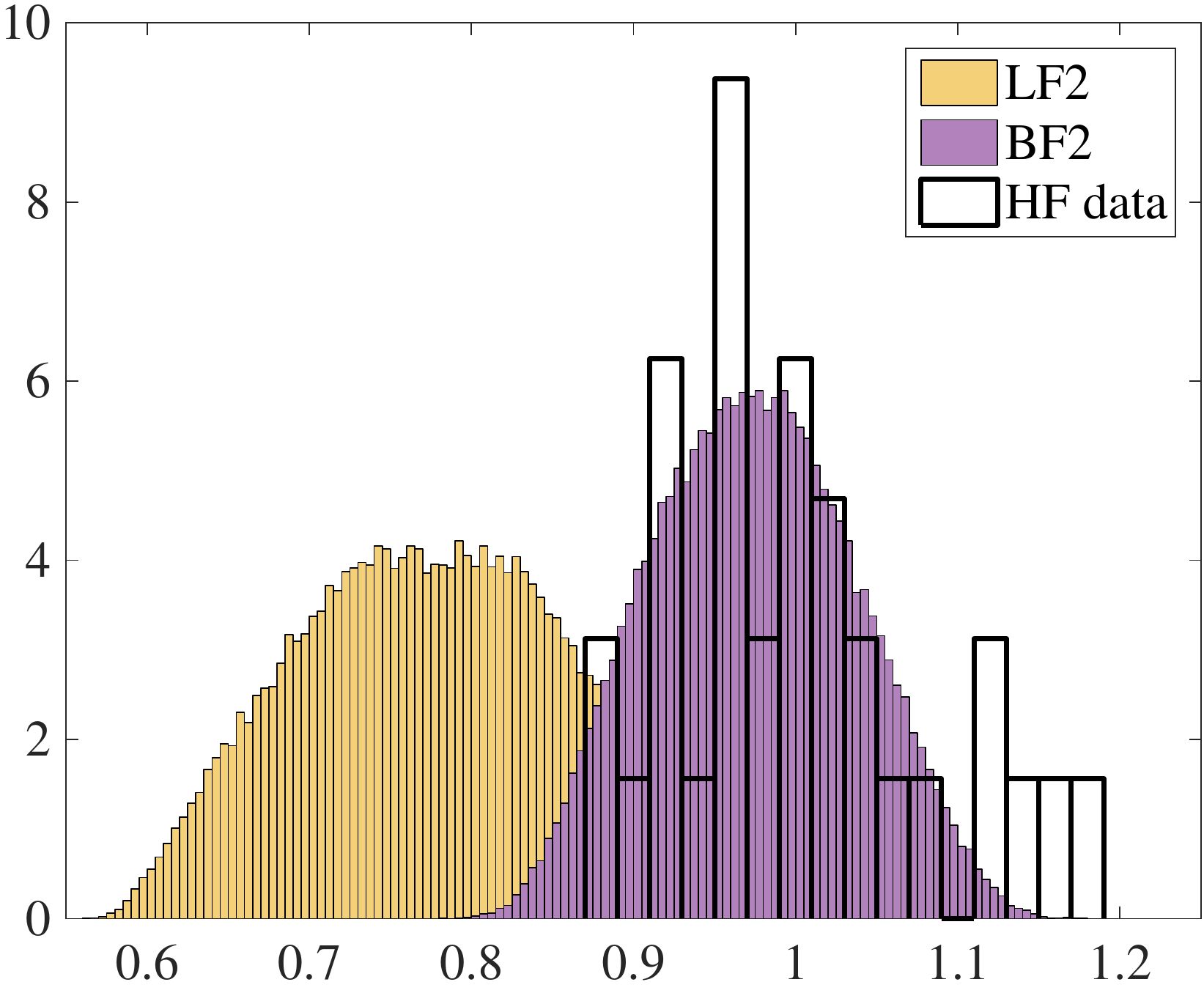}
        \caption{LF2 and BF2}
    \end{subfigure}
   \caption{Normalized histogram of the total normalized heat flux through the $\Delta x= 0.3L$ plane based on sparse PCE of (a) LF1 and BF1 and (b) LF2 and BF2 realizations. \label{fig:exhf_pdf}}
\end{figure}

To estimate the QoI PDFs, histograms of the LF and BF data are generated from $100,000$ sparse PCE samples. Fig. \ref{fig:exhf_pdf} provides these normalized histograms for the heat flux QoI to compare with available HF data. Fig. \ref{fig:exhf_pdf} (a) provides data derived from the LF1 and BF1 models, while Fig. \ref{fig:exhf_pdf} (b) provides the histograms from LF2 and BF2 data. We observe that the BF histograms more closely follow the histogram of the HF data. 

\begin{figure}[H]
    \centering
        \begin{subfigure}[b]{0.48\textwidth}
    \centering
        \includegraphics[width=.7\textwidth]{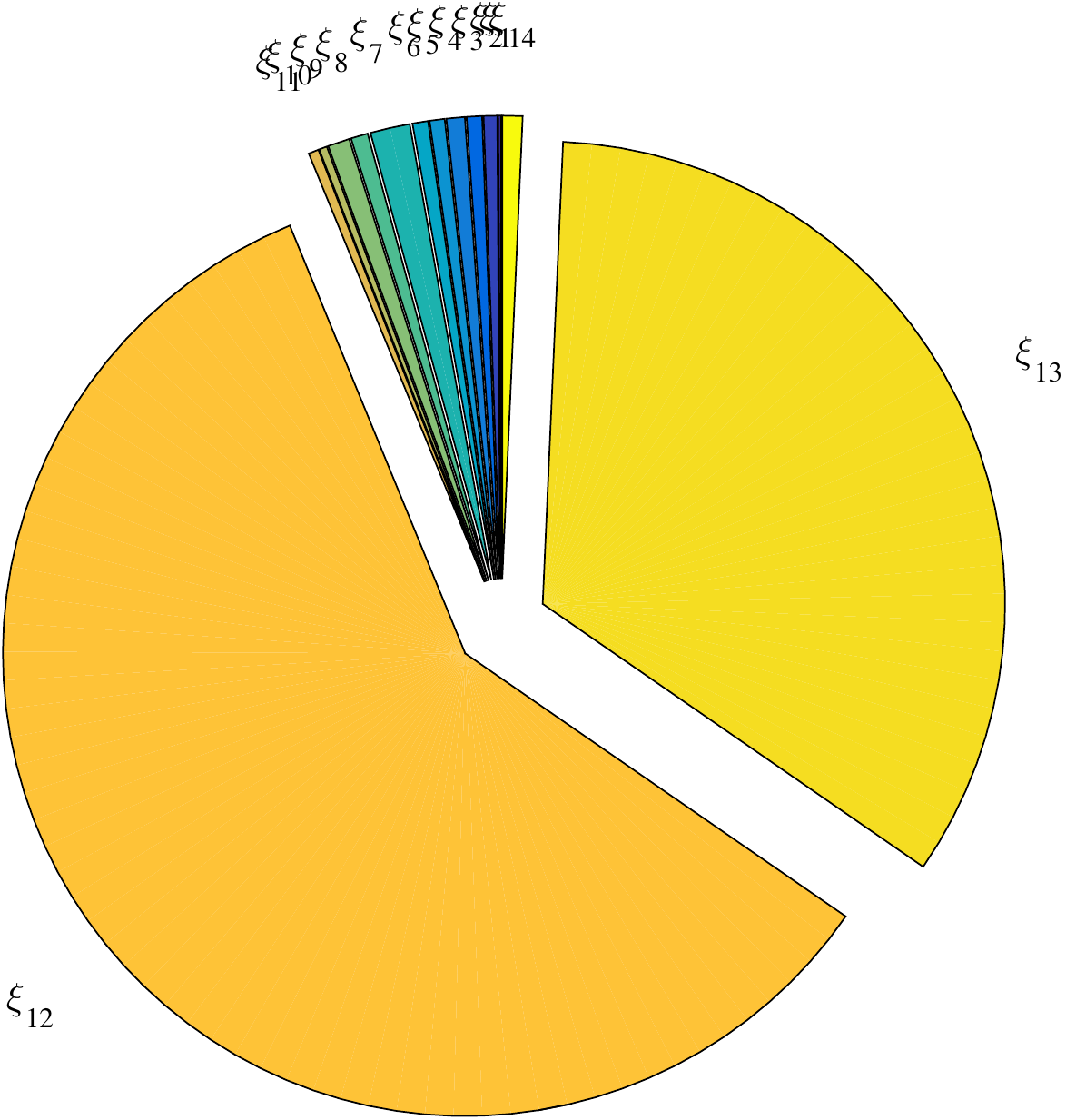}
        \caption{LF1}
    \end{subfigure}
    \begin{subfigure}[b]{0.48\textwidth}
    \centering
        \includegraphics[width=.7\textwidth]{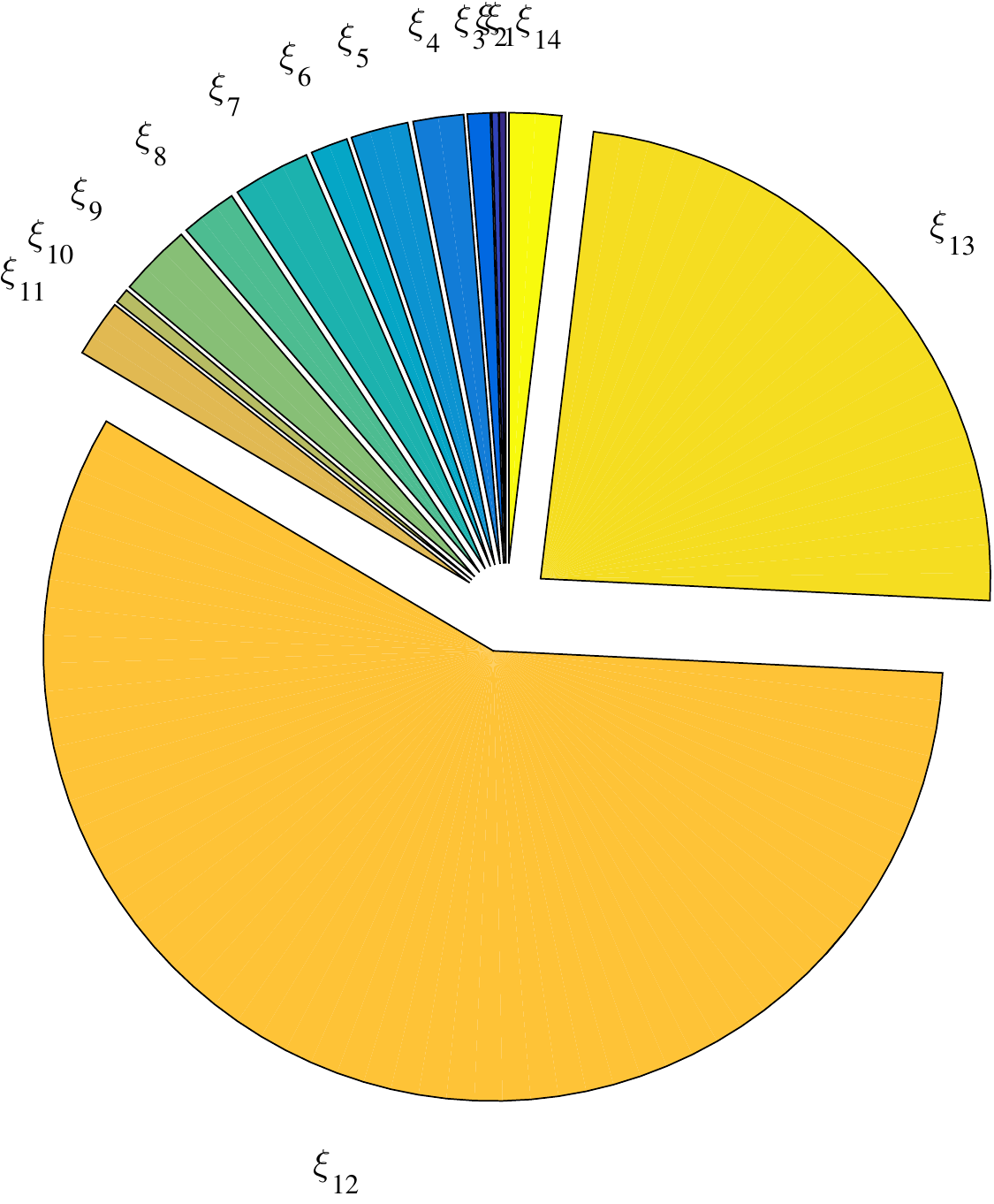}
        \caption{BF1}
    \end{subfigure}
    \begin{subfigure}[b]{0.48\textwidth}
    \centering
        \includegraphics[width=.7\textwidth]{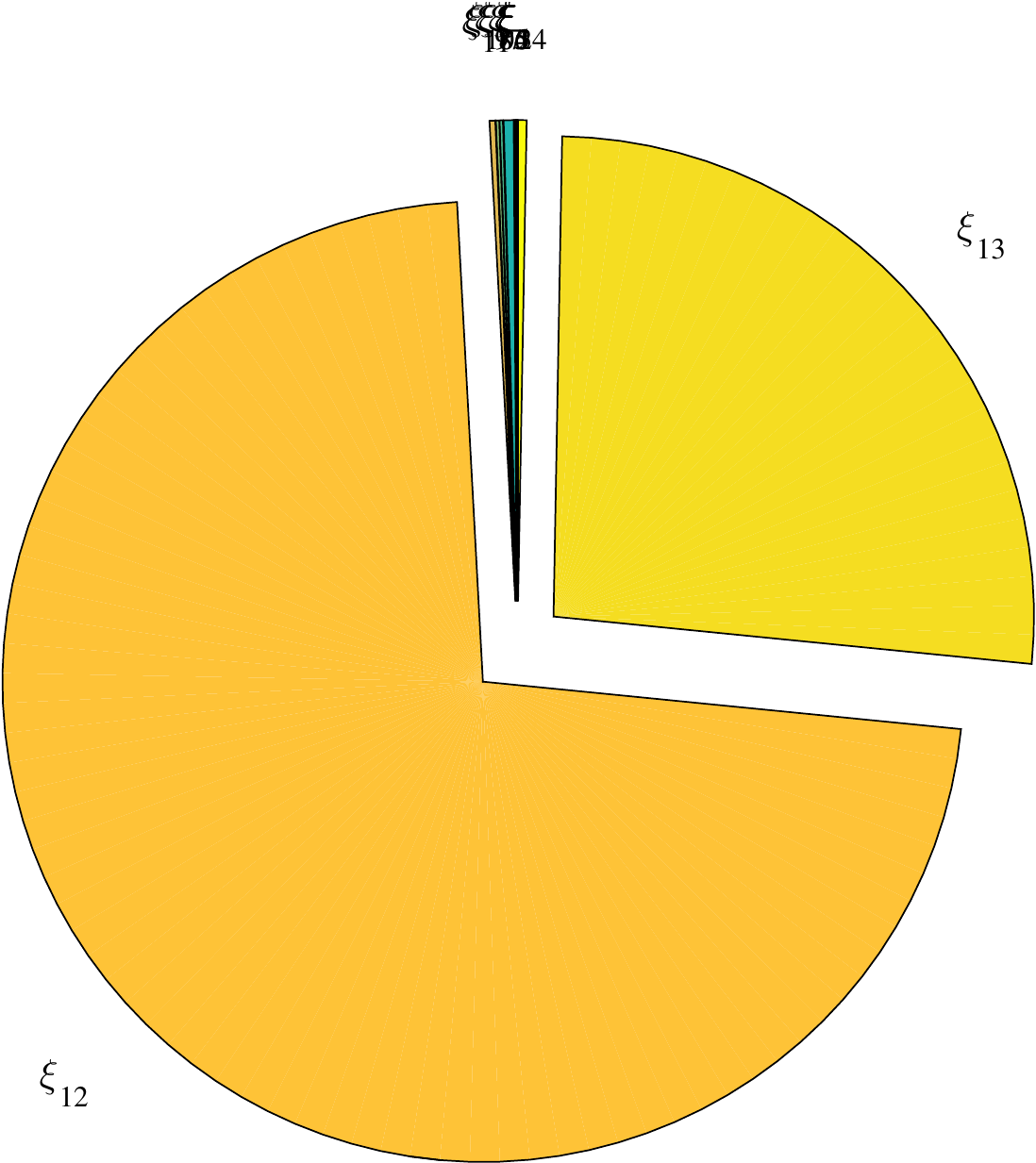}
        \caption{LF2}
    \end{subfigure}
    \begin{subfigure}[b]{0.48\textwidth}
    \centering
        \includegraphics[width=.7\textwidth]{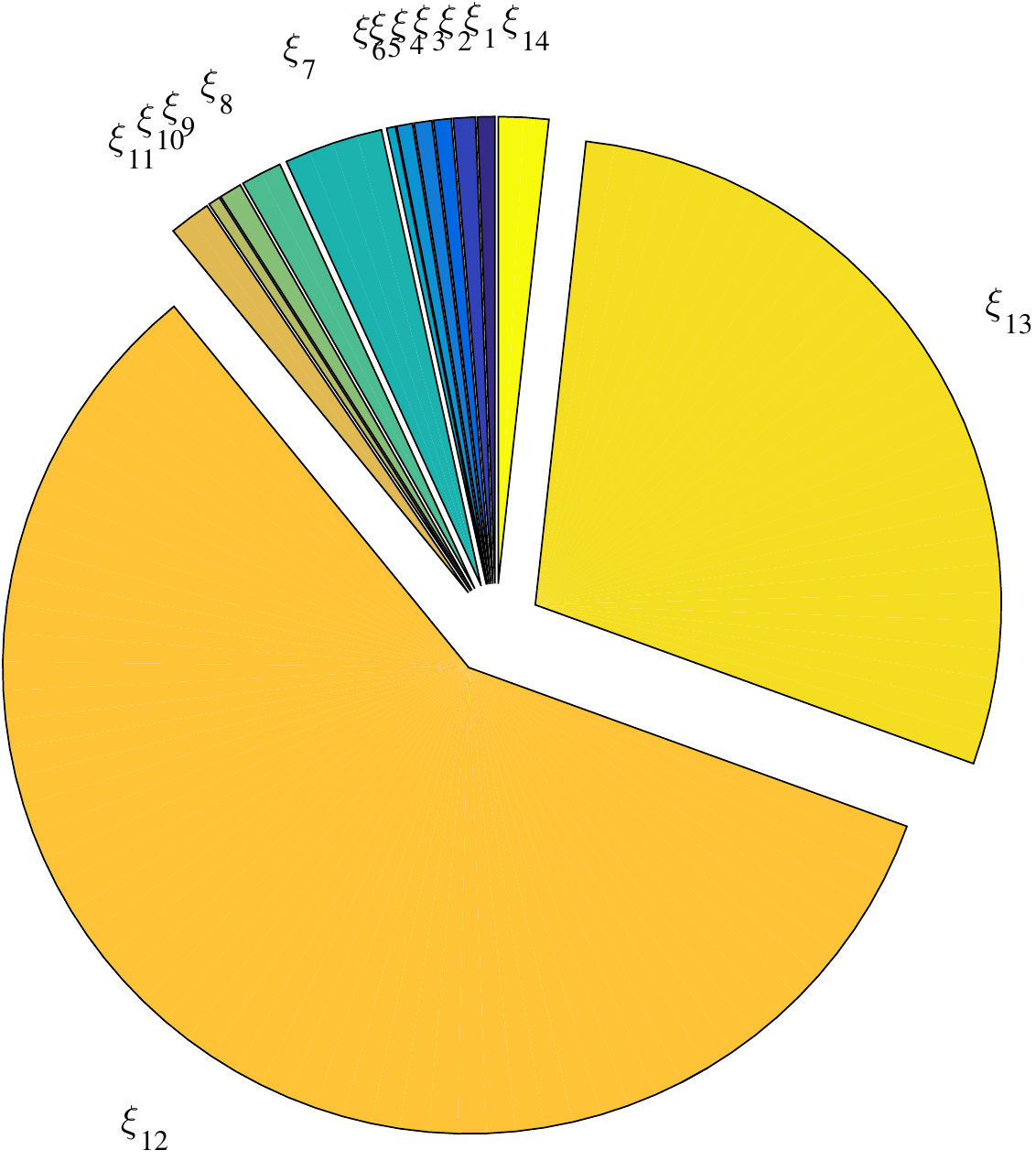}
        \caption{BF2}
    \end{subfigure}
   \caption{Importance of input parameters calculated from PCE coefficients on the (a) LF1, (b) BF1, (c) LF2, and (d) BF2 heat flux data. Starting with parameter $\xi_1$ at the top position, importance of each $\xi_i$ is provided in counterclockwise order with respect to increasing $i$, with corresponding description provided in Table \ref{tab:uncertainties}. As determined from all four model surrogates, heat flux from the radiated wall ($\xi _{12}$) and the opposite wall ($\xi _{13}$) to the fluid contribute the most to the heat flux variability.  \label{fig:exhf_sa}}
\end{figure}

For the final set of results the total Sobol' indices are calculated from the PCE coefficients of the LF and BF heat flux samples. The relative contribution of each parameter to the variance of the estimated heat flux is displayed in Fig. \ref{fig:exhf_sa} (a)-(d). The Sobol' indices determined from all four sets of PCE coefficients clearly indicate that the inputs representing the uncertainty in the heat fluxes from the radiated wall ($\xi _{12}$) and its opposite wall ($\xi _{13}$) contribute the most to the heat flux variation. Comparatively, the remaining uncertain parameters are best distinguished from the BF1 results of Fig. \ref{fig:exhf_sa} (b) based on its accurate CoV estimate. In particular, the BF1 result indicates that inputs $\xi _{i}$ with $i= 4-9, 11$ are next in terms of contribution to the total QoI variance, with about $1\%$ of the variance, and $\xi _{i}$ with $i= 1-3, 10$ have the least important contribution, with $< 0.1\%$ of the total variance. 

As the results of this section have shown, both BF approximations are significantly more accurate than either LF approximation, with the sparse PCE surrogate of the BF1 approximation most accurately estimating the HF heat flux data. A discussion of cost comparisons between the five models may be found in Section~\ref{sec:cost}.

\subsection{QoI \# 2: Spatially-Averaged and Point Estimates of $\Delta T/T_0$ Along Profile at Probe Location}\label{sec:qoi_t}

Next, time-averaged $\Delta T/T_{0} = (\langle T\rangle-T_{0})/T_{0}$ along the profile at the probe location is considered. The BF approximation employs realizations $\bm u $ of the change in temperature $\Delta T/ T_{0}$ along this profile. After investigating the estimation of the full temperature profile for all models, focus is placed on estimating the spatial mean of $\Delta T /T_{0}$ along the profile, as well as $\Delta T /T_{0}$ at three points along the profile, namely, $y/W= 0.5$, $y/W= 0.1$, and $y/W= 0.05$ at the probe location. 

\begin{figure}[h]
    \centering
    \begin{subfigure}[b]{0.48\textwidth}
    \centering
        \includegraphics[width=.92\textwidth]{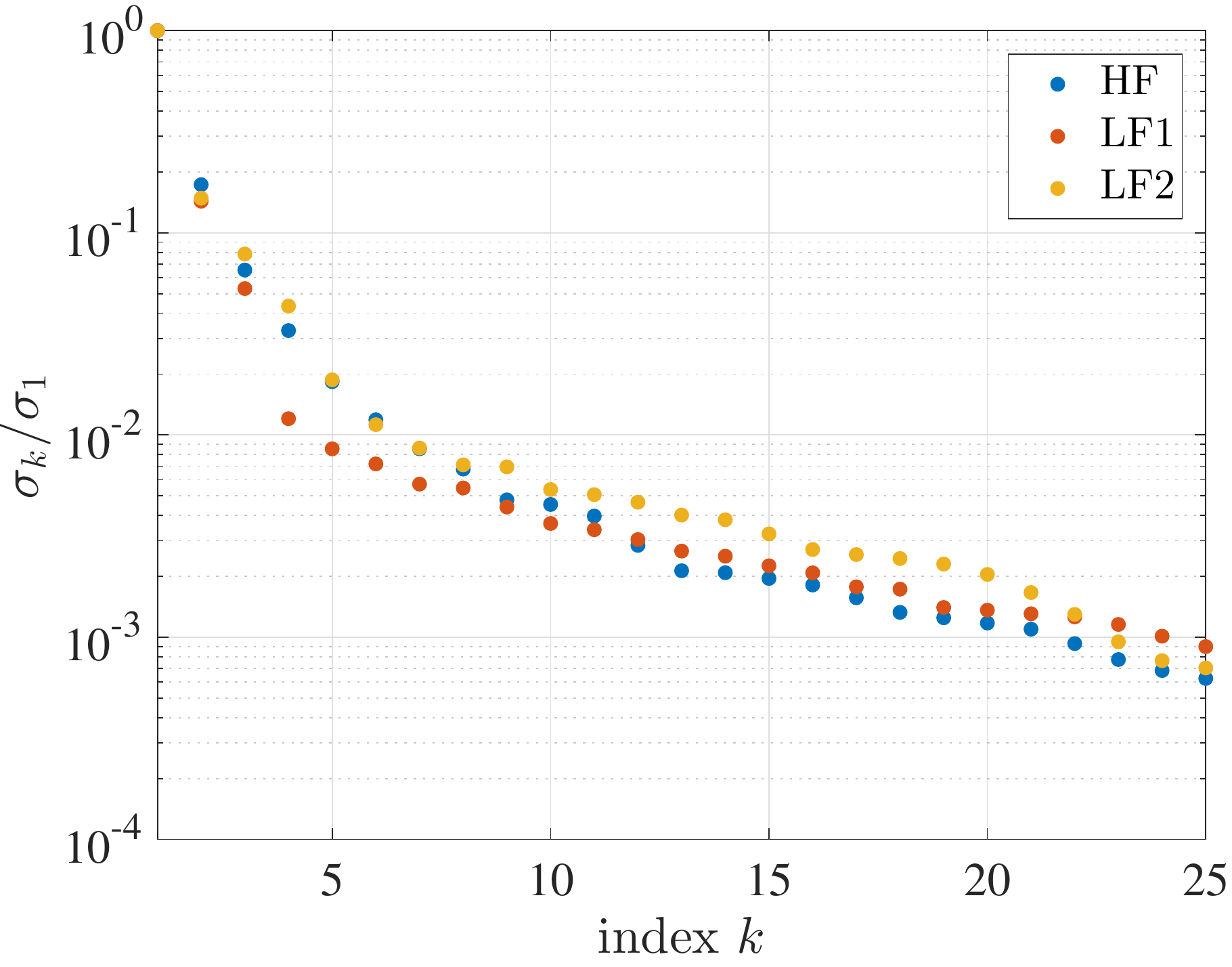}
        \caption{}
    \end{subfigure}
    \begin{subfigure}[b]{0.48\textwidth}
    \centering
        \includegraphics[width=.9\textwidth]{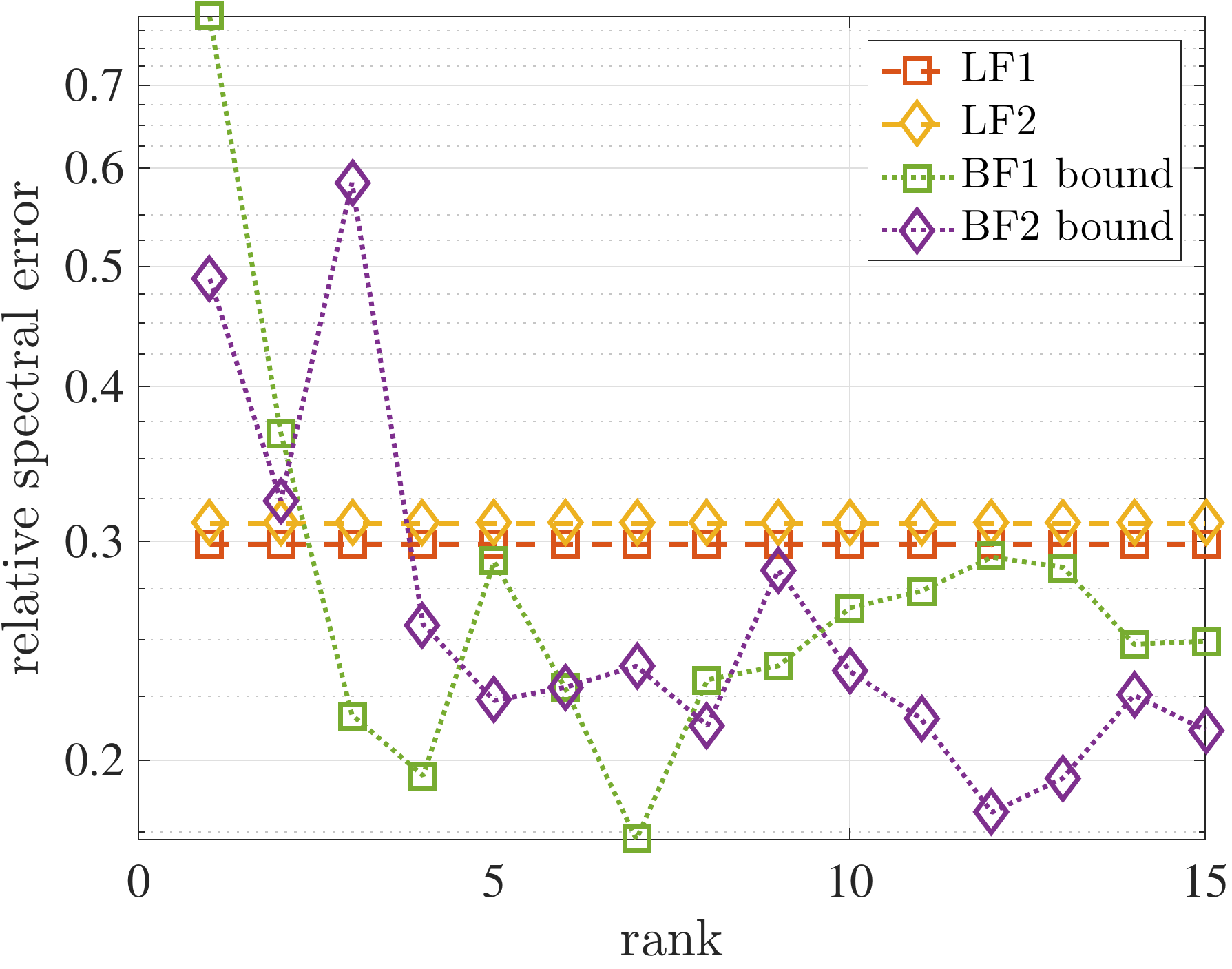}
        \caption{}
    \end{subfigure}
   \caption{(a) Decay of normalized singular values of LF and HF matrices, using available $\Delta T/T_{0}$ data along the profile at the probe location. (b) Error bound estimates of the rank $r$ BF approximations for $\Delta T/T_{0}$ as a function of $r$. For comparison, the LF1 and LF2 relative spectral errors are included as well. \label{fig:ext_sing}} 
\end{figure}

Similar to the heat flux QoI, an optimal rank of the BF approximation must be first identified. 
Fig. \ref{fig:ext_sing} (a) shows the decay of the singular values of the LF and HF matrices, indicating that an approximation rank $r\ge 4$ accurately represents the LF1 and LF2 data. Fig. \ref{fig:ext_sing} (b) displays the BF approximation error bound and the calculated relative spectral errors (see~(\ref{eq:rel_spect_error})) for both LF data as a function of rank $r$. For these error bound estimates, $R = r+2$ simulations are used to estimate the values of $\hat{\epsilon}(\tau)$. As in QoI \#1 above, the error bound estimate for BF1 levels out for the conservative value of $r=6$ (notice the larger value of the error estimate for $r=5$).  We therefore set the BF1 approximation rank to $r=6$. While the BF2 bound suggests a rank of $r=5$, for the interest of a simpler presentation, we chose $r=6$ for BF2 rank as well. 

\begin{figure}[H]
    \centering
    \begin{subfigure}[b]{0.48\textwidth}
    \centering
        \includegraphics[width=.9\textwidth]{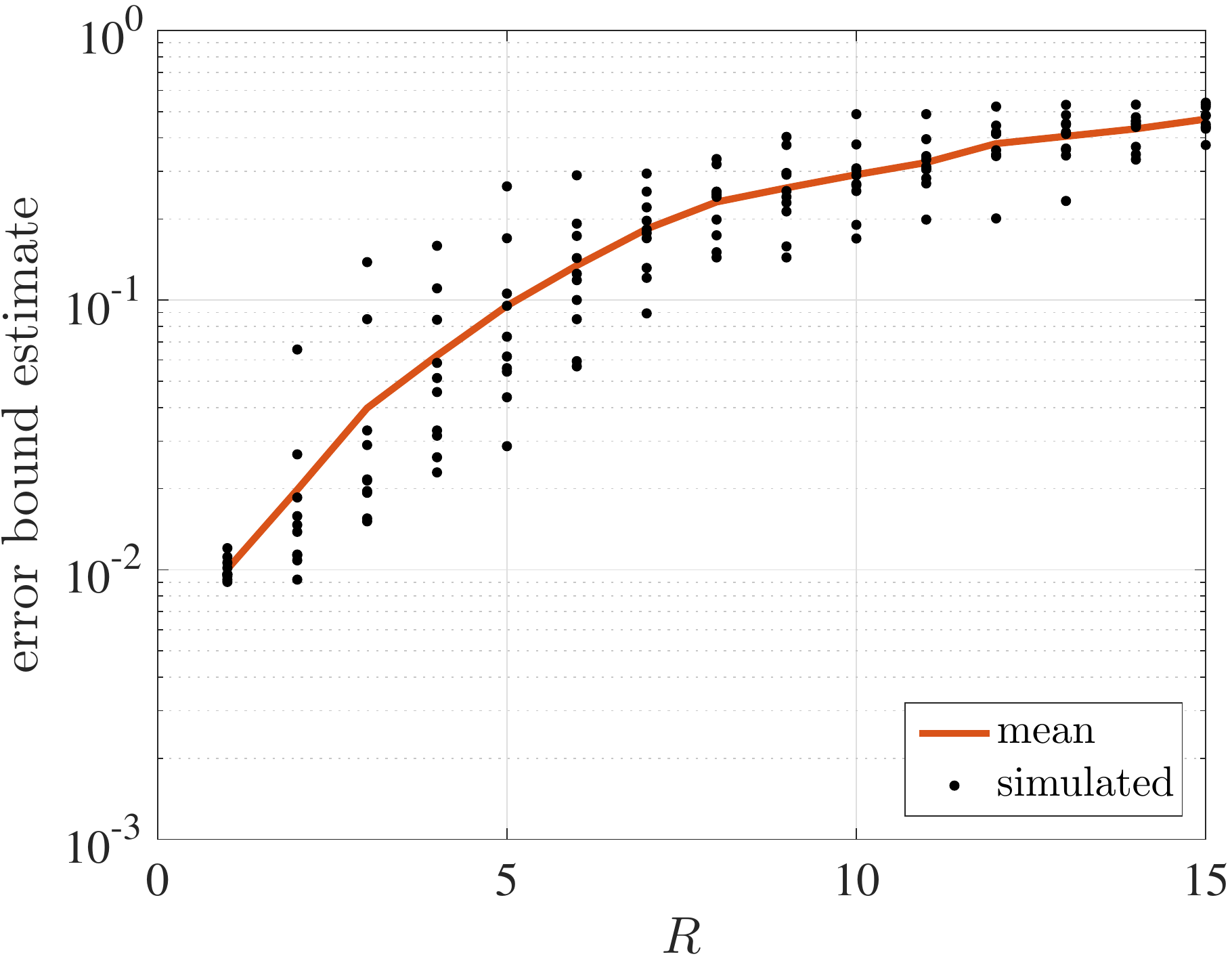}
        \caption{BF1}
    \end{subfigure}
    \begin{subfigure}[b]{0.48\textwidth}
    \centering
        \includegraphics[width=.9\textwidth]{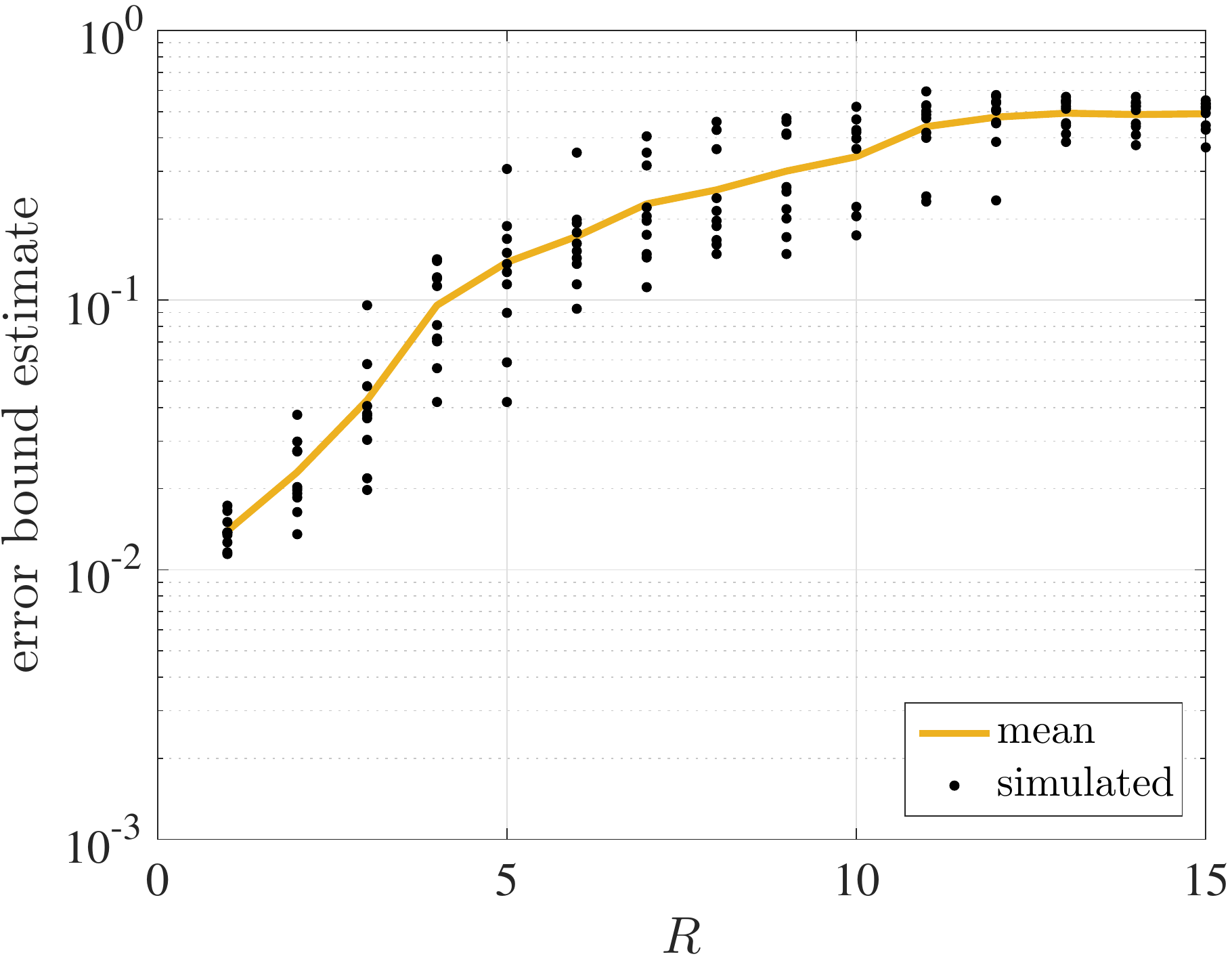}
        \caption{BF2}
    \end{subfigure}
   \caption{Error bound estimate from~(\ref{eq:bound}) using $\hat{\epsilon}(\tau)$ and rank $r=6$ for varying values of $R$, for (a) BF1 and (b) BF2 models. Values are based on $10$ different sets of $R$ columns, where column selection not fully independent due to limited number of available HF samples.\label{fig:ext_bound}}
\end{figure}

\begin{figure}[H]
    \centering
    \begin{subfigure}[b]{0.48\textwidth}
    \centering
        \includegraphics[width=.9\textwidth]{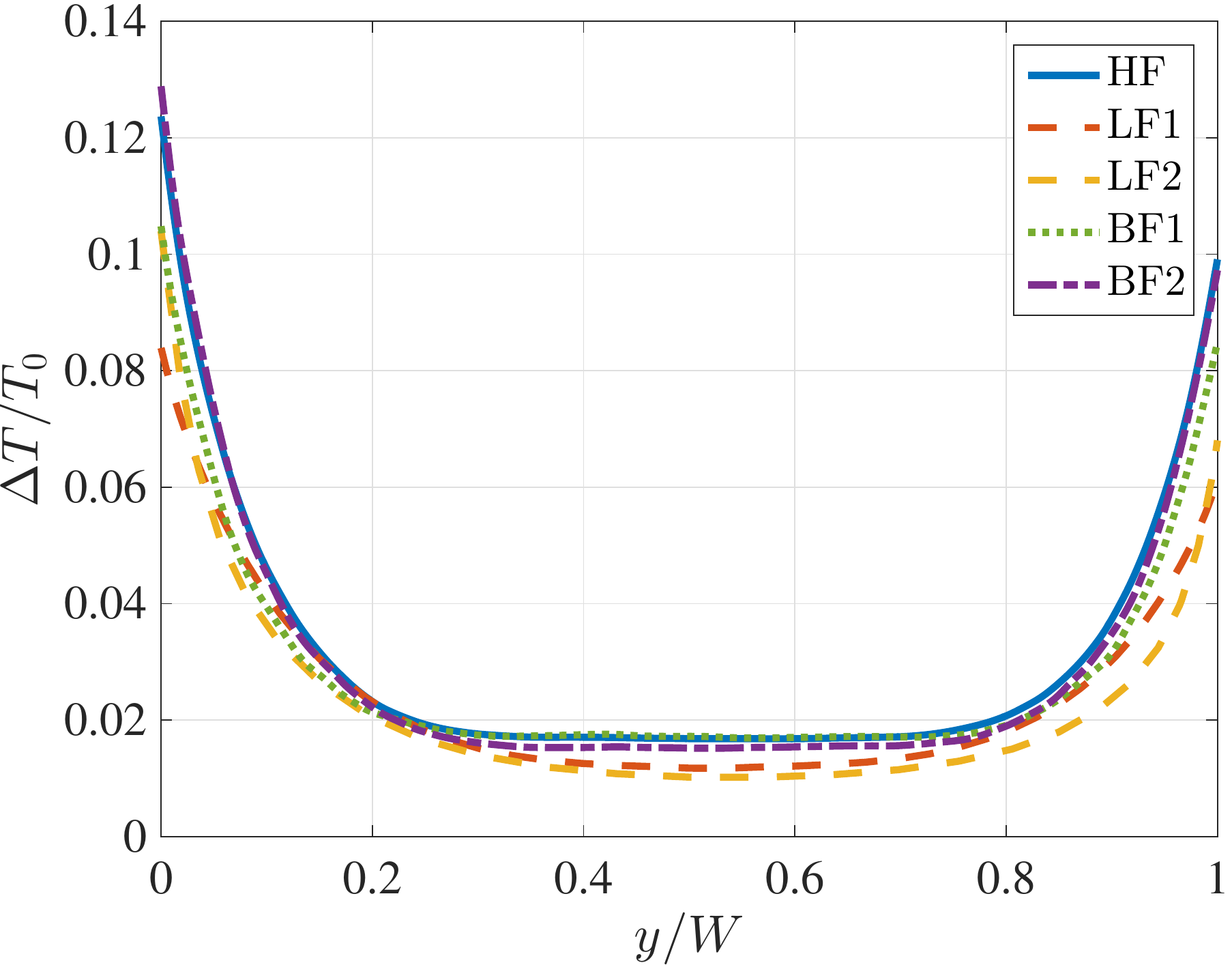}
        \caption{}
    \end{subfigure}
    \begin{subfigure}[b]{0.48\textwidth}
    \centering
        \includegraphics[width=.9\textwidth]{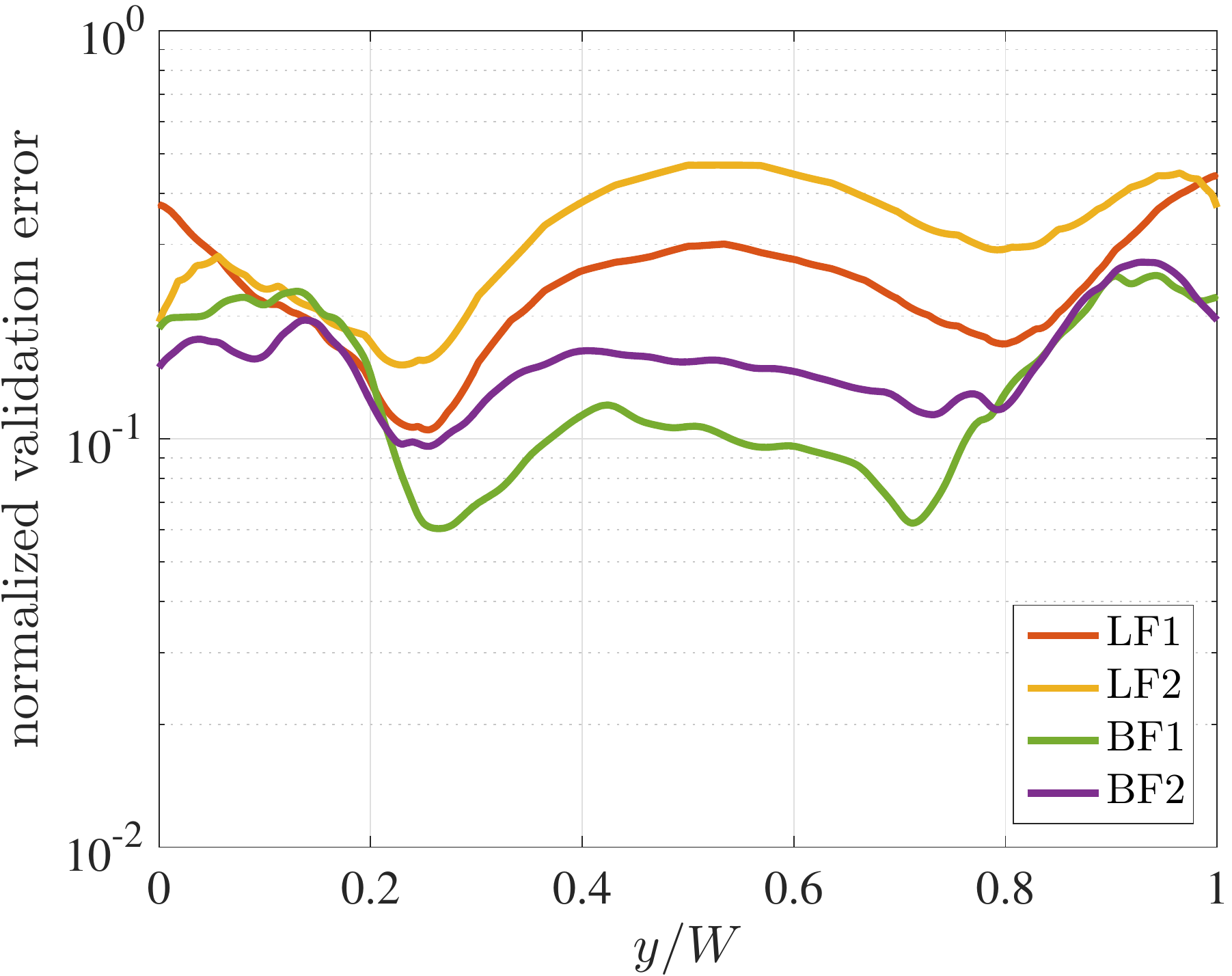}
        \caption{}
    \end{subfigure}
   \caption{(a) Average $\Delta T/T_{0}$ profile calculated from available simulations for all five models with rank $r=6$ for the BF approximations. (b) Error estimates of $\Delta T/T_{0}$ (from (\ref{eq:rel_l2_error})) at each point along the profile. \label{fig:ext_profile}}
\end{figure}

Fig.~\ref{fig:ext_bound} provides the error bound estimates for (a) the BF1 model and (b) the BF2 model, as a function of $R$. Single points indicate individual error bound calculations from $R$ random columns of $\bm U_H$ and $\bm U_L$, out of $26$ total columns (and thus not completely independent). The solid line provides the average of these $10$ points at each value of $R$. Notice that the average bound estimate for BF1 results in both Figs.~\ref{fig:ext_bound} (a) and (b) suggest larger values of $R$ -- compared to $R=r+2$ used in the results of this section -- lead to more accurate BF error estimates. In both cases for $R=15$ the estimated error is larger than the corresponding LF model errors. Naively speaking, this implies that the BF approximation may not lead to significant error improvement over the LF models, unlike in Section~\ref{sec:qoi_hf} for the heat flux data. Fig.~\ref{fig:ext_profile} (b) confirms this observation. 

To compare the abilities of the LF and BF models to reconstruct the $\Delta T/T_{0}$ temperature profile at the probe location, the results from Fig. \ref{fig:ext_profile} are considered. Fig. \ref{fig:ext_profile} (a) displays the average $\Delta T/T_{0}$ temperature profile derived from the LF, BF, and HF models. At most points along the profile, the mean BF1 and BF2 $\Delta T/T_{0}$ are observed to be more accurate than the mean of either LF model. To quantify this error, Fig. \ref{fig:ext_profile} (b) provides the $\ell_2$ error (see~(\ref{eq:rel_l2_error})) evaluated at each point along the profile. While the BF models are more accurate than the LF models at most points, the interior of the profile exhibits the greatest improvement. In the following, the focus is placed on four different QoIs of $\Delta T/T_{0}$: the spatial mean of $\Delta T$ along the profile, and $\Delta T/T_{0}$ quantities at points $y/W=0.5$, $y/W=0.1$, and $y/W=0.05$ along the profile.

\begin{figure}[H]
    \centering
            \begin{subfigure}[b]{0.48\textwidth}
    \centering
        \includegraphics[width=.9\textwidth]{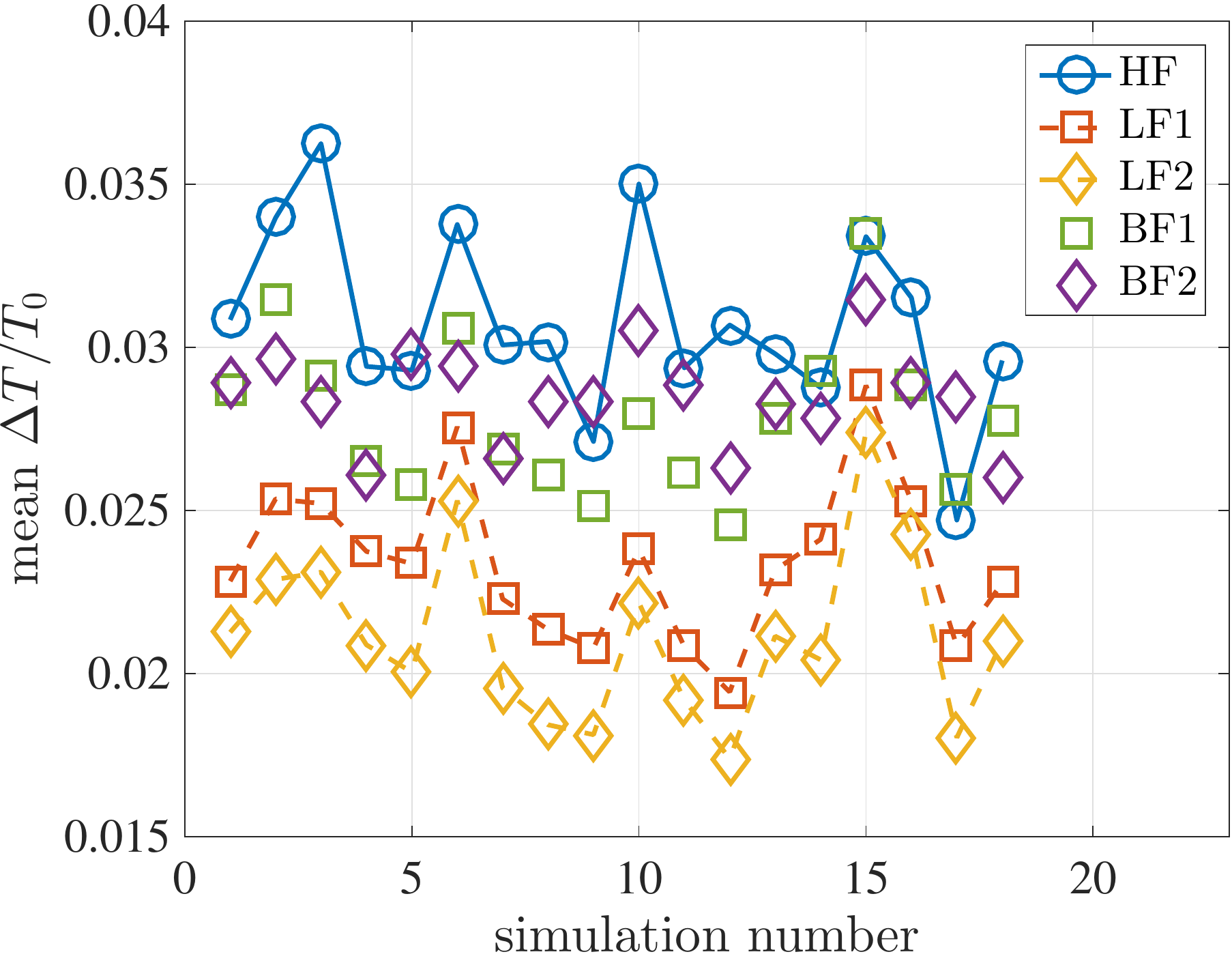}
        \caption{}
    \end{subfigure}
    \begin{subfigure}[b]{0.48\textwidth}
    \centering
        \includegraphics[width=.9\textwidth]{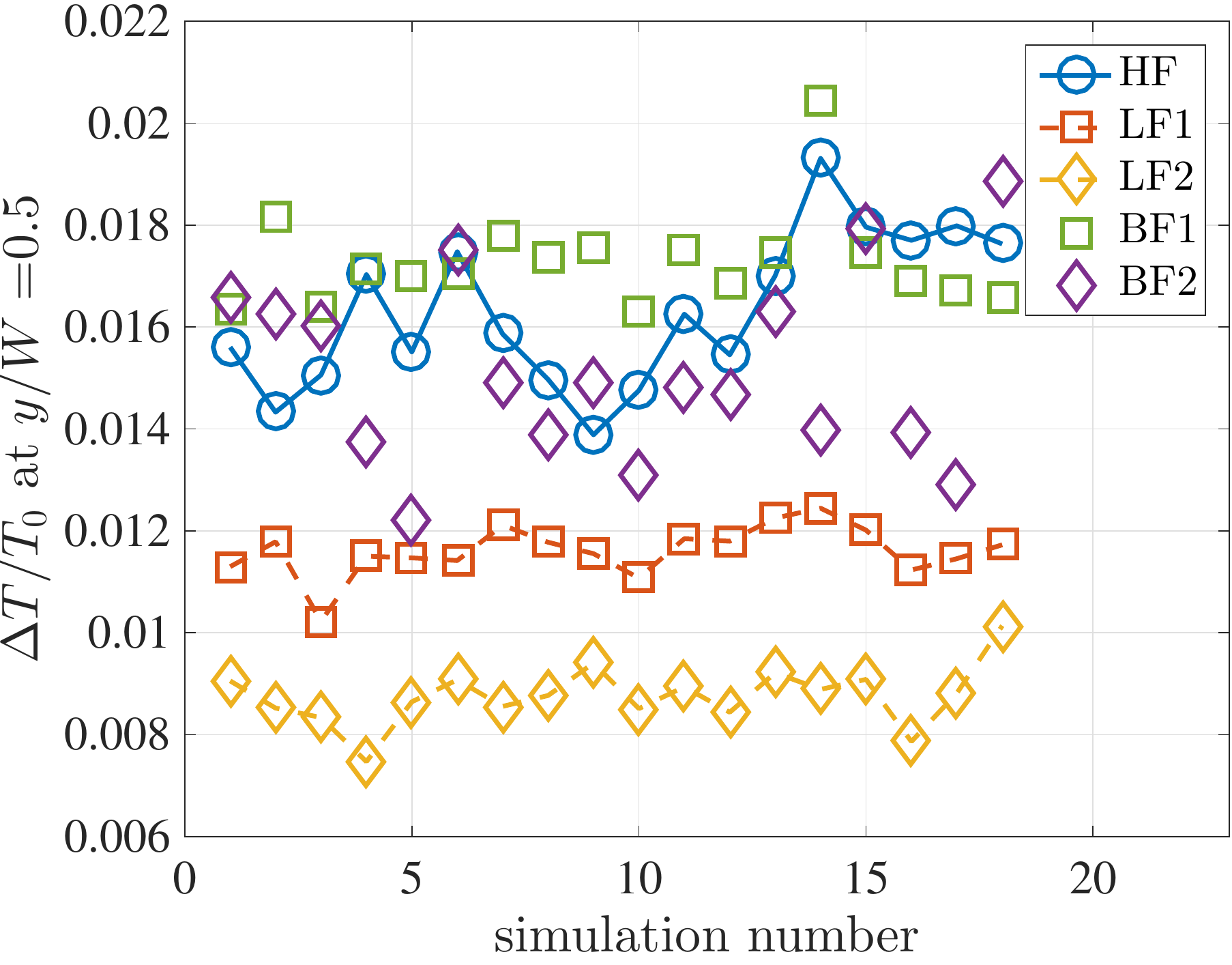}
        \caption{}
    \end{subfigure}
    \begin{subfigure}[b]{0.48\textwidth}
    \centering
        \includegraphics[width=.9\textwidth]{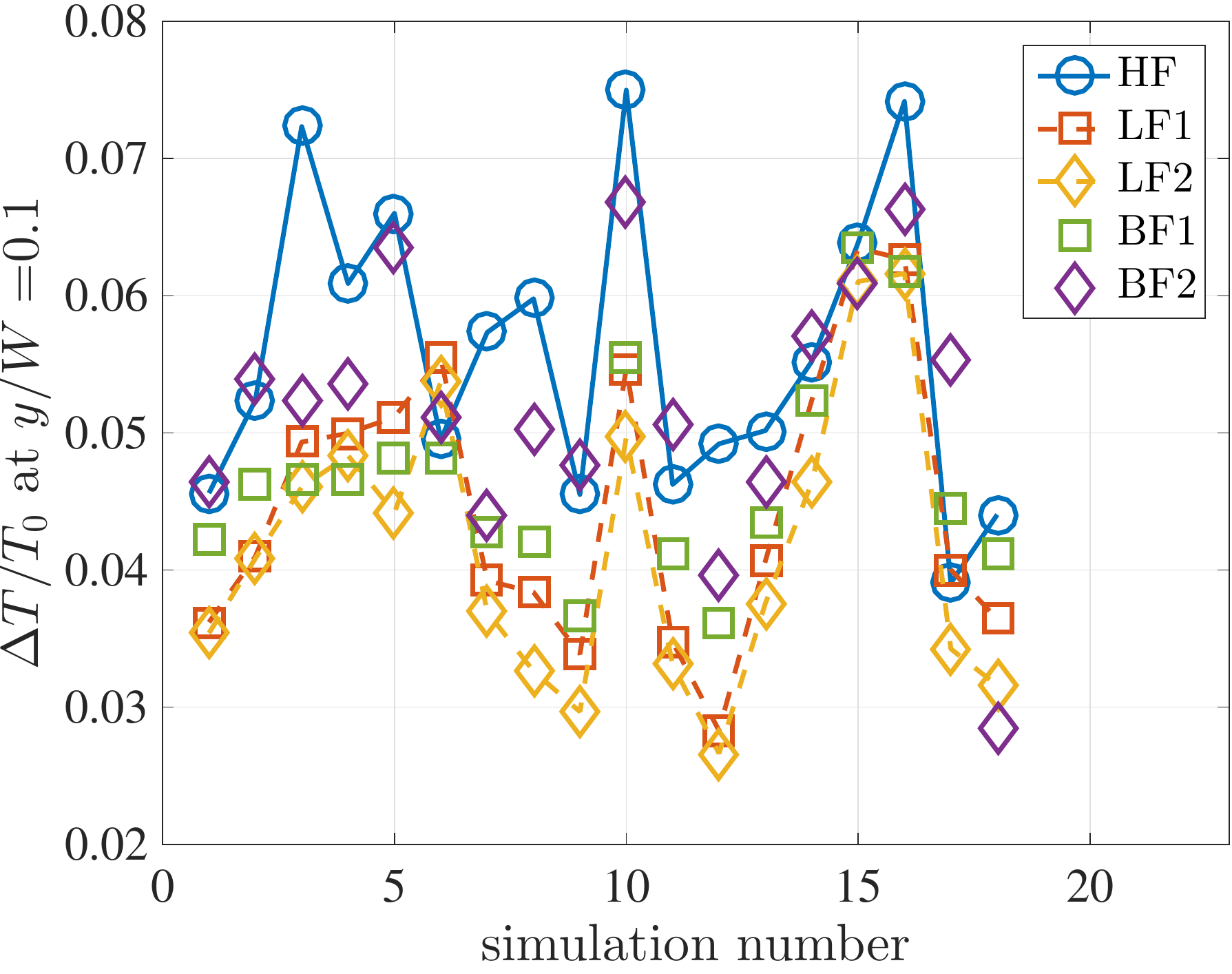}
        \caption{}
    \end{subfigure}
    \begin{subfigure}[b]{0.48\textwidth}
    \centering
        \includegraphics[width=.9\textwidth]{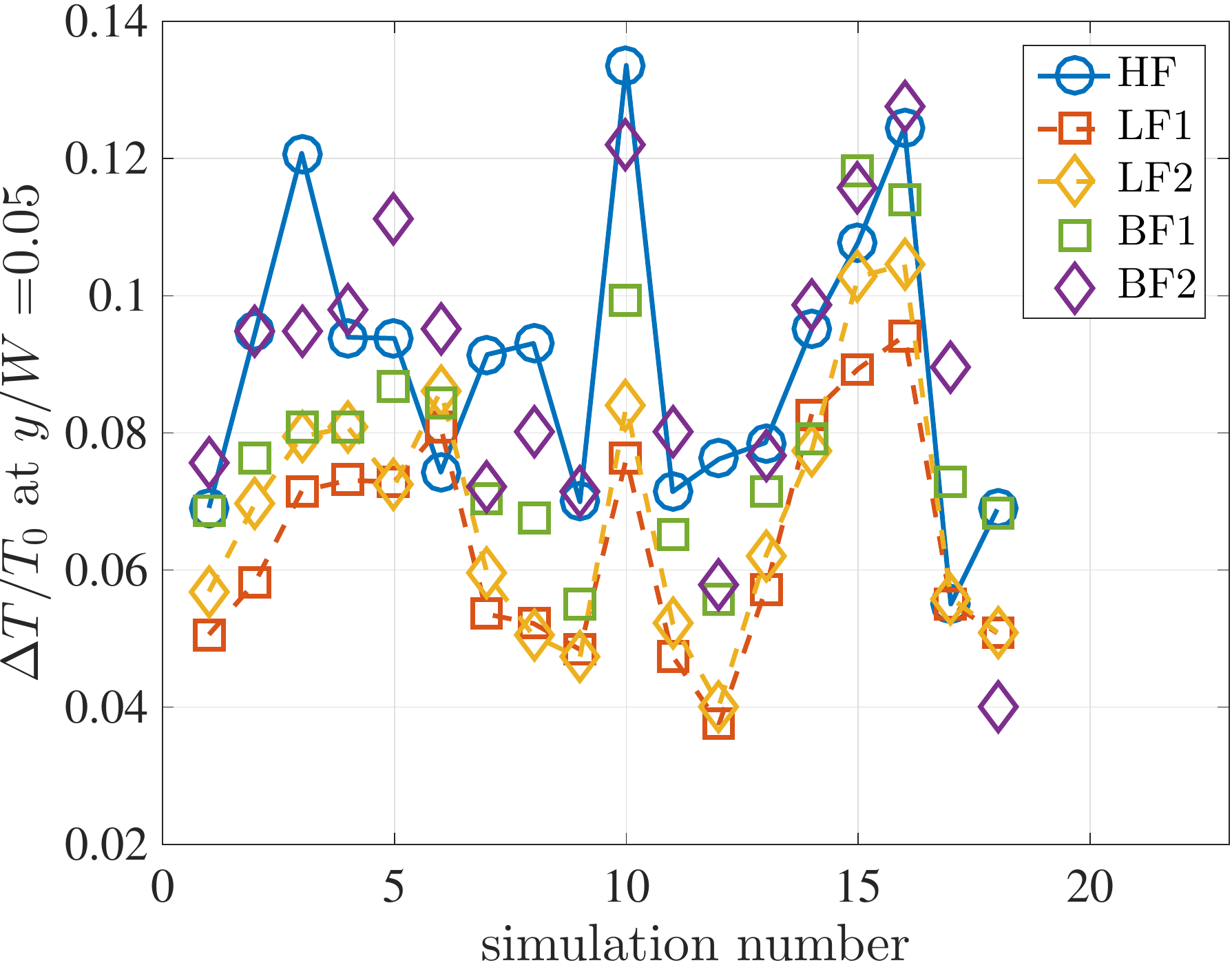}
        \caption{}
    \end{subfigure}
   \caption{$\Delta T/T_0$ values of $17$ independent simulations for the QoI of (a) spatial mean along profile at probe location, and points $y/W=0.5$ (b), $y/W=0.1$ (c), and $y/W=0.05$ (d), along probe profile. Simulated values are from the five different models, where the BF approximation is of rank $r=6$. \label{fig:ext_simulations}}
\end{figure}

Figs~\ref{fig:ext_simulations} (a)-(d) provide the simulated values of the four $\Delta T/T_{0}$ QoIs when using the HF, LF, and BF models. Note these do not include the data used for the BF approximation basis. Fig. \ref{fig:ext_simulations} (a) and (b), which provide the mean value of $\Delta T/T_{0}$ and the value of $\Delta T/T_{0}$ at $y/W=0.5$, respectively, show improved performance of the BF approximation compared to the LF models for all simulations. With regards to the $\ell_2$ error of the QoIs, the mean values are about $2\times$ more accurate and the values at $y/W=0.5$ are $3\times$ more accurate. On the other hand, in Fig. \ref{fig:ext_simulations} (c) and (d), which provide the values of $\Delta T/T_{0}$ at $y/W=0.1$ and $y/W=0.05$, respectively, error improvement does not appear to be significant. For both QoIs, the BF approximations have a smaller $\ell_2$ error than their corresponding LF models; however, the gain is smaller than $2\times$. These results suggest that the approximation performs well for the mean and $\Delta T/T_{0}$ QoIs near the interior of the profile, but accuracy decays for values closer to the walls.

While prior temperature results were calculated directly from the simulated values, to estimate the moments and PDFs, as well as to perform sensitivity analysis, surrogates of the LF and BF models are formed via sparse PCE approximations. For each of the four temperature QoIs, the mean and CoV estimates determined from the sparse PCE coefficients are provided in Table \ref{tab:pce_t} (a)-(d), as well as the relative error between each QoI mean and the HF QoI mean. For each subtable in Table \ref{tab:pce_t}, the BF mean estimates are more accurate than either of the LF estimates, by a factor of $3-11$. Between the two BF approximations, BF2 consistently has a small error. Aside from Table \ref{tab:pce_t} (b), in which BF1 is more accurate ($2\%$ error vs. $4\%$ error), the BF2 errors are the smallest for each subtable. In terms of the QoI CoV estimates, the values of all models are comparable to that of the HF data. In Table \ref{tab:pce_t} (a) and (b) the CoV results are similar to those of the heat flux results; specifically, since the BF CoV is the same order of magnitude of the respective errors, it is not necessarily distinguishable and thus can't completely be relied on as an estimate for the HF CoV. However, as the BF approximations exhibit small errors and CoV values, the mean estimates are reliable. 
On the other hand, the BF CoV estimates of Table \ref{tab:pce_t} (c) and (d) are larger than the corresponding error, and therefore are more dependable than the estimates in Table \ref{tab:pce_t} (a) and (b). Because of this, stronger conclusions may be made when performing global sensitivity analysis.
\begin{table}[H]
    \begin{subtable}{.5\linewidth}
      \centering
\begin{tabular}{| l | ll | l |} 
\hline            
\textbf{Model} & \textbf{Mean} & \textbf{Rel.} & \textbf{CoV}   \\ [0.5ex]     
\textbf{Fidelity} & \textbf{QoI} & \textbf{Error} &   \\ [0.5ex] 
\hline                 
  \hline 
 LF1 &$0.024$ & $21.7$\%  &$0.12$  \\ [0.5ex] 
BF1&$0.028$  &   $8.11$\% &$0.11$\\ [0.5ex] 
LF2 &$0.022$ & $28$ \% &$0.13$\\ [0.5ex] 
BF2 &$0.029$ & $3.3$\%&$0.07$\\ [0.5ex] 
\hline
\hline
HF &$ 0.031$   &  -           & $0.10$\\ [0.5ex] 
\hline
\end{tabular}
\caption{Spatial mean $\Delta T/T_0$}
    \end{subtable}
    \begin{subtable}{.5\linewidth}
      \centering
\begin{tabular}{| l | ll | l |} 
\hline            
\textbf{Model} & \textbf{Mean} & \textbf{Rel.} & \textbf{CoV}   \\ [0.5ex]     
\textbf{Fidelity} & \textbf{QoI} & \textbf{Error} &   \\ [0.5ex] 
\hline                 
  \hline 
 LF1 &$0.012$ & $31$\%  &$0.04$  \\ [0.5ex] 
BF1&$0.017$  &   $2.0$\% &$0.06$\\ [0.5ex] 
LF2 &$0.010$ & $40$ \% &$0.09$\\ [0.5ex] 
BF2 &$0.016$ & $4.4$\%&$0.10$\\ [0.5ex] 
\hline
\hline
HF &$0.017$   &  -           & $0.08$\\ [0.5ex] 
\hline
\end{tabular}
\caption{$\Delta T/T_0$ at $y/W=0.5$}
    \end{subtable} %
        \begin{subtable}{.5\linewidth}
      \centering
\begin{tabular}{| l | ll | l |} 
\hline            
\textbf{Model} & \textbf{Mean} & \textbf{Rel.} & \textbf{CoV}   \\ [0.5ex]     
\textbf{Fidelity} & \textbf{QoI} & \textbf{Error} &   \\ [0.5ex] 
\hline                 
  \hline 
 LF1 &$0.046$ & $16.1$\%  &$0.25$  \\ [0.5ex] 
BF1&$0.047$  &   $14.7$\% &$0.21$\\ [0.5ex] 
LF2 &$0.042$ & $24.0$ \% &$0.27$\\ [0.5ex] 
BF2 &$0.052$ & $5.1$\%&$0.20$\\ [0.5ex] 
\hline
\hline
HF &$0.055$   &  -           & $0.20$\\ [0.5ex] 
\hline
\end{tabular}
\caption{$\Delta T/T_0$ at $y/W=0.1$}
    \end{subtable}
        \begin{subtable}{.5\linewidth}
      \centering
\begin{tabular}{| l | ll | l |} 
\hline            
\textbf{Model} & \textbf{Mean} & \textbf{Rel.} & \textbf{CoV}   \\ [0.5ex]     
\textbf{Fidelity} & \textbf{QoI} & \textbf{Error} &   \\ [0.5ex] 
\hline                 
  \hline 
 LF1 &$0.066$ & $27.3$\%  &$0.28$  \\ [0.5ex] 
BF1&$0.081$  & $11.1$\% &$0.24$\\ [0.5ex] 
LF2 &$0.071$ & $22.6$ \% &$0.30$\\ [0.5ex] 
BF2 &$0.088$ & $2.8$\%&$0.27$\\ [0.5ex] 
\hline
\hline
HF &$0.091$   &  -           & $0.25$\\ [0.5ex] 
\hline
\end{tabular}
        \caption{$\Delta T/T_0$ at $y/W=0.05$}
    \end{subtable}
    \caption{Statistics from sparse PCE\label{tab:pce_t}}
\end{table}

\begin{figure}[H]
    \centering
    \begin{subfigure}[b]{0.48\textwidth}
    \centering
        \includegraphics[trim = 15mm 70mm 20mm  70mm, clip, width=.7\textwidth]{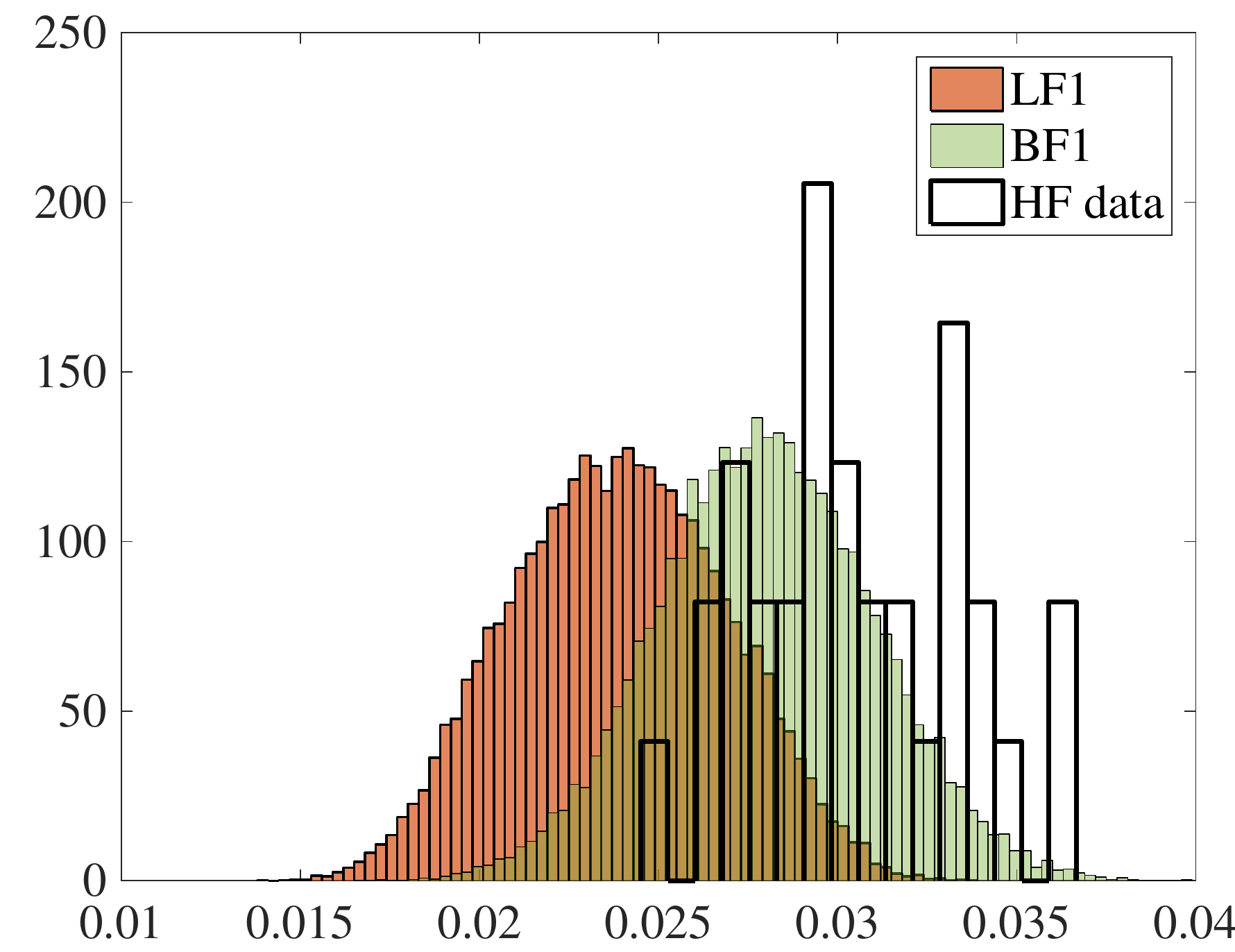}
        \caption{LF1 and BF1 for mean $\Delta T/T_0$}
    \end{subfigure}
    \begin{subfigure}[b]{0.48\textwidth}
    \centering
        \includegraphics[trim = 15mm 70mm 20mm  70mm, clip, width=.7\textwidth]{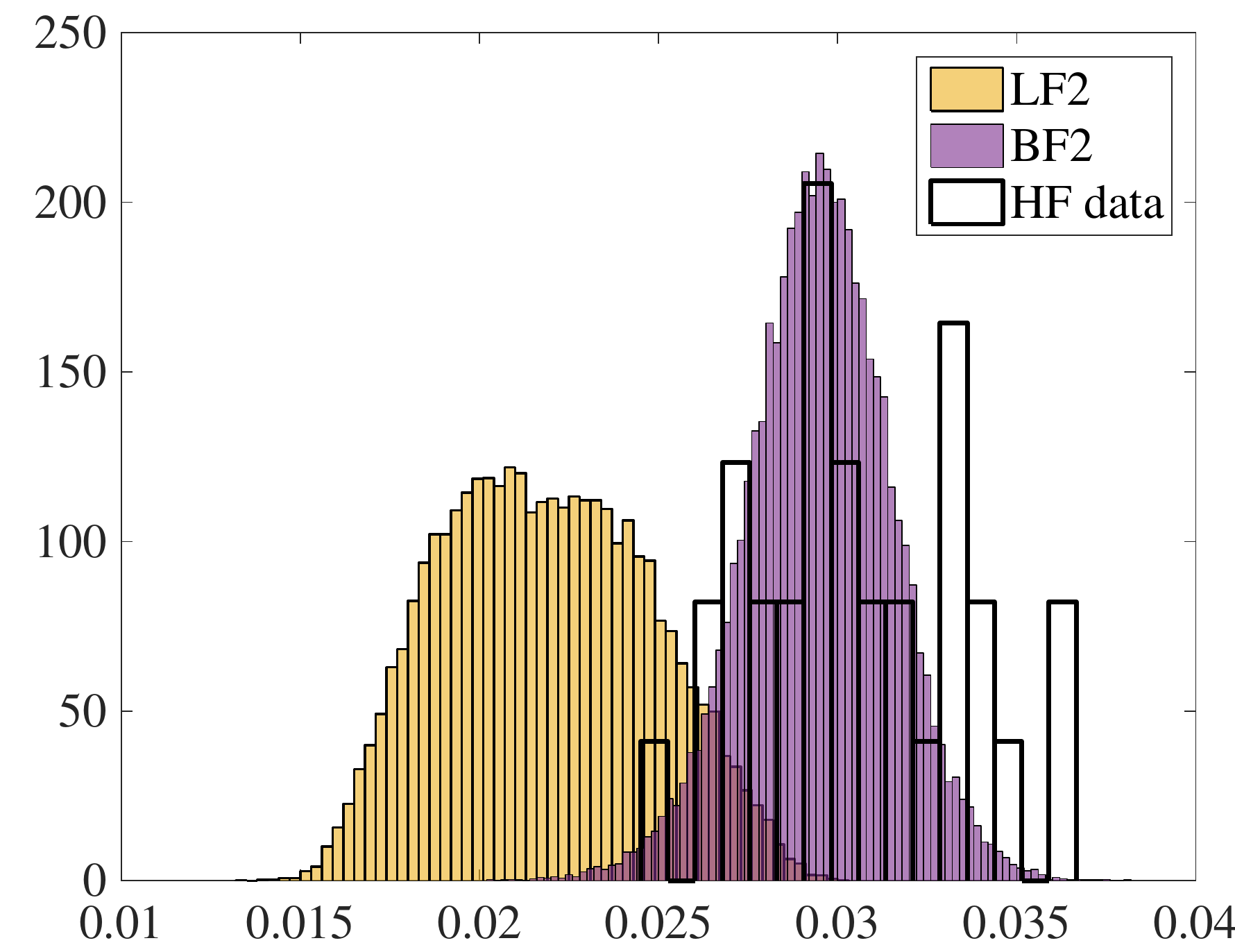}
        \caption{LF2 and BF2 for mean $\Delta T/T_0$}
    \end{subfigure}
        \begin{subfigure}[b]{0.48\textwidth}
    \centering
        \includegraphics[trim = 15mm 70mm 20mm  70mm, clip, width=.7\textwidth]{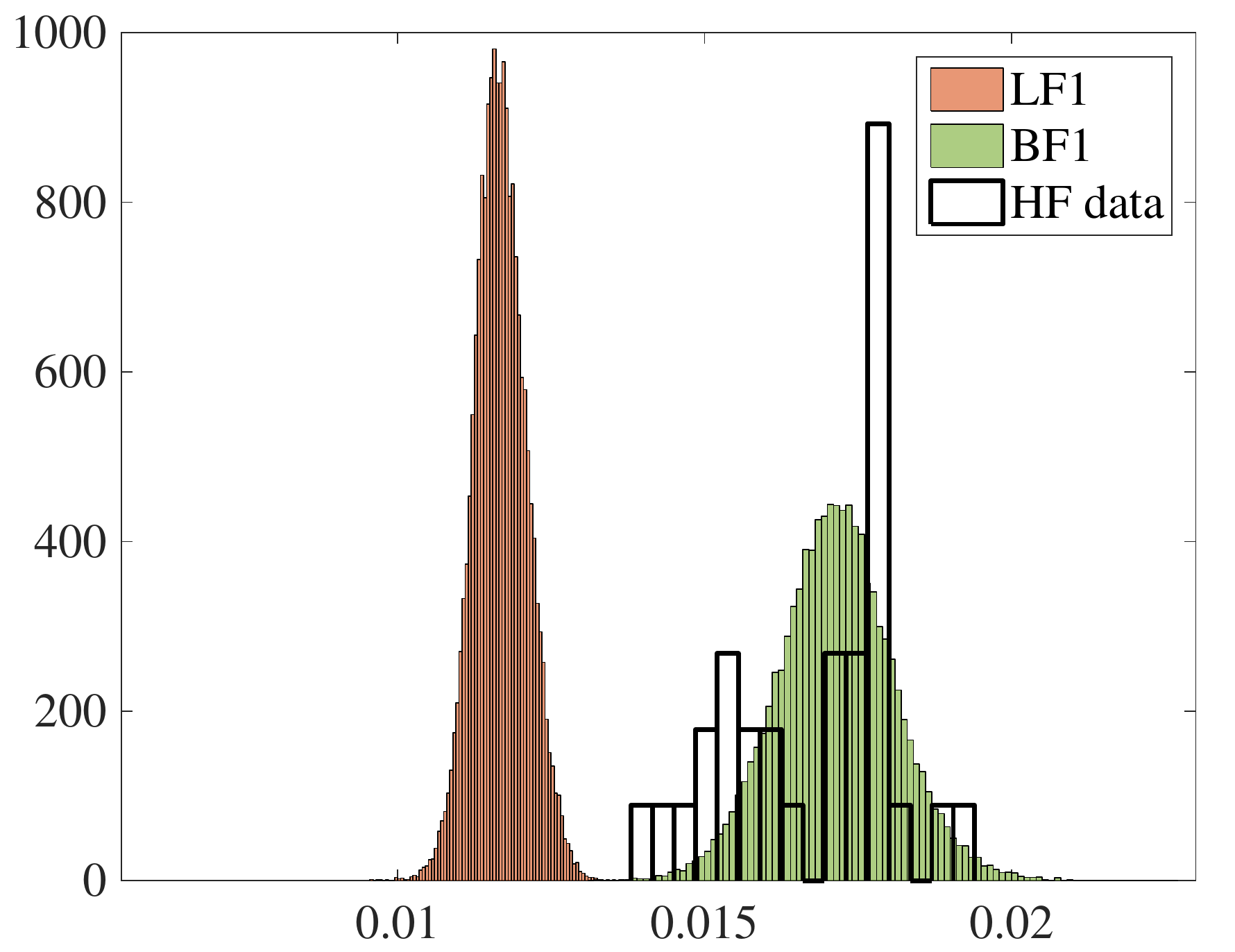}
        \caption{LF1 and BF1 for $\Delta T/T_0$  at $y/W=0.5$}
    \end{subfigure}
    \begin{subfigure}[b]{0.48\textwidth}
    \centering
        \includegraphics[trim = 15mm 70mm 20mm  70mm, clip, width=.7\textwidth]{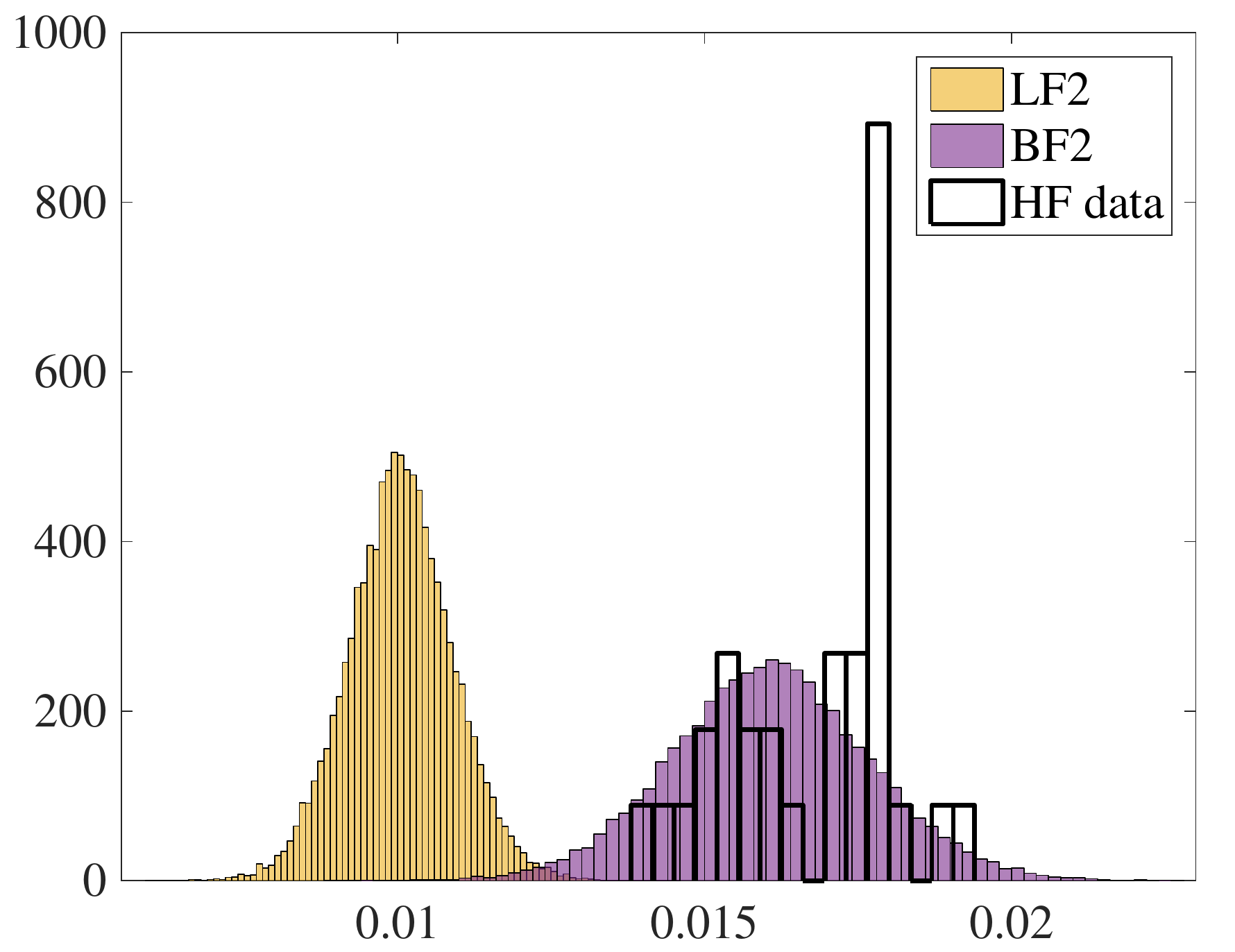}
        \caption{LF2 and BF2 for $\Delta T/T_0$  at $y/W=0.5$}
    \end{subfigure}
        \begin{subfigure}[b]{0.48\textwidth}
    \centering
        \includegraphics[trim = 15mm 70mm 20mm  70mm, clip, width=.7\textwidth]{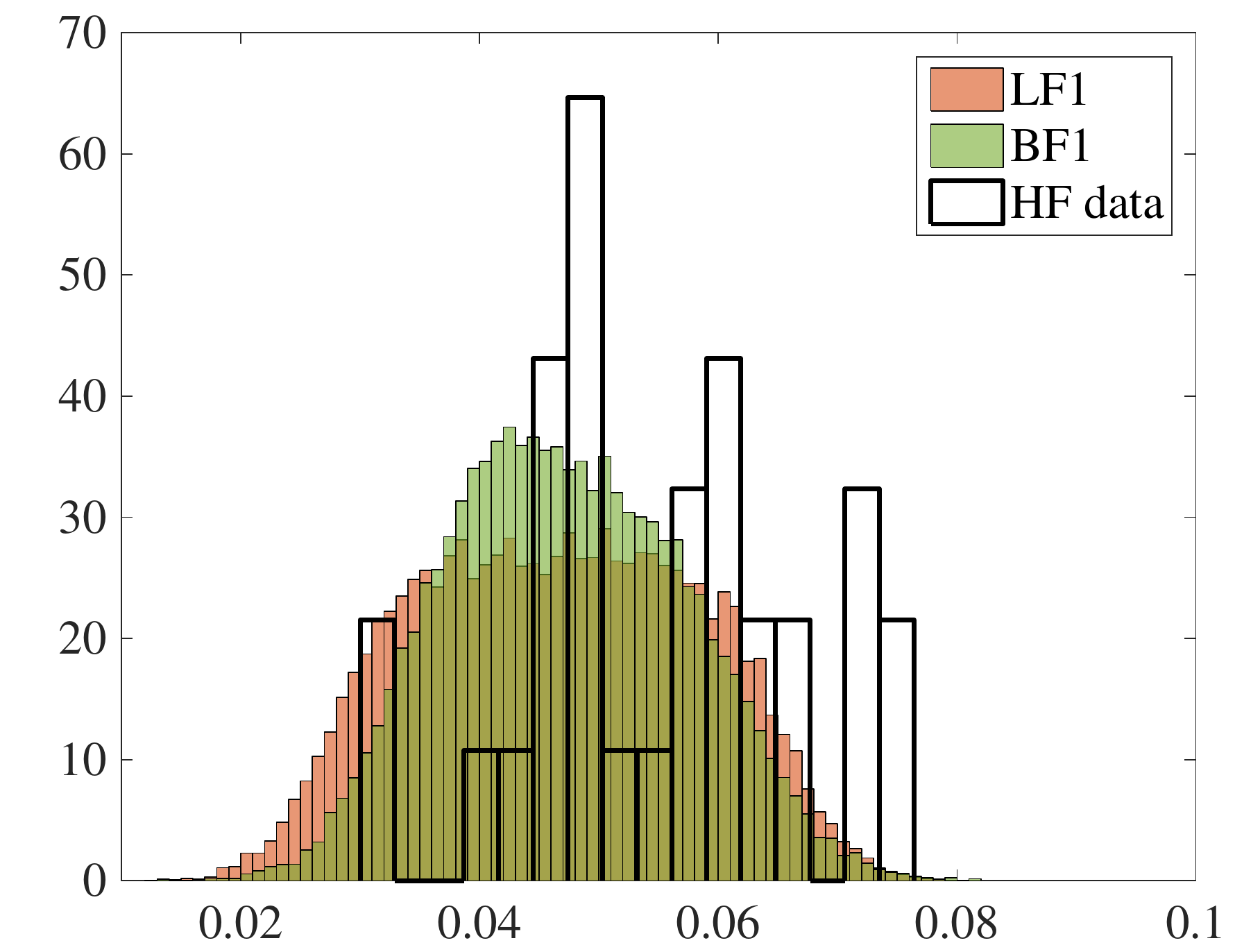}
        \caption{LF1 and BF1 for $\Delta T/T_0$  at $y/W=0.1$}
    \end{subfigure}
    \begin{subfigure}[b]{0.48\textwidth}
    \centering
        \includegraphics[trim = 15mm 70mm 20mm  70mm, clip, width=.7\textwidth]{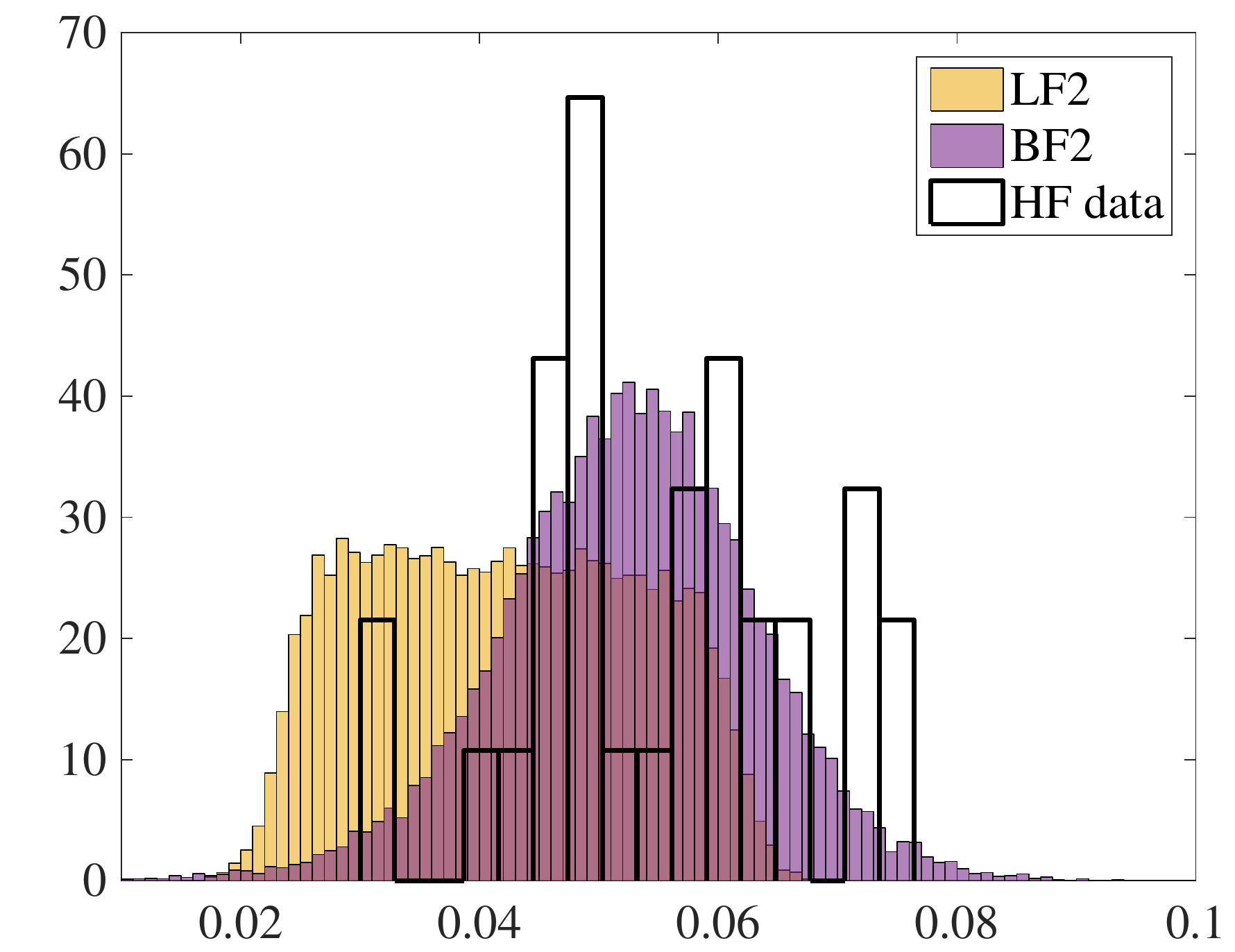}
        \caption{LF2 and BF2 for $\Delta T/T_0$  at $y/W=0.1$}
    \end{subfigure}
        \begin{subfigure}[b]{0.48\textwidth}
    \centering
        \includegraphics[trim = 15mm 70mm 20mm  70mm, clip, width=.7\textwidth]{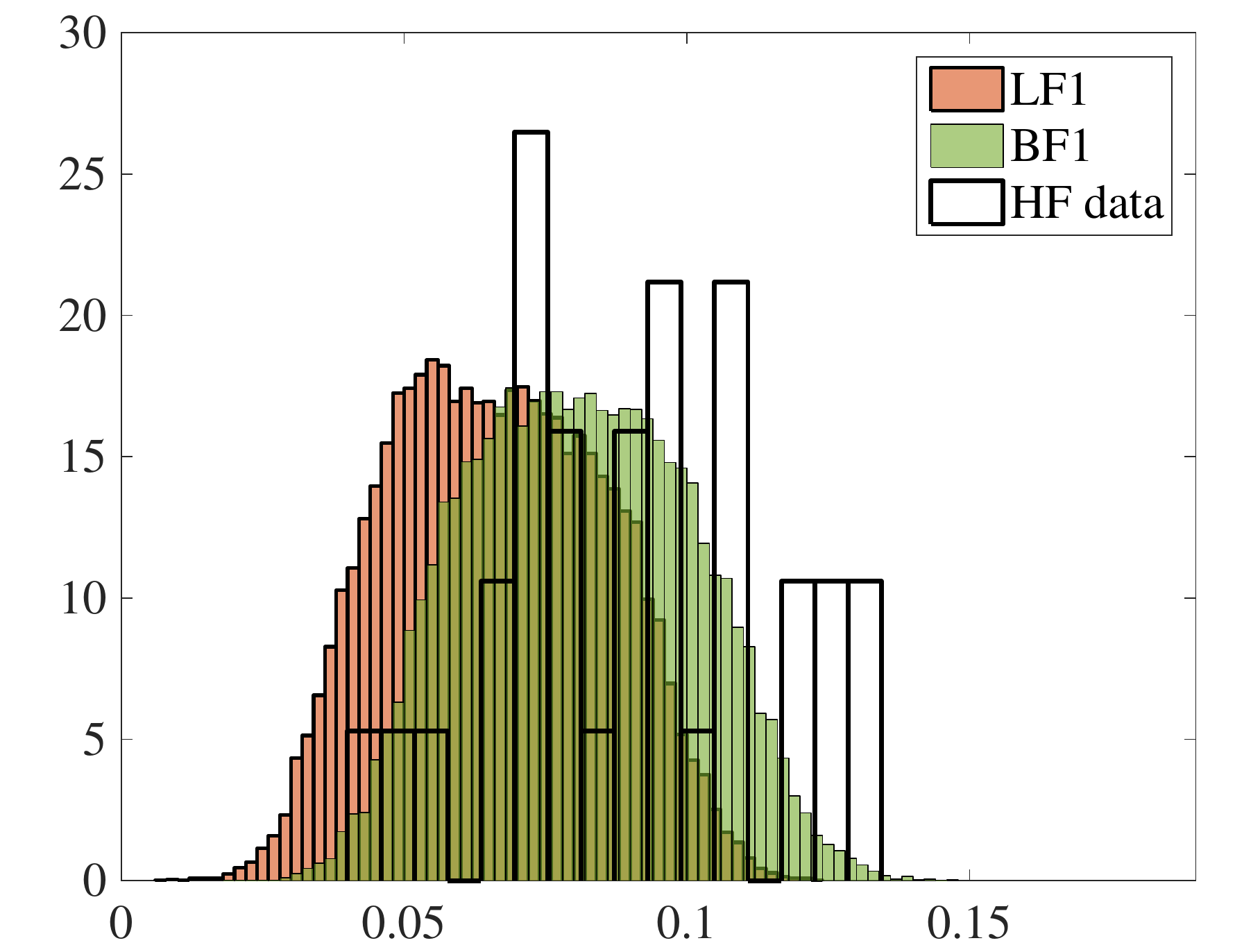}
        \caption{LF1 and BF1 for $\Delta T/T_0$  at $y/W=0.05$}
    \end{subfigure}
    \begin{subfigure}[b]{0.48\textwidth}
    \centering
        \includegraphics[trim = 15mm 70mm 20mm  70mm, clip, width=.7\textwidth]{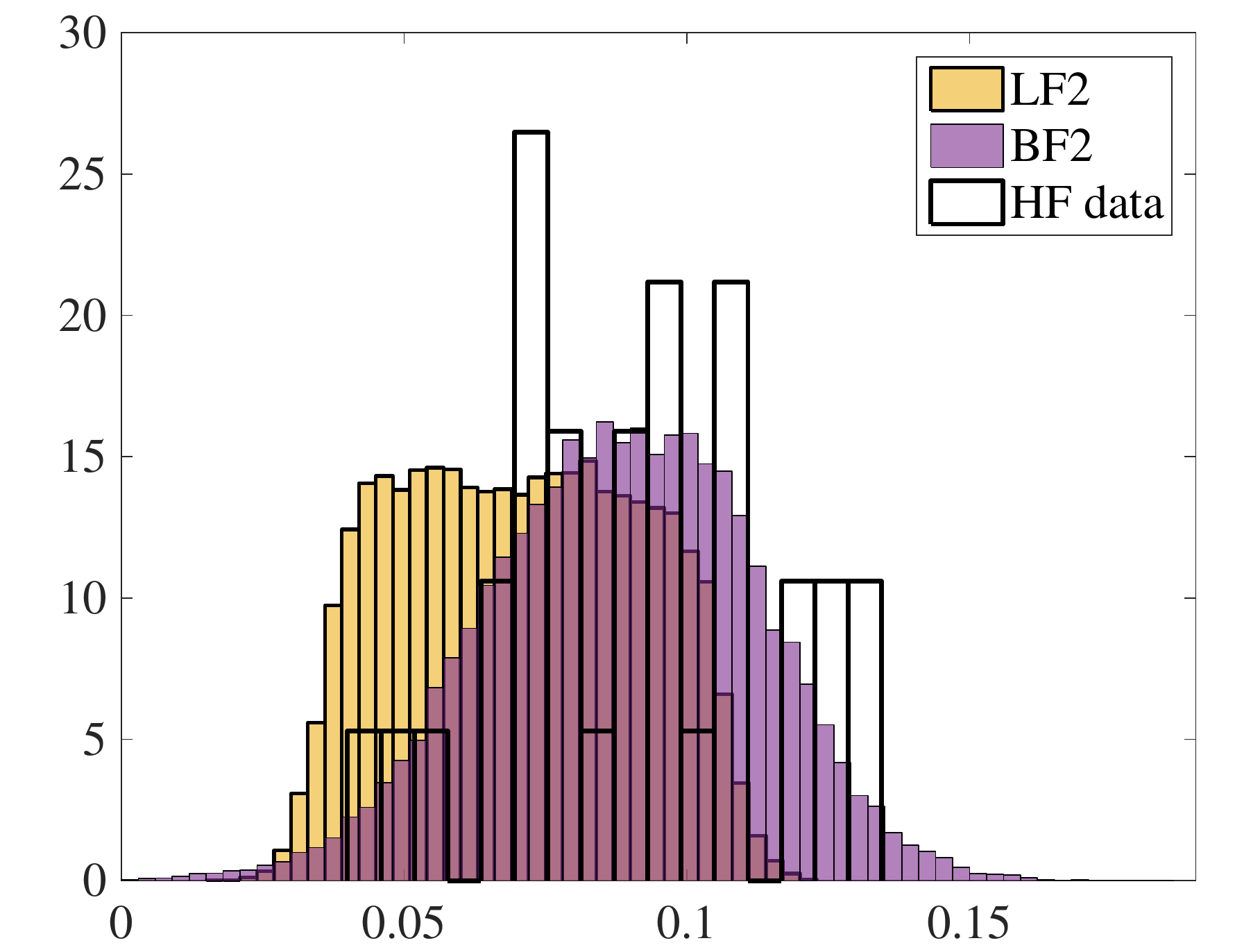}
        \caption{LF2 and BF2 for $\Delta T/T_0$  at $y/W=0.05$}
    \end{subfigure}
   \caption{Normalized histograms of the LF1 and BF1 surrogate models (left column) and LF2 and BF2 surrogate models (right column) for the four $\Delta T/T_0$ QoIs: mean ((a) and (b)), $y/W=0.5$ ((c) and (d)), $y/W=0.1$ ((e) and (f)), and $y/W=0.05$ ((g) and (h)). Histograms formed from 25,000 samples of the sparse PCE surrogates. \label{fig:ext_pdf}}
\end{figure}

To estimate the PDFs of the four $\Delta T/T_{0}$ QoIs, histograms of the LF and BF sparse PCE surrogates are formed using $25,000$ samples. These results are provided in Fig. \ref{fig:ext_pdf}, where the left column results are associated with LF1 and BF1 surrogates, and the right column results are associated with LF2 and BF2 surrogates. For comparison available simulated HF data are provided as well. 
With the exception of Fig. \ref{fig:ext_pdf} (e) and (g), where the LF1 and BF1 results are closely overlaid, all of the histograms show the BF results more accurately approximating the HF data than the LF results. 
This significance is observed more so for Fig. \ref{fig:ext_pdf} (a)-(d), which displays the QoI histograms of the spatial mean of $\Delta T/T_{0}$ and $\Delta T/T_{0}$ at the center of the profile. In Fig. \ref{fig:ext_pdf} (c)-(h), which shows the $\Delta T/T_{0}$ QoIs at specific points along the profile, the BF surrogates are consistently accurate. The LF surrogates are the least accurate in the center of the profile (Fig. \ref{fig:ext_pdf} (c) and (d)), and improve for QoI point estimates near the wall (Fig. \ref{fig:ext_pdf} (e)-(h)). 

\begin{figure}[h]
    \centering
        \begin{subfigure}[b]{0.48\textwidth}
    \centering
        \includegraphics[width=.7\textwidth]{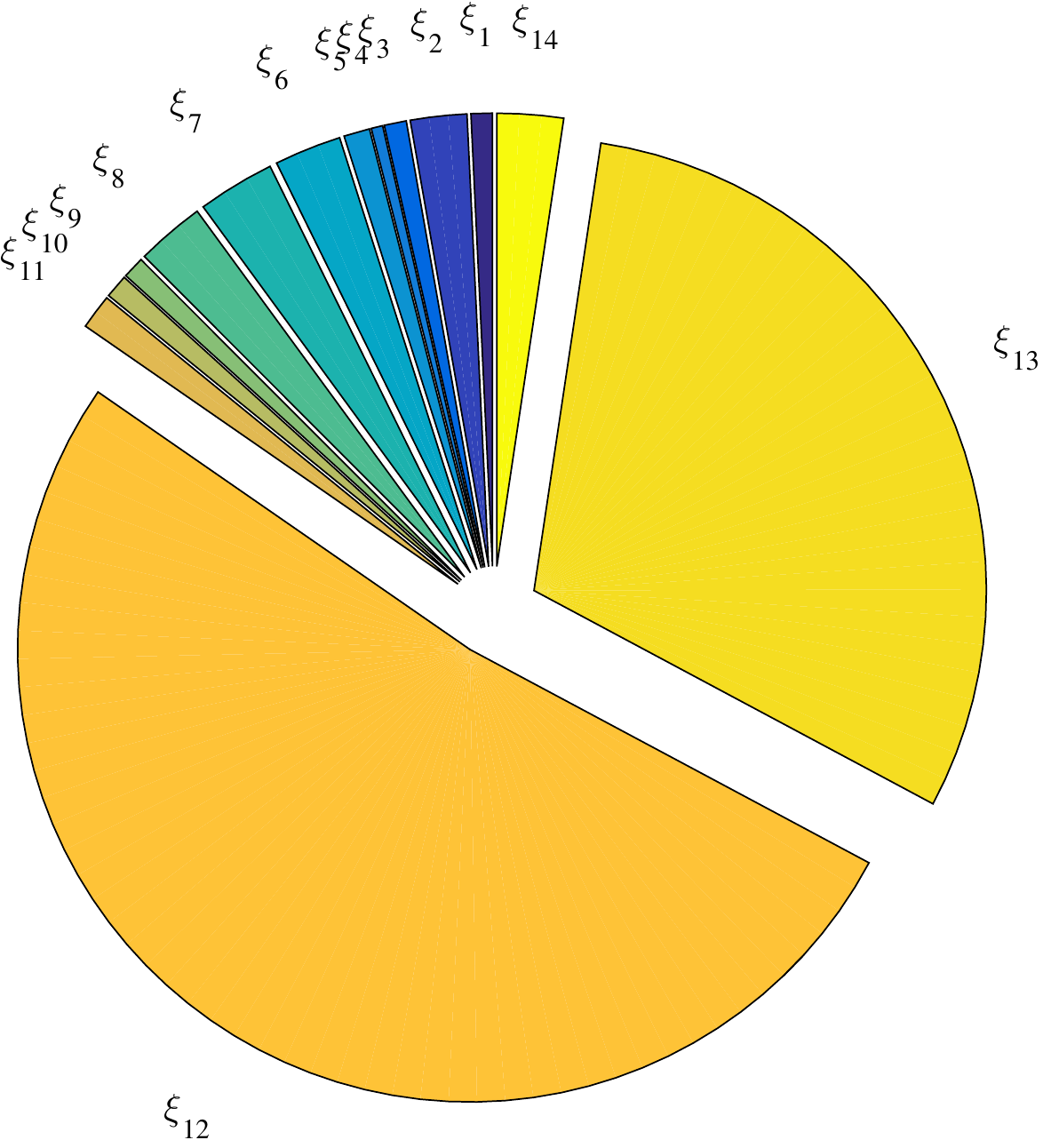}
        \caption{LF1 mean}
    \end{subfigure}
    \begin{subfigure}[b]{0.48\textwidth}
    \centering
        \includegraphics[width=.7\textwidth]{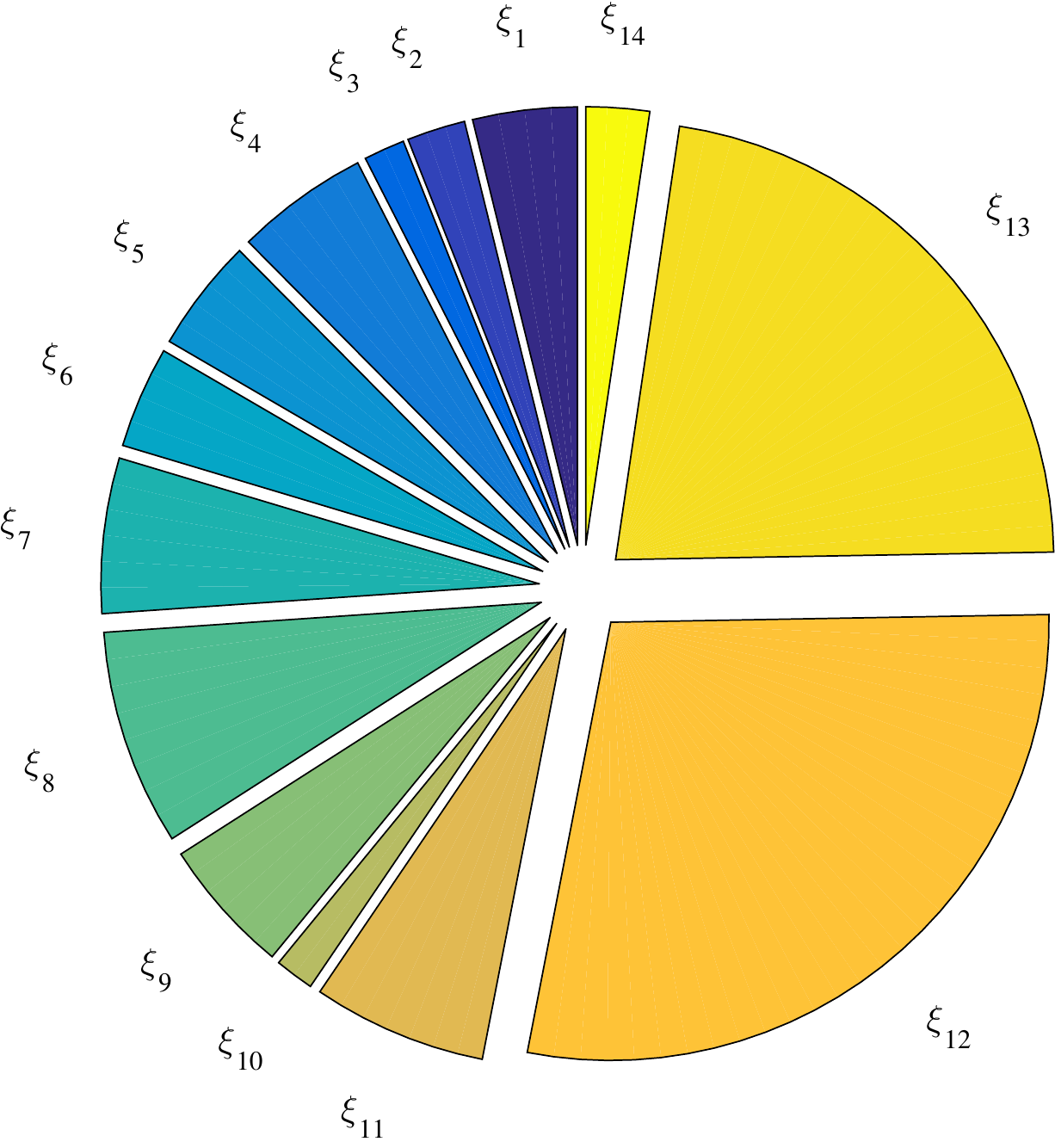}
        \caption{BF1 mean}
    \end{subfigure}
        \begin{subfigure}[b]{0.48\textwidth}
    \centering
        \includegraphics[width=.6\textwidth]{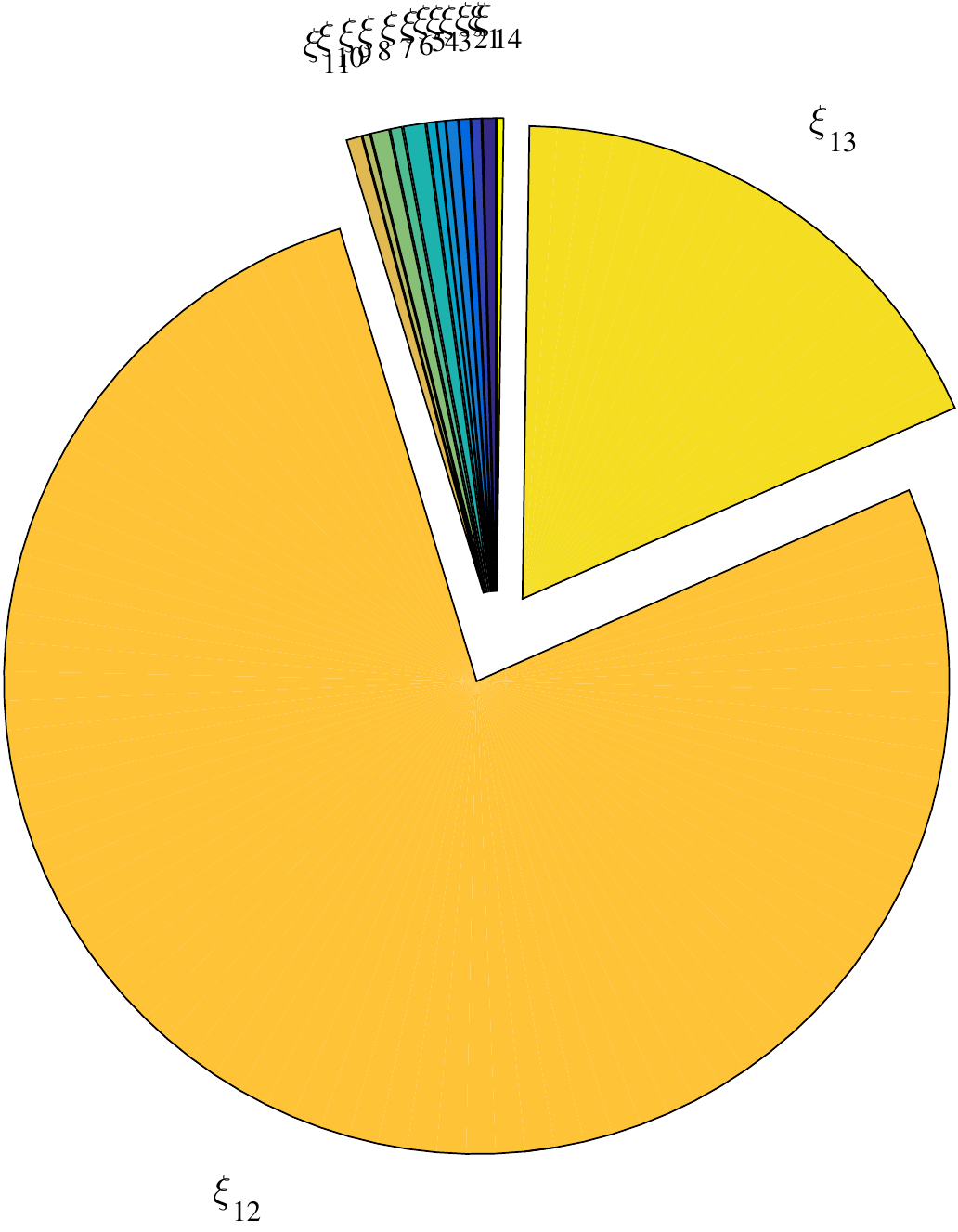}
        \caption{LF2 mean}
    \end{subfigure}
    \begin{subfigure}[b]{0.48\textwidth}
    \centering
        \includegraphics[width=.7\textwidth]{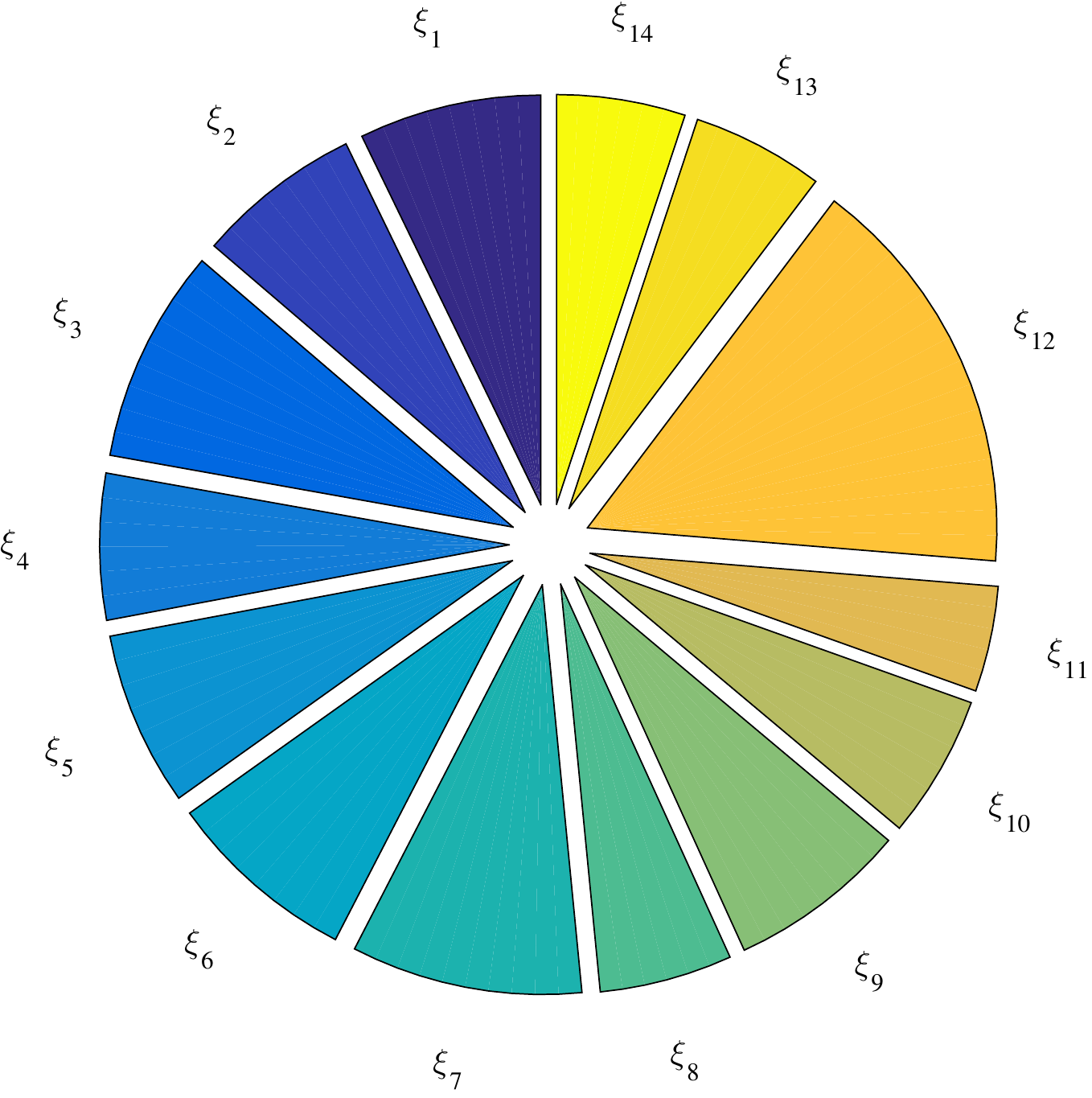}
        \caption{BF2 mean}
    \end{subfigure}
   \caption{Importance of input parameters for the spatial mean $\Delta T/T_{0}$ QoI from sparse PCE coefficients of the (a) LF1 (b) BF1, (c) LF2, and (d) BF2 surrogate models. Starting with parameter $\xi_1$ at the top position, importance of each $\xi_i$ is provided in counterclockwise order with respect to increasing $i$, with corresponding description provided in Table \ref{tab:uncertainties}.  \label{fig:ext_sa_mean}}
\end{figure}

The last result considered is global sensitivity analysis via Sobol' indices, as calculated from the sparse PCE coefficients. From these estimates, two sets of results are presented: comparisons between the four models and comparisons between the four QoIs. To compare the four models,
sensitivity analysis is completed for the spatial mean $\Delta T/T_{0}$ QoI. Fig. \ref{fig:ext_sa_mean} provides the decomposition of important parameters from the (a) LF1, (b) BF1, (c) LF2, and (d) BF2 models. All pie charts suggest that heat flux from the radiated wall to the fluid ($\xi _{12}$) is the most important parameter affecting the QoI variance. However, it is important to note the results of Table \ref{tab:pce_t} (a); specifically, the LF1 and LF2 errors are larger than the corresponding CoV estimates, indicating that the associated sensitivity analysis is not necessarily reliable. The BF1 and BF2 data, on the other hand, have an error that is smaller than the CoV, but on the same order of magnitude. As such, Fig. \ref{fig:ext_sa_mean} (b) and (c) show that the heat flux from the radiated wall and opposite wall to the fluid ($\xi _{12}$ and $\xi _{13}$, respectively) are the two most important input parameters. Further conclusions cannot be made with regard to the remaining parameters as the CoV estimates of Table \ref{tab:pce_t} (a) are of the same order as the error.

\begin{figure}[h]
    \centering 
    \begin{subfigure}[b]{0.48\textwidth}.pdf
    \centering
        \includegraphics[width=.7\textwidth]{pie_BF2_probe_profile_avg_T_mean.pdf}
        \caption{BF2, spatial mean}
    \end{subfigure}  
    \begin{subfigure}[b]{0.48\textwidth}
    \centering
        \includegraphics[width=.7\textwidth]{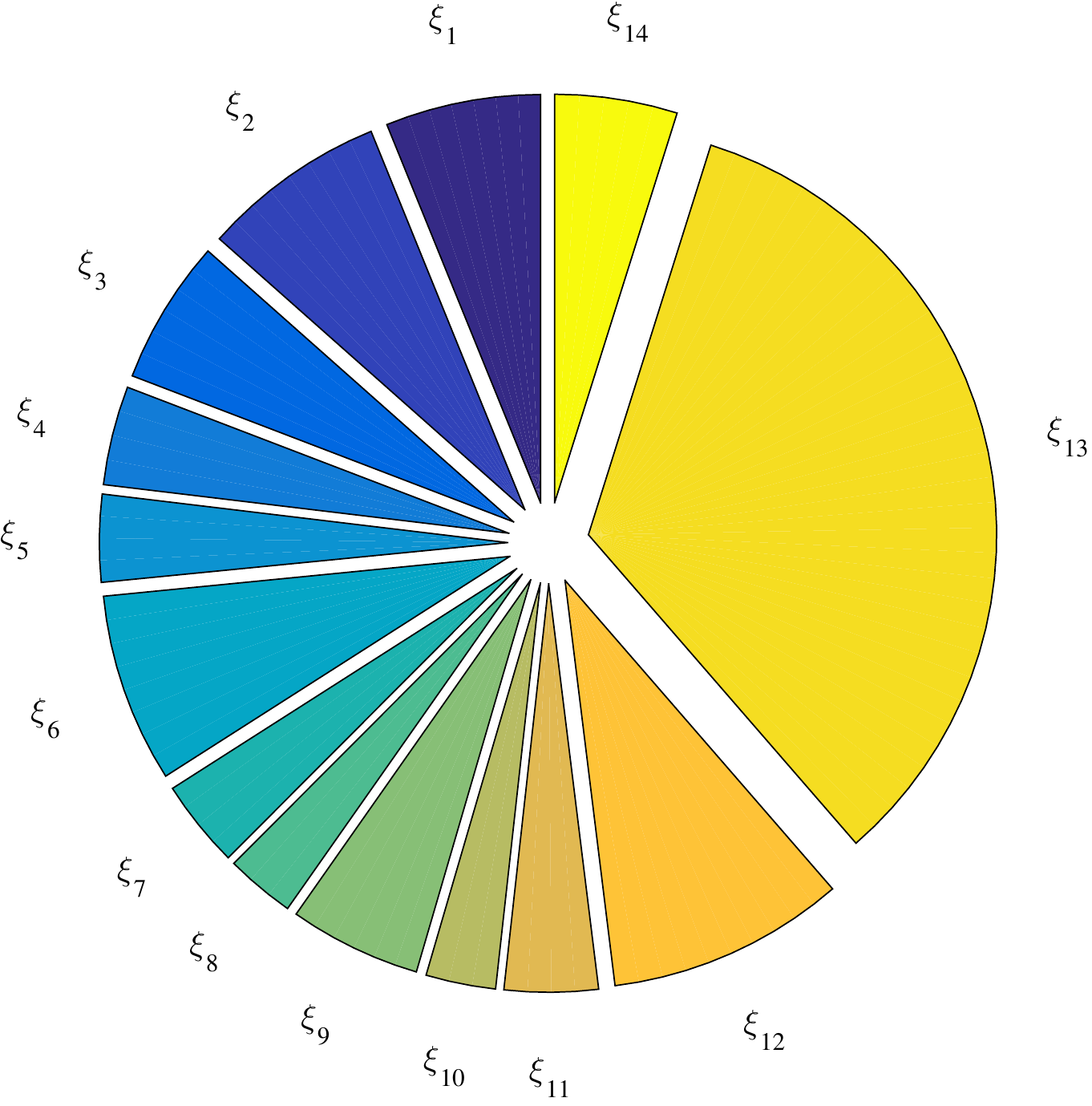}
        \caption{BF2, $y/W=0.5$}
    \end{subfigure}   
    \begin{subfigure}[b]{0.48\textwidth}
    \centering
        \includegraphics[width=.7\textwidth]{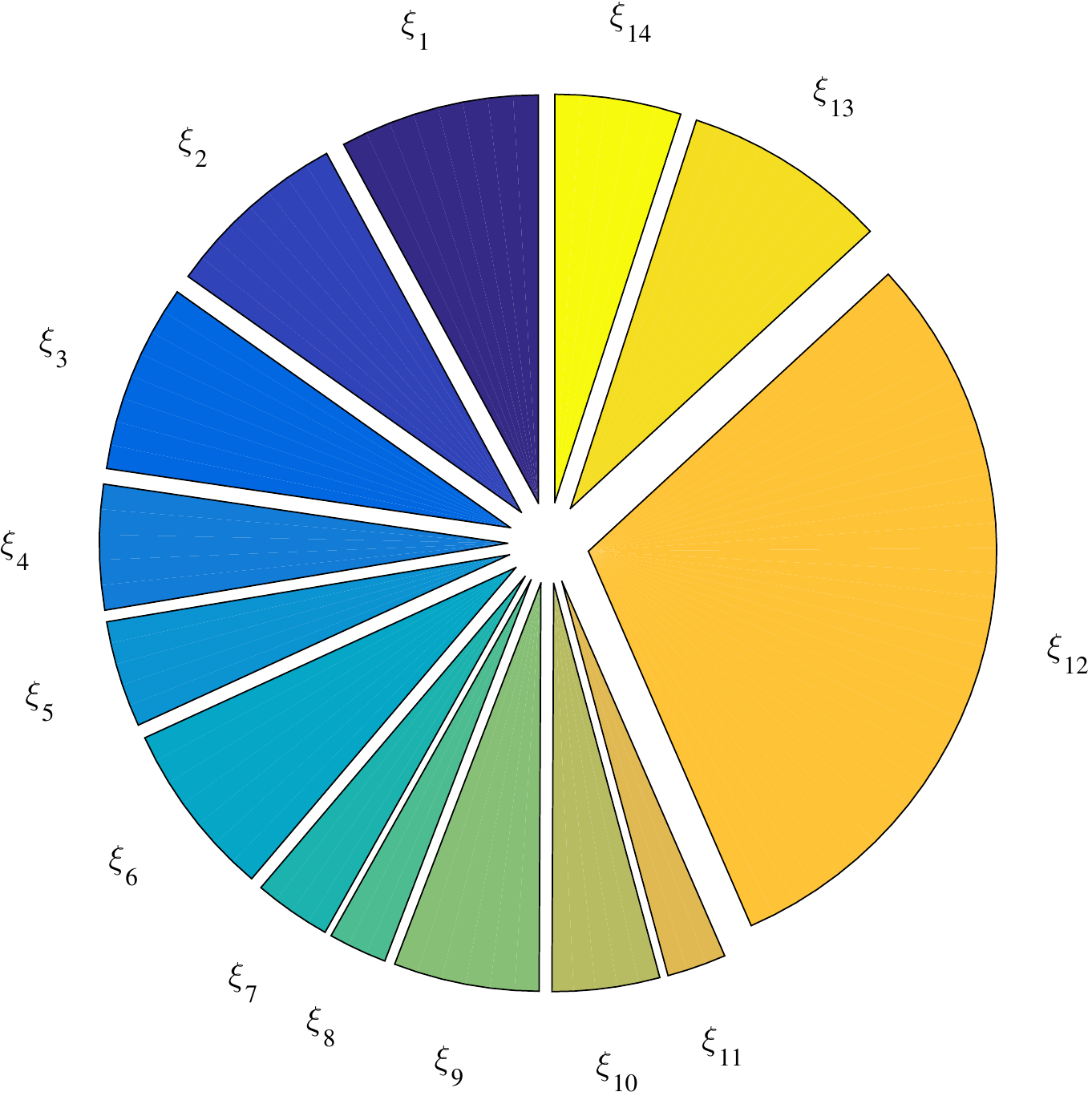}
        \caption{BF2, $y/W=0.1$}
    \end{subfigure}
    \begin{subfigure}[b]{0.48\textwidth}
    \centering
        \includegraphics[width=.7\textwidth]{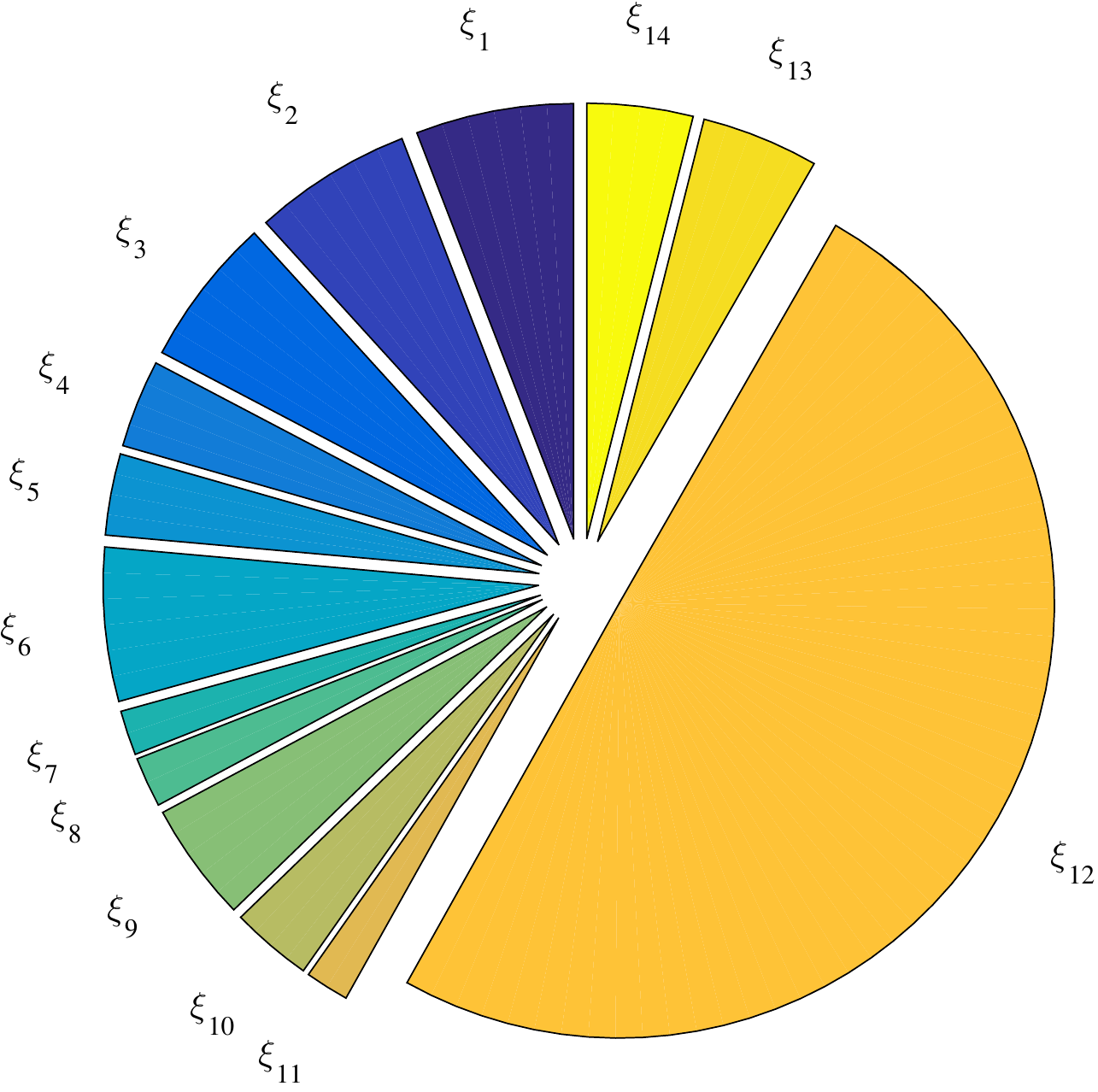}
        \caption{BF2, $y/W=0.05$}
    \end{subfigure}
   \caption{Importance of input parameters from sparse PCE coefficients from the BF2 model for the $\Delta T/T_{0}$ QoIs of (a) spatial mean along profile at probe location (b) $y/W = 0.5$ along profile at probe location, (c) $y/W=0.1$ along profile at probe location, and (d) $y/W=0.05$ along profile at probe location. Starting with parameter $\xi_1$ at the top position, importance of each $\xi_i$ is provided in counterclockwise order with respect to increasing $i$, with corresponding description provided in Table \ref{tab:uncertainties}. \label{fig:ext_sa_bf2}}
\end{figure}

Based on the moment estimations of Table \ref{tab:pce_t} and histograms of Fig. \ref{fig:ext_pdf}, it is clear that the BF approximations provide improved information of the HF QoIs compared to the corresponding LF models. As a consequence, Fig. \ref{fig:ext_sa_bf2} provides the importance of input parameters of the four QoIs as determined by the BF2 sparse PCE coefficients. The BF2 model is selected as it consistently has a low error (see Table \ref{tab:pce_t}), and does so with lower cost than the BF1 surrogate model. Fig. \ref{fig:ext_sa_bf2} (a)-(d) shows how the importance of parameters changes with the four QoIs. 
From Fig. \ref{fig:ext_sa_bf2} (a)-(d), the heat flux from the radiated wall to the fluid ($\xi _{12}$) is the most important parameter for the mean $\Delta T/T_{0}$ QoI, as well as the two point QoIs near the radiated wall (Figs. \ref{fig:ext_sa_bf2} (a), (b), and (c), respectively); in contrast, heat flux from the opposite wall to the fluid ($\xi _{13}$) is the most important parameter for the $\Delta T/T_{0}$ QoI in the middle of the profile (see Fig.~\ref{fig:ext_sa_bf2} (d)). 
This suggests that, over the whole profile, variations in the heat flux from the radiated wall will greatly affect the $\Delta T/T_{0}$ values, but more so at points close to the radiated wall. For points further from this wall, the variations in this heat flux will play less of a role in the variations of $\Delta T/T_{0}$.

In terms of the remaining parameters, the results of Table \ref{tab:pce_t} (a) and (b) indicate that there is lack of sufficient precision in the variance estimate to conclude the importance of the remaining parameters for Fig. \ref{fig:ext_sa_bf2} (a) and (b). However, the CoV estimates in \ref{tab:pce_t} (c) and (d) are significantly larger than the corresponding error estimates for the BF2 models, allowing for further conclusions to be made from the sensitivity analysis results. Specifically, the data of Fig. \ref{fig:ext_sa_bf2} (c) and (d) show that the remaining parameters contribute equally to the variance of the $\Delta T/T_{0}$ QoIs near the radiated wall. 

The $\Delta T/T_{0}$ results of this section show that the BF approximations provide a more accurate reduced model than either of the LF models, with greatest improvements for mean $\Delta T/T_{0}$ and $\Delta T/T_{0}$ at the interior of the profile. In addition, theoretical error results guarantee that the BF approximations will be at least as accurate as the corresponding LF data. As will be shown in the following, the cost of this approximation is close to that of the LF models when many simulations are required.

\subsection{Computational Cost Comparisons of the Five Models}\label{sec:cost}
\begin{figure}[h]
    \centering
        \includegraphics[width=.42\textwidth]{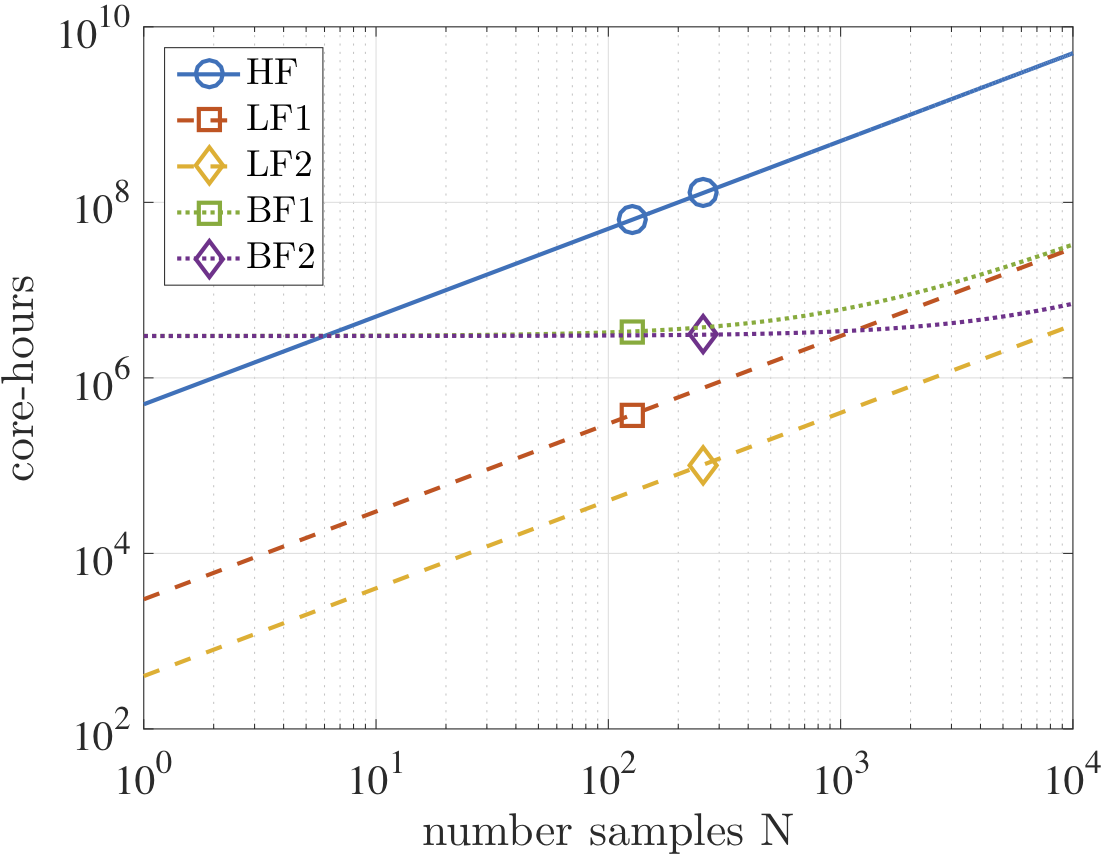}
   \caption{Number of core-hours on Mira (ALCF)~\cite{Mira-O} to obtain $N$ simulated values from each model. Markers provide number of core-hours for $N=128$ and $N=256$, to provide the cost of the simulated values generated for this study. \label{fig:exhf_cost}}
\end{figure}

The final component necessary to justify this BF approximation is to perform a cost comparison of all five models. Fig. \ref{fig:exhf_cost} provides the approximate number of core-hours needed to generate $N$ converged simulations from the HF, LF, and rank $r=6$ BF models, extrapolated to large values of $N$. The number of simulations generated in this study are indicated by markers. 
For the BF1, the cost to generate $N=128$ samples is $20\times$ less expensive than the HF model, and for the BF2, the cost to generate $N=256$ samples is $50\times$ less expensive than the HF model. In comparison, the LF1 and LF2 models are $170\times$ and $1300\times$ less expensive, respectively; however, as shown in the results, they are poor approximations to the HF data. 
As the $r$ HF simulations greatly affect the cost of the BF approximation, significant cost reduction is observed for larger values of $N$. For $\mathcal{O}(10^3)$ samples, which corresponds to values of $N$ that are of interest in the context of the application studied in this work, the computational cost of the BF models more closely aligns with the cost of the LF models. While $N=10^3$ HF simulations is approximately $500$M core-hours, obtaining the equivalent number of simulations is approximately $6$M and $3.5$M core-hours for the BF1 and BF2 approximations, respectively. This drastic cost improvement, without a significant loss of accuracy that is observed with the LF models, makes the BF approximation a powerful tool to perform UQ for large-scale problems.

\section{Conclusions}\label{sec:conclusions}

As the results of this work show, when solutions to large-scale parametric/stochastic problems exhibit a low-rank structure, i.e., lend themselves to a reduced basis representation, models with lower fidelities (cheaper to simulate) may be utilized to generate accurate approximations of the solution with significantly lower computational cost.  An instance of such a BF approximation was utilized to quantify the uncertainty in thermal solutions of interest (time-averaged heat flux and temperature near the outflow boundary) of a particle-based solar receiver model involving turbulence, particle transport,
\lluis{and radiative heat transfer,}
an example of large-scale, multi-physics systems. The sources of uncertainty included particle properties (\lluis{restitution coefficient}, Nusselt number, absorption and scattering \lluis{efficiencies}), thermal boundary conditions,  mass loading ratio, among others. A single HF simulation of the problem,
\lluis{consisting of DNS flow model, Lagrangian particle transport,}
and highly resolved DOM discretization of the
\lluis{radiative transfer equations,}
requires about half a million core-hours. This, in conjunction with the relatively large number of uncertain inputs, renders standard UQ techniques infeasible. 

To tackle the computational cost issue, two lower-fidelity models of the problem were constructed based on coarsening the fluid and DOM grids, as well as by reducing the number of particles. These models were used to identify a reduced basis and an interpolation rule for the thermal QoIs of the HF model. As a result, only a small number of HF model realizations (e.g., six) at selected samples of uncertain parameters were needed to generate BF samples of the QoIs, which were in turn used for a PCE-based estimation of the QoI statistics and sensitivity analysis.  All simulations were performed using the same computational code and in a black box fashion. The errors in predicting the QoIs and their statistics via these BF models were computed and compared to those of their LF counterpart. It was observed that the BF solutions, while requiring a small number of HF realizations, were considerably (as high as 100$\times$ for the mean of heat flux) more accurate than the LF estimates. The adopted BF strategy features an error bound which uses a small number of HF realizations (along with LF samples) to estimate the error with respect to the HF model. The efficacy of the bound in determining the number of required HF samples and estimating the BF error was also demonstrated on the thermal QoIs. 

\lluis{The construction of the BF models enabled us to efficiently carry out sensitivity analyses via PCE-based Sobol' indices.}
\lluis{The results indicate that the thermal QoIs considered in this work are extremely sensitive to the heat fluxes from the walls to the gas-particle mixture resulting from the interaction between radiation and non-ideal transparent walls.}
\lluis{The importance of this effect is amplified in wall-bounded, particle-laden turbulent flows as turbophoretic mechanisms tend to accumulate particles at the walls where the importance of non-ideal transmissivity is the highest.}
\lluis{The analysis also extracted that uncertainty in particle properties, which is typically disregarded in computational studies of irradiated particle-laden turbulence, plays an important role in the solution outcome that may result in large variability in, otherwise, robust first- and second-order statistics.}

Ongoing and future work focuses on exploring the performance of the BF approximation under more challenging physical conditions in terms of turbulent intensities, particle loading, and radiation intensity. Of notable interest is also the assessment of the approach in predicting higher order statistics, e.g., particles and temperature fluctuations, and more complex physical phenomena, such as particle clustering and turbulence modulation.


\section*{Acknowledgments}

This work was funded by the United States Department of Energy's National Nuclear Security Administration under the Predictive Science Academic Alliance Program II at Stanford University, Grant DE-NA-0002373.
The fifth author acknowledges funding by the US Department of Energy's Office of Science Advanced Scientific Computing Research, Award DE-SC0006402 and National Science Foundation Grant CMMI-145460. Sandia National Laboratories is a multimission laboratory managed and operated by National Technology \& Engineering Solutions of Sandia, LLC, a wholly owned subsidiary of Honeywell International Inc., for the U.S. Department of Energy’s National Nuclear Security Administration under contract DE-NA0003525. This paper describes objective technical results and analysis. Any subjective views or opinions that might be expressed in the paper do not necessarily represent the views of the U.S. Department of Energy or the United States Government. 

An award of computer time was provided by the ASCR Leadership Computing Challenge program. This research used resources of the Argonne Leadership Computing Facility, which is a Department of Energy's Office of Science User Facility supported under contract DE-AC02-06CH11357. This research also used resources of the Oak Ridge Leadership Computing Facility, which is a Department of Energy's Office of Science User Facility supported under contract DE-AC05-00OR22725.

\section*{References}\label{sec:ref}
\bibliographystyle{model1-num-names}
\bibliography{bib_psaap}

\begin{thebibliography}{66}
\expandafter\ifx\csname natexlab\endcsname\relax\def\natexlab#1{#1}\fi
\providecommand{\bibinfo}[2]{#2}
\ifx\xfnm\relax \def\xfnm[#1]{\unskip,\space#1}\fi
\bibitem[{Psa(2017)}]{PsaapII-O}
\bibinfo{title}{{E}xascale {C}omputing {E}ngineering {C}enter. {P}redictive
  {S}cience {A}cademic {A}lliance {P}rogram {(PSAAP) II}, {S}tanford
  {U}niversity}, \bibinfo{year}{2017}.
\bibitem[{Veron(2015)}]{Veron2015-A}
\bibinfo{author}{F.~Veron},
\newblock \bibinfo{title}{Ocean spray},
\newblock \bibinfo{journal}{Annual Review of Fluid Mechanics}
  \bibinfo{volume}{47} (\bibinfo{year}{2015}) \bibinfo{pages}{507--538}.
\bibitem[{Shaw(2003)}]{Shaw2003-A}
\bibinfo{author}{R.~A. Shaw},
\newblock \bibinfo{title}{Particle-turbulence interactions in atmospheric
  clouds},
\newblock \bibinfo{journal}{Annual Review of Fluid Mechanics}
  \bibinfo{volume}{35} (\bibinfo{year}{2003}) \bibinfo{pages}{183--227}.
\bibitem[{Tieszen(2001)}]{Tieszen2001-A}
\bibinfo{author}{S.~R. Tieszen},
\newblock \bibinfo{title}{On the fluid mechanics of fires},
\newblock \bibinfo{journal}{Annual Review of Fluid Mechanics}
  \bibinfo{volume}{33} (\bibinfo{year}{2001}) \bibinfo{pages}{67--92}.
\bibitem[{Lasheras and Hopfinger(2000)}]{Lasheras2000-A}
\bibinfo{author}{J.~C. Lasheras}, \bibinfo{author}{E.~J. Hopfinger},
\newblock \bibinfo{title}{Liquid jet instability and atomization in a coaxial
  gas stream},
\newblock \bibinfo{journal}{Annual Review of Fluid Mechanics}
  \bibinfo{volume}{32} (\bibinfo{year}{2000}) \bibinfo{pages}{275--308}.
\bibitem[{Raman and Fox(2016)}]{Raman2016-A}
\bibinfo{author}{V.~Raman}, \bibinfo{author}{R.~O. Fox},
\newblock \bibinfo{title}{Modeling of fine-particle formation in turbulent
  flames},
\newblock \bibinfo{journal}{Annual Review of Fluid Mechanics}
  \bibinfo{volume}{48} (\bibinfo{year}{2016}) \bibinfo{pages}{159--190}.
\bibitem[{Ho(2017)}]{Ho2017-A}
\bibinfo{author}{C.~K. Ho},
\newblock \bibinfo{title}{Advances in central receivers for concentrating solar
  applications},
\newblock \bibinfo{journal}{Solar Energy} \bibinfo{volume}{152}
  (\bibinfo{year}{2017}) \bibinfo{pages}{38--56}.
\bibitem[{Balachandar and Eaton(2010)}]{Balachandar2010-A}
\bibinfo{author}{S.~Balachandar}, \bibinfo{author}{J.~K. Eaton},
\newblock \bibinfo{title}{Turbulent dispersed multiphase flow},
\newblock \bibinfo{journal}{Annual Review of Fluid Mechanics}
  \bibinfo{volume}{42} (\bibinfo{year}{2010}) \bibinfo{pages}{111--133}.
\bibitem[{Caporaloni et~al.(1975)Caporaloni, Tampieri, Trombetti, and
  Vittori}]{Caporaloni1975-A}
\bibinfo{author}{M.~Caporaloni}, \bibinfo{author}{F.~Tampieri},
  \bibinfo{author}{F.~Trombetti}, \bibinfo{author}{O.~Vittori},
\newblock \bibinfo{title}{Transfer of particles in nonisotropic air
  turbulence},
\newblock \bibinfo{journal}{Journal of Atmospheric Sciences}
  \bibinfo{volume}{32} (\bibinfo{year}{1975}) \bibinfo{pages}{565--568}.
\bibitem[{Squires and Eaton(1991)}]{Squires1991-A}
\bibinfo{author}{K.~D. Squires}, \bibinfo{author}{J.~K. Eaton},
\newblock \bibinfo{title}{Preferential concentration of particles by
  turbulence},
\newblock \bibinfo{journal}{Physics of Fluids} \bibinfo{volume}{3}
  (\bibinfo{year}{1991}) \bibinfo{pages}{1169--1178}.
\bibitem[{Wang and Squires(1996)}]{Wang1996-A}
\bibinfo{author}{Q.~Wang}, \bibinfo{author}{K.~D. Squires},
\newblock \bibinfo{title}{Large eddy simulation of particle-laden turbulent
  channel flow},
\newblock \bibinfo{journal}{Physics of Fluids} \bibinfo{volume}{8}
  (\bibinfo{year}{1996}) \bibinfo{pages}{1207--1223}.
\bibitem[{Sardina et~al.(2012)Sardina, Schlatter, Brandt, and
  Picano}]{Sardina2012-A}
\bibinfo{author}{G.~Sardina}, \bibinfo{author}{P.~Schlatter},
  \bibinfo{author}{L.~Brandt}, \bibinfo{author}{F.~Picano},
\newblock \bibinfo{title}{Wall accumulation and spatial localization in
  particle-laden wall flows},
\newblock \bibinfo{journal}{Journal of Fluid Mechanics} \bibinfo{volume}{699}
  (\bibinfo{year}{2012}) \bibinfo{pages}{50--78}.
\bibitem[{Zamansky et~al.(2014)Zamansky, Coletti, Massot, and
  Mani}]{Zamansky2014-A}
\bibinfo{author}{R.~Zamansky}, \bibinfo{author}{F.~Coletti},
  \bibinfo{author}{M.~Massot}, \bibinfo{author}{A.~Mani},
\newblock \bibinfo{title}{Radiation induces turbulence in particle-laden
  fluids},
\newblock \bibinfo{journal}{Physics of Fluids} \bibinfo{volume}{26}
  (\bibinfo{year}{2014}) \bibinfo{pages}{071701}.
\bibitem[{Frankel et~al.(2016)Frankel, Pouransari, Coletti, and
  Mani}]{Frankel2016-A}
\bibinfo{author}{A.~Frankel}, \bibinfo{author}{H.~Pouransari},
  \bibinfo{author}{F.~Coletti}, \bibinfo{author}{A.~Mani},
\newblock \bibinfo{title}{Settling of heated particles in homogeneous
  turbulence},
\newblock \bibinfo{journal}{Journal of Fluid Mechanics} \bibinfo{volume}{792}
  (\bibinfo{year}{2016}) \bibinfo{pages}{869--893}.
\bibitem[{Pouransari and Mani(2017)}]{Pouransari2017-A}
\bibinfo{author}{H.~Pouransari}, \bibinfo{author}{A.~Mani},
\newblock \bibinfo{title}{Effects of preferential concentration on heat
  transfer in particle-based solar receivers},
\newblock \bibinfo{journal}{Journal of Solar Energy Engineering}
  \bibinfo{volume}{139} (\bibinfo{year}{2017}) \bibinfo{pages}{021008}.
\bibitem[{Rahmani et~al.(2018)Rahmani, Geraci, Iaccarino, and
  Mani}]{Rahmani2018-A}
\bibinfo{author}{M.~Rahmani}, \bibinfo{author}{G.~Geraci},
  \bibinfo{author}{G.~Iaccarino}, \bibinfo{author}{A.~Mani},
\newblock \bibinfo{title}{Effects of particle polydispersity on radiative heat
  transfer in particle-laden turbulent flows},
\newblock \bibinfo{journal}{Int. J. Multiph. Flow} \bibinfo{volume}{104}
  (\bibinfo{year}{2018}) \bibinfo{pages}{42--59}.
\bibitem[{Mathelin and Hussaini(2003)}]{Mathelin03}
\bibinfo{author}{L.~Mathelin}, \bibinfo{author}{M.~Y. Hussaini},
  \bibinfo{title}{A stochastic collocation algorithm for uncertainty analysis},
  \bibinfo{type}{Technical Report} \bibinfo{number}{NASA 1.26:212153;
  NASA/CR-2003-212153}, NASA Langley Research Center, \bibinfo{year}{2003}.
\bibitem[{Xiu and Hesthaven(2005)}]{Xiu05}
\bibinfo{author}{D.~Xiu}, \bibinfo{author}{J.~S. Hesthaven},
\newblock \bibinfo{title}{High-order collocation methods for differential
  equations with random inputs},
\newblock \bibinfo{journal}{SIAM Journal on Scientific Computing}
  \bibinfo{volume}{27} (\bibinfo{year}{2005}) \bibinfo{pages}{1118--1139}.
\bibitem[{Ghanem and Spanos(2002)}]{Ghanem03}
\bibinfo{author}{R.~Ghanem}, \bibinfo{author}{P.~Spanos},
  \bibinfo{title}{Stochastic finite elements: a spectral approach},
  \bibinfo{publisher}{Dover}, \bibinfo{year}{2002}.
\bibitem[{Xiu and Karniadakis(2002)}]{Xiu02}
\bibinfo{author}{D.~Xiu}, \bibinfo{author}{G.~M. Karniadakis},
\newblock \bibinfo{title}{The {W}iener-{A}skey polynomial chaos for stochastic
  differential equations},
\newblock \bibinfo{journal}{SIAM Journal on Scientific Computing}
  \bibinfo{volume}{24} (\bibinfo{year}{2002}) \bibinfo{pages}{619--644}.
\bibitem[{Doostan and Owhadi(2011)}]{Doostan11a}
\bibinfo{author}{A.~Doostan}, \bibinfo{author}{H.~Owhadi},
\newblock \bibinfo{title}{A non-adapted sparse approximation of {PDEs} with
  stochastic inputs},
\newblock \bibinfo{journal}{Journal of Computational Physics}
  \bibinfo{volume}{230} (\bibinfo{year}{2011}) \bibinfo{pages}{3015--3034}.
\bibitem[{Brandt(1977)}]{Brandt77}
\bibinfo{author}{A.~Brandt},
\newblock \bibinfo{title}{Multi-level adaptive solutions to boundary-value
  problems},
\newblock \bibinfo{journal}{Mathematics of computation} \bibinfo{volume}{31}
  (\bibinfo{year}{1977}) \bibinfo{pages}{333--390}.
\bibitem[{Briggs et~al.(2000)Briggs, Henson, and McCormick}]{Briggs00}
\bibinfo{author}{W.~L. Briggs}, \bibinfo{author}{V.~E. Henson},
  \bibinfo{author}{S.~F. McCormick},
\newblock \bibinfo{title}{A multigrid tutorial}  (\bibinfo{year}{2000}).
\bibitem[{Peherstorfer et~al.(2016)Peherstorfer, Willcox, and
  Gunzburger}]{Peherstorfer16}
\bibinfo{author}{B.~Peherstorfer}, \bibinfo{author}{K.~Willcox},
  \bibinfo{author}{M.~Gunzburger},
\newblock \bibinfo{title}{Survey of multifidelity methods in uncertainty
  propagation, inference, and optimization}  (\bibinfo{year}{2016}).
\bibitem[{Fern{\'a}ndez-Godino et~al.(2016)Fern{\'a}ndez-Godino, Park, Kim, and
  Haftka}]{Fernandez16}
\bibinfo{author}{M.~G. Fern{\'a}ndez-Godino}, \bibinfo{author}{C.~Park},
  \bibinfo{author}{N.~H. Kim}, \bibinfo{author}{R.~T. Haftka},
\newblock \bibinfo{title}{Review of multi-fidelity models},
\newblock \bibinfo{journal}{arXiv preprint arXiv:1609.07196}
  (\bibinfo{year}{2016}).
\bibitem[{Asmussen and Glynn(2007)}]{Asmussen07}
\bibinfo{author}{S.~Asmussen}, \bibinfo{author}{P.~Glynn},
  \bibinfo{title}{Stochastic simulation: algorithms and analysis},
  volume~\bibinfo{volume}{57}, \bibinfo{publisher}{Springer Science \& Business
  Media}, \bibinfo{year}{2007}.
\bibitem[{Heinrich(2001)}]{Heinrich01}
\bibinfo{author}{S.~Heinrich},
\newblock \bibinfo{title}{Multilevel \uppercase{M}onte \uppercase{C}arlo
  methods},
\newblock in: \bibinfo{booktitle}{Large-scale scientific computing},
  \bibinfo{publisher}{Springer}, \bibinfo{year}{2001}, pp.
  \bibinfo{pages}{58--67}.
\bibitem[{Giles(2008)}]{Giles08}
\bibinfo{author}{M.~B. Giles},
\newblock \bibinfo{title}{Multilevel \uppercase{M}onte \uppercase{C}arlo path
  simulation},
\newblock \bibinfo{journal}{Operations Research} \bibinfo{volume}{56}
  (\bibinfo{year}{2008}) \bibinfo{pages}{607--617}.
\bibitem[{Cliffe et~al.(2011)Cliffe, Giles, Scheichl, and
  Teckentrup}]{Cliffe11}
\bibinfo{author}{K.~A. Cliffe}, \bibinfo{author}{M.~B. Giles},
  \bibinfo{author}{R.~Scheichl}, \bibinfo{author}{A.~L. Teckentrup},
\newblock \bibinfo{title}{Multilevel \uppercase{M}onte \uppercase{C}arlo
  methods and applications to elliptic pdes with random coefficients},
\newblock \bibinfo{journal}{Computing and Visualization in Science}
  \bibinfo{volume}{14} (\bibinfo{year}{2011}) \bibinfo{pages}{3--15}.
\bibitem[{Barth et~al.(2011)Barth, Schwab, and Zollinger}]{Barth11}
\bibinfo{author}{A.~Barth}, \bibinfo{author}{C.~Schwab},
  \bibinfo{author}{N.~Zollinger},
\newblock \bibinfo{title}{Multi-level \uppercase{M}onte \uppercase{C}arlo
  finite element method for elliptic pdes with stochastic coefficients},
\newblock \bibinfo{journal}{Numerische Mathematik} \bibinfo{volume}{119}
  (\bibinfo{year}{2011}) \bibinfo{pages}{123--161}.
\bibitem[{Teckentrup et~al.(2013)Teckentrup, Scheichl, Giles, and
  Ullmann}]{Teckentrup13}
\bibinfo{author}{A.~L. Teckentrup}, \bibinfo{author}{R.~Scheichl},
  \bibinfo{author}{M.~B. Giles}, \bibinfo{author}{E.~Ullmann},
\newblock \bibinfo{title}{Further analysis of multilevel \uppercase{M}onte
  \uppercase{C}arlo methods for elliptic pdes with random coefficients},
\newblock \bibinfo{journal}{Numerische Mathematik} \bibinfo{volume}{125}
  (\bibinfo{year}{2013}) \bibinfo{pages}{569--600}.
\bibitem[{Speight(2009)}]{Speight09}
\bibinfo{author}{A.~Speight},
\newblock \bibinfo{title}{A multilevel approach to control variates},
\newblock \bibinfo{journal}{Journal of Computational Finance, Forthcoming}
  (\bibinfo{year}{2009}).
\bibitem[{Nobile and Tesei(2015)}]{Nobile15}
\bibinfo{author}{F.~Nobile}, \bibinfo{author}{F.~Tesei},
\newblock \bibinfo{title}{A multi level \uppercase{M}onte \uppercase{C}arlo
  method with control variate for elliptic pdes with log-normal coefficients},
\newblock \bibinfo{journal}{Stochastic Partial Differential Equations: Analysis
  and Computations} \bibinfo{volume}{3} (\bibinfo{year}{2015})
  \bibinfo{pages}{398--444}.
\bibitem[{Vidal-Codina et~al.(2015)Vidal-Codina, Nguyen, Giles, and
  Peraire}]{Vidal15}
\bibinfo{author}{F.~Vidal-Codina}, \bibinfo{author}{N.~C. Nguyen},
  \bibinfo{author}{M.~B. Giles}, \bibinfo{author}{J.~Peraire},
\newblock \bibinfo{title}{A model and variance reduction method for computing
  statistical outputs of stochastic elliptic partial differential equations},
\newblock \bibinfo{journal}{Journal of Computational Physics}
  \bibinfo{volume}{297} (\bibinfo{year}{2015}) \bibinfo{pages}{700--720}.
\bibitem[{Geraci et~al.(2015)Geraci, Eldred, and Iaccarino}]{Geraci15}
\bibinfo{author}{G.~Geraci}, \bibinfo{author}{M.~Eldred},
  \bibinfo{author}{G.~Iaccarino},
\newblock \bibinfo{title}{A multifidelity control variate approach for the
  multilevel \uppercase{M}onte \uppercase{C}arlo technique},
\newblock \bibinfo{journal}{Center for Turbulence Research Annual Research
  Briefs, Stanford University}  (\bibinfo{year}{2015}).
\bibitem[{Fairbanks et~al.(2017)Fairbanks, Doostan, Ketelsen, and
  Iaccarino}]{Fairbanks17}
\bibinfo{author}{H.~R. Fairbanks}, \bibinfo{author}{A.~Doostan},
  \bibinfo{author}{C.~Ketelsen}, \bibinfo{author}{G.~Iaccarino},
\newblock \bibinfo{title}{A low-rank control variate for multilevel
  \uppercase{M}onte \uppercase{C}arlo simulation of high-dimensional uncertain
  systems},
\newblock \bibinfo{journal}{Journal of Computational Physics}
  \bibinfo{volume}{341} (\bibinfo{year}{2017}) \bibinfo{pages}{121--139}.
\bibitem[{Peherstorfer et~al.(2016)Peherstorfer, Cui, Marzouk, and
  Willcox}]{Peherstorfer16b}
\bibinfo{author}{B.~Peherstorfer}, \bibinfo{author}{T.~Cui},
  \bibinfo{author}{Y.~Marzouk}, \bibinfo{author}{K.~Willcox},
\newblock \bibinfo{title}{Multifidelity importance sampling},
\newblock \bibinfo{journal}{Computer Methods in Applied Mechanics and
  Engineering} \bibinfo{volume}{300} (\bibinfo{year}{2016})
  \bibinfo{pages}{490--509}.
\bibitem[{Kennedy and O'Hagan(2000)}]{Kennedy00}
\bibinfo{author}{M.~C. Kennedy}, \bibinfo{author}{A.~O'Hagan},
\newblock \bibinfo{title}{Predicting the output from a complex computer code
  when fast approximations are available},
\newblock \bibinfo{journal}{Biometrika} \bibinfo{volume}{87}
  (\bibinfo{year}{2000}) \bibinfo{pages}{1--13}.
\bibitem[{Qian et~al.(2006)Qian, Seepersad, Joseph, Allen, and Wu}]{Qian06}
\bibinfo{author}{Z.~Qian}, \bibinfo{author}{C.~C. Seepersad},
  \bibinfo{author}{V.~R. Joseph}, \bibinfo{author}{J.~K. Allen},
  \bibinfo{author}{C.~F.~J. Wu},
\newblock \bibinfo{title}{Building surrogate models based on detailed and
  approximate simulations},
\newblock \bibinfo{journal}{Journal of Mechanical Design} \bibinfo{volume}{128}
  (\bibinfo{year}{2006}) \bibinfo{pages}{668--677}.
\bibitem[{Laurenceau and Sagaut(2008)}]{Laurenceau08}
\bibinfo{author}{J.~Laurenceau}, \bibinfo{author}{P.~Sagaut},
  \bibinfo{title}{Building efficient response surfaces of aerodynamic functions
  with kriging and cokriging}, \bibinfo{year}{2008}.
\bibitem[{Narayan et~al.(2014)Narayan, Gittelson, and Xiu}]{Narayan14}
\bibinfo{author}{A.~Narayan}, \bibinfo{author}{C.~Gittelson},
  \bibinfo{author}{D.~Xiu},
\newblock \bibinfo{title}{A stochastic collocation algorithm with multifidelity
  models},
\newblock \bibinfo{journal}{SIAM Journal on Scientific Computing}
  \bibinfo{volume}{36} (\bibinfo{year}{2014}) \bibinfo{pages}{A495--A521}.
\bibitem[{Zhu et~al.(2014)Zhu, Narayan, and Xiu}]{Zhu14}
\bibinfo{author}{X.~Zhu}, \bibinfo{author}{A.~Narayan},
  \bibinfo{author}{D.~Xiu},
\newblock \bibinfo{title}{Computational aspects of stochastic collocation with
  multifidelity models},
\newblock \bibinfo{journal}{SIAM/ASA Journal on Uncertainty Quantification}
  \bibinfo{volume}{2} (\bibinfo{year}{2014}) \bibinfo{pages}{444--463}.
\bibitem[{Doostan et~al.(2016)Doostan, Geraci, and Iaccarino}]{Doostan16}
\bibinfo{author}{A.~Doostan}, \bibinfo{author}{G.~Geraci},
  \bibinfo{author}{G.~Iaccarino},
\newblock \bibinfo{title}{A bi-fidelity approach for uncertainty quantification
  of heat transfer in a rectangular ribbed channel},
\newblock in: \bibinfo{booktitle}{ASME Turbo Expo 2016: Turbomachinery
  Technical Conference and Exposition}, \bibinfo{organization}{American Society
  of Mechanical Engineers}, pp. \bibinfo{pages}{V02CT45A031--V02CT45A031}.
\bibitem[{Skinner et~al.(2017)Skinner, Doostan, Peters, Evans, and
  Jansen}]{Skinner17}
\bibinfo{author}{R.~W. Skinner}, \bibinfo{author}{A.~Doostan},
  \bibinfo{author}{E.~L. Peters}, \bibinfo{author}{J.~A. Evans},
  \bibinfo{author}{K.~E. Jansen},
\newblock \bibinfo{title}{An evaluation of bi-fidelity modeling efficiency on a
  general family of naca airfoils},
\newblock in: \bibinfo{booktitle}{35th AIAA Applied Aerodynamics Conference},
  p. \bibinfo{pages}{3260}.
\bibitem[{Hampton et~al.(2018)Hampton, Fairbanks, Narayan, and
  Doostan}]{Hampton2018practical}
\bibinfo{author}{J.~Hampton}, \bibinfo{author}{H.~R. Fairbanks},
  \bibinfo{author}{A.~Narayan}, \bibinfo{author}{A.~Doostan},
\newblock \bibinfo{title}{Practical error bounds for a non-intrusive
  bi-fidelity approach to parametric/stochastic model reduction},
\newblock \bibinfo{journal}{Journal of Computational Physics}
  \bibinfo{volume}{368} (\bibinfo{year}{2018}) \bibinfo{pages}{315--332}.
\bibitem[{Villafa{\~n}e et~al.(2017)Villafa{\~n}e, Banko, Elkins, and
  Eaton}]{Villafane2017-A}
\bibinfo{author}{L.~Villafa{\~n}e}, \bibinfo{author}{A.~Banko},
  \bibinfo{author}{C.~Elkins}, \bibinfo{author}{J.~K. Eaton},
\newblock \bibinfo{title}{Gas heating by radiation absorbing inertial particles
  in a turbulent duct flow},
\newblock \bibinfo{journal}{CTR Annu. Res. Briefs}  (\bibinfo{year}{2017})
  \bibinfo{pages}{35--47}.
\bibitem[{Frankel and Iaccarino(2017)}]{Frankel2017-A}
\bibinfo{author}{A.~Frankel}, \bibinfo{author}{G.~Iaccarino},
\newblock \bibinfo{title}{Efficient control variates for uncertainty
  quantification of radiation transport},
\newblock \bibinfo{journal}{Journal of Quantitative Spectroscopy and Radiative
  Transfer} \bibinfo{volume}{189} (\bibinfo{year}{2017})
  \bibinfo{pages}{398--406}.
\bibitem[{Jofre et~al.(2018)Jofre, Domino, and Iaccarino}]{Jofre2018-A}
\bibinfo{author}{L.~Jofre}, \bibinfo{author}{S.~P. Domino},
  \bibinfo{author}{G.~Iaccarino},
\newblock \bibinfo{title}{A framework for characterizing structural uncertainty
  in large-eddy simulation closures},
\newblock \bibinfo{journal}{Flow Turbulence and Combustion}
  \bibinfo{volume}{100} (\bibinfo{year}{2018}) \bibinfo{pages}{341--363}.
\bibitem[{Mir(2017)}]{Mira-O}
\bibinfo{title}{{M}ira, {A}rgonne {L}eadership {C}omputing {F}acility},
  \bibinfo{year}{2017}.
\bibitem[{Sutherland(1893)}]{Sutherland1893-A}
\bibinfo{author}{W.~Sutherland},
\newblock \bibinfo{title}{{LII}. the viscosity of gases and molecular force},
\newblock \bibinfo{journal}{Philosophical Magazine} \bibinfo{volume}{36}
  (\bibinfo{year}{1893}) \bibinfo{pages}{507--531}.
\bibitem[{Weast et~al.(1989)Weast, Astle, and Beyer}]{Weast1989-B}
\bibinfo{author}{R.~C. Weast}, \bibinfo{author}{M.~J. Astle},
  \bibinfo{author}{W.~H. Beyer}, \bibinfo{title}{Handbook of chemistry and
  physics}, \bibinfo{publisher}{CRC Press}, \bibinfo{year}{1989}.
\bibitem[{Maxey and Riley(1983)}]{Maxey1983-A}
\bibinfo{author}{M.~R. Maxey}, \bibinfo{author}{J.~J. Riley},
\newblock \bibinfo{title}{Equation of motion for a small rigid sphere in a
  nonuniform flow},
\newblock \bibinfo{journal}{Physics of Fluids} \bibinfo{volume}{26}
  (\bibinfo{year}{1983}) \bibinfo{pages}{883--889}.
\bibitem[{Esmaily et~al.(2018)Esmaily, Jofre, Mani, and
  Iaccarino}]{Esmaily2018-A}
\bibinfo{author}{M.~Esmaily}, \bibinfo{author}{L.~Jofre},
  \bibinfo{author}{A.~Mani}, \bibinfo{author}{G.~Iaccarino},
\newblock \bibinfo{title}{A scalable geometric multigrid solver for
  nonsymmetric elliptic systems with application to variable-density flows},
\newblock \bibinfo{journal}{Journal of Computational Physics}
  \bibinfo{volume}{357} (\bibinfo{year}{2018}) \bibinfo{pages}{142--158}.
\bibitem[{Yang and Hunt(2006)}]{Yang2006-A}
\bibinfo{author}{F.-L. Yang}, \bibinfo{author}{M.~L. Hunt},
\newblock \bibinfo{title}{Dynamics of particle-particle collisions in a viscous
  liquid},
\newblock \bibinfo{journal}{Physics of Fluids} \bibinfo{volume}{18}
  (\bibinfo{year}{2006}) \bibinfo{pages}{121506}.
\bibitem[{Brenner(1962)}]{Brenner1962-A}
\bibinfo{author}{H.~Brenner},
\newblock \bibinfo{title}{Effect of finite boundaries on the {S}tokes
  resistance of an arbitrary particle},
\newblock \bibinfo{journal}{Journal of Fluid Mechanics} \bibinfo{volume}{12}
  (\bibinfo{year}{1962}) \bibinfo{pages}{35--48}.
\bibitem[{Ganguli and Lele(2018{\natexlab{a}})}]{Ganguli2018a-A}
\bibinfo{author}{S.~Ganguli}, \bibinfo{author}{S.~K. Lele},
\newblock \bibinfo{title}{Drag of a heated sphere at low {R}eynolds numbers.
  {P}art {I}. {A}bsence of buoyancy},
\newblock \bibinfo{journal}{J. Fluid Mech.} \bibinfo{volume}{Under review}
  (\bibinfo{year}{2018}{\natexlab{a}}).
\bibitem[{Ganguli and Lele(2018{\natexlab{b}})}]{Ganguli2018b-A}
\bibinfo{author}{S.~Ganguli}, \bibinfo{author}{S.~K. Lele},
\newblock \bibinfo{title}{Drag of a heated sphere at low {R}eynolds numbers.
  {P}art {II}. {P}resence of buoyancy},
\newblock \bibinfo{journal}{J. Fluid Mech.} \bibinfo{volume}{Under review}
  (\bibinfo{year}{2018}{\natexlab{b}}).
\bibitem[{Farbar et~al.(2017)Farbar, Boyd, and Esmaily-Moghadam}]{Farbar2017-A}
\bibinfo{author}{E.~Farbar}, \bibinfo{author}{I.~D. Boyd},
  \bibinfo{author}{M.~Esmaily-Moghadam},
\newblock \bibinfo{title}{{M}onte {C}arlo modeling of radiative heat transfer
  in particle-laden flow},
\newblock \bibinfo{journal}{Journal of Quantitative Spectroscopy and Radiative
  Transfer} \bibinfo{volume}{184} (\bibinfo{year}{2017})
  \bibinfo{pages}{146--160}.
\bibitem[{Subramaniam(2013)}]{Subramaniam2013-A}
\bibinfo{author}{S.~Subramaniam},
\newblock \bibinfo{title}{{L}agrangian-{E}ulerian methods for multiphase
  flows},
\newblock \bibinfo{journal}{Progress in Energy and Combustion Science}
  \bibinfo{volume}{39} (\bibinfo{year}{2013}) \bibinfo{pages}{215--245}.
\bibitem[{Amsden et~al.(1989)Amsden, O'Rourke, and Butler}]{Amsden1989-TR}
\bibinfo{author}{A.~A. Amsden}, \bibinfo{author}{P.~J. O'Rourke},
  \bibinfo{author}{T.~D. Butler},
\newblock \bibinfo{title}{{KIVA}-{II}: a computer program for chemically
  reactive flows with sprays},
\newblock \bibinfo{journal}{LA-12503-MS, LANL}  (\bibinfo{year}{1989}).
\bibitem[{Gu and Eisenstat(1996)}]{Gu96}
\bibinfo{author}{M.~Gu}, \bibinfo{author}{S.~C. Eisenstat},
\newblock \bibinfo{title}{Efficient algorithms for computing a strong
  rank-revealing \uppercase{QR} factorization},
\newblock \bibinfo{journal}{SIAM Journal on Scientific Computing}
  \bibinfo{volume}{17} (\bibinfo{year}{1996}) \bibinfo{pages}{848--869}.
\bibitem[{Cheng et~al.(2005)Cheng, Gimbutas, Martinsson, and Rokhlin}]{Cheng05}
\bibinfo{author}{H.~Cheng}, \bibinfo{author}{Z.~Gimbutas},
  \bibinfo{author}{P.~G. Martinsson}, \bibinfo{author}{V.~Rokhlin},
\newblock \bibinfo{title}{On the compression of low rank matrices},
\newblock \bibinfo{journal}{SIAM Journal on Scientific Computing}
  \bibinfo{volume}{26} (\bibinfo{year}{2005}) \bibinfo{pages}{1389--1404}.
\bibitem[{Martinsson et~al.(2011)Martinsson, Rokhlin, and
  Tygert}]{Martinsson11}
\bibinfo{author}{P.~G. Martinsson}, \bibinfo{author}{V.~Rokhlin},
  \bibinfo{author}{M.~Tygert},
\newblock \bibinfo{title}{A randomized algorithm for the decomposition of
  matrices},
\newblock \bibinfo{journal}{Applied and Computational Harmonic Analysis}
  \bibinfo{volume}{30} (\bibinfo{year}{2011}) \bibinfo{pages}{47--68}.
\bibitem[{Hampton and Doostan(2015)}]{Hampton15a}
\bibinfo{author}{J.~Hampton}, \bibinfo{author}{A.~Doostan},
\newblock \bibinfo{title}{Compressive sampling of polynomial chaos expansions:
  Convergence analysis and sampling strategies},
\newblock \bibinfo{journal}{Journal of Computational Physics}
  \bibinfo{volume}{280} (\bibinfo{year}{2015}) \bibinfo{pages}{363--386}.
\bibitem[{Sobol'(2001)}]{Sobol01}
\bibinfo{author}{I.~M. Sobol'},
\newblock \bibinfo{title}{Global sensitivity indices for nonlinear mathematical
  models and their {M}onte {C}arlo estimates},
\newblock \bibinfo{journal}{Math. Comput. Simul.} \bibinfo{volume}{55}
  (\bibinfo{year}{2001}) \bibinfo{pages}{271--280s}.
\bibitem[{Sudret(2008)}]{Sudret08}
\bibinfo{author}{B.~Sudret},
\newblock \bibinfo{title}{Global sensitivity analysis using polynomial chaos
  expansions},
\newblock \bibinfo{journal}{Reliability Engineering \& System Safety}
  \bibinfo{volume}{93} (\bibinfo{year}{2008}) \bibinfo{pages}{964--979}.

\end{thebibliography}

\end{document}